\documentclass[11pt]{article}
\usepackage{arxiv}
\usepackage{xspace}

\newcommand{\rev}[1]{{#1}}
\newcommand{\reviewrev}[1]{{#1}}
\newcommand{\decisionrev}[1]{{#1}}
\usepackage{tikz}
\usetikzlibrary{arrows.meta,decorations.pathreplacing,positioning,calc}
\tikzset{
  query circuit/.style={
    x=1cm,y=1cm,line width=.55pt,font=\normalsize,
    every node/.style={outer sep=0pt}
  },
  query gate/.style={draw,fill=white,inner sep=0pt},
  algorithm gate/.style={
    query gate,minimum width=.95cm,minimum height=2.20cm
  },
  oracle gate/.style={
    query gate,minimum width=1.30cm,minimum height=2.20cm
  },
  classical output/.style={
    double distance=1.4pt,-{Latex[length=4pt,width=4pt]}
  },
  state cut/.style={densely dashed,draw=black!60,line width=.45pt},
  pics/query measurement/.style={code={
    \node[algorithm gate,minimum width=1.10cm] at (0,0) {};
    \draw (-.32,.08) arc[start angle=180,end angle=0,radius=.32];
    \draw[-{Latex[length=2.4pt,width=2.1pt]}]
      (0,.03) -- (.21,.47);
    \node at (0,-.48) {$\{\Pi_\rho\}$};
  }}
}
\newcommand{\querycut}[4]{%
  \draw[state cut] (#1,#2)--(#1,#3);
  \node[font=\small,anchor=north,inner sep=3pt] at (#1,#2) {$#4$};
}

\newcommand{\1}{\mathbf{1}}
\newcommand{\E}{\mathbb{E}}
\newcommand{\Z}{\mathbb{Z}}

\newcommand{\Supp}[1]{{\operatorname{Supp}\left( #1\right)}}
\newcommand{\norm}[1]{{\left\lVert #1 \right\rVert}}
\newcommand{\abs}[1]{{\left\lvert #1\right\rvert}}

\newcommand{\Tr}{{\mathrm{Tr}}}

\makeatletter
\DeclareRobustCommand{\problemname}[1]{%
  {\upshape
   \ifboolexpr{test {\ifdefstring{\f@series}{b}} or
               test {\ifdefstring{\f@series}{bx}}}
     {#1}{\scshape #1}}%
}
\makeatother
\newcommand{\kColl}{\problemname{\ensuremath{k}-to-one Testing}\xspace}
\newcommand{\kDist}{\problemname{Search \ensuremath{k}-Distinctness}\xspace}
\newcommand{\kFind}{\problemname{Promised \ensuremath{k}-to-one Finding}\xspace}
\newcommand{\permBucketFind}{\problemname{Permutation Bucket Finding}\xspace}
\title{Quantum Lower Bounds for Collision Problems}
\author{
 Pan Peng\thanks{School of Computer Science and Technology,
 University of Science and Technology of China.
 Email: \href{mailto:ppeng@ustc.edu.cn}{ppeng@ustc.edu.cn}.}
 \and
 Jingyu Wu\thanks{School of Computer Science and Technology,
 University of Science and Technology of China.
 Email: \href{mailto:wujingyu@mail.ustc.edu.cn}{wujingyu@mail.ustc.edu.cn}.}
}
\date{}
\hypersetup{pdftitle={Quantum Lower Bounds for Collision Problems},
 pdfauthor={Pan Peng and Jingyu Wu}}
\begin{document}
\pagenumbering{roman}
\maketitle
\begin{abstract}
We establish tight lower bounds on the adaptive and non-adaptive
quantum query complexities of \kFind on an exactly $k$-to-one input
string $x\in\mathbb Z_q^n$, for every fixed $k\ge2$ and every feasible
alphabet size $q\ge n/k$. The lower bounds are
$\Omega_k(n^{1/2-1/(2(2^k-1))})$ for unrestricted adaptive algorithms
and $\Omega_k(n^{1-1/k})$ for one-round non-adaptive algorithms.
The former matches the adaptive quantum algorithm of Liu and
Zhandry~\cite{liu2019finding}, whereas the latter matches classical
birthday sampling even when the parallel queries are arbitrarily
entangled. \decisionrev{We also establish} the non-adaptive lower bound
$\Omega(\sqrt{n/k})$ for \kColl when $q\ge n$, resolving an open
question posed by Koiran, Landes, Portier, and
Yao~\cite{koiran2010adversary}.

Our proofs generate the balanced input from a uniformly random
permutation and use compressed permutation oracles to describe the
information learned by the algorithm. The adaptive proof bounds the
growth of partial collisions through successive queries. The one-round
proof instead analyzes the state after the parallel queries directly,
bounding the probability that the compressed database contains a full
collision. In both proofs, a short verification step accounts for
correct outputs that include unrecorded positions. For the decision
problem, we also compare one-round states with independent and distinct
output labels.
\end{abstract}
\thispagestyle{empty}
\clearpage
\tableofcontents
\clearpage
\pagenumbering{arabic}

\section{Introduction}
\label{sec:introduction}

Collision problems are fundamental to
quantum query complexity. Given oracle access to a string
\[
 x=(x_1,\ldots,x_n)\in\mathbb Z_q^n,
\]
a \emph{$k$-collision} is a set of $k$ distinct indices
$i_1,\ldots,i_k$ such that $x_{i_1}=\cdots=x_{i_k}$.
Throughout this paper, $k\ge2$ is a fixed constant, $n$ is the
input length, and $q$ is the alphabet size.
The string partitions $[n]$ into blocks of equal-valued positions;
collision problems ask us to find these hidden blocks or distinguish
their possible structures.

One of the most basic versions of the collision problems is \kFind. Here $n=kN$, and the input
$x\in\mathbb Z_q^n$ is promised to be \emph{exactly $k$-to-one}:
every symbol that appears in $x$ appears exactly $k$ times.
Equivalently, its equality partition consists of $N$ blocks of size $k$.
The task is to output all $k$ positions in any one block.
A related decision problem, \kColl, asks whether the input is
injective or exactly $k$-to-one, under the promise that one of these
cases holds. Note that it suffices to find a pair of equal entries to rule out the
injective case, whereas \kFind requires a full block of $k$ positions.

Collision finding ($k=2$) and multicollision finding ($k\ge3$)
capture a basic difficulty in query computation:
learning one input value need not reveal where to find another occurrence.
These problems model attacks on hash functions and related cryptographic
constructions, where an adversary seeks several inputs with the same output
\cite{brassard1997quantum,joux2004multicollisions,liu2019finding,zhandry2019record}.
They have also motivated quantum algorithms and
lower-bound methods, including the polynomial method
\cite{aaronson2004quantum,kutin2005quantum,ambainis2005polynomial},
quantum walks \cite{ambainis2007quantum,magniez2007quantum},
learning graphs \cite{belovs2012learning}, adversary methods
\cite{belovs2018adversary}, and compressed-oracle and Fourier methods
\cite{zhandry2019record,belovs2026tight}.
The exact $k$-to-one promise lets us study full multicollisions when
many of them exist. One example is a truncated permutation: discarding $d$ output
bits gives an exactly $2^d$-to-one function\rev{, and the algorithm
sees only the truncated output}. Such truncation is used in
constructions of pseudorandom functions~\cite{hall1998building}.
\reviewrev{There is also a graph interpretation: draw an edge from
each position $i$ to its value $x_i$.}
A $k$-collision is then a set of $k$ edges with the same endpoint.
This relates collision finding to testing for directed stars
\cite{apers2025quantum,peng2027property}.

\paragraph{\rev{Query count and adaptivity.}}
Collision problems provide a natural setting for separating the benefits of
quantum computation from those of adaptivity. Classically, non-adaptive
sampling is already optimal even among adaptive randomized algorithms: for fixed $k$, finding a complete block of a
$k$-to-one input requires $\Theta_k(n^{1-1/k})$ queries, whereas distinguishing
an injective input from a $k$-to-one input requires
$\Theta(\sqrt{n/k})$ queries.
Appendix~\ref{app:classical-bounds} gives self-contained proofs.

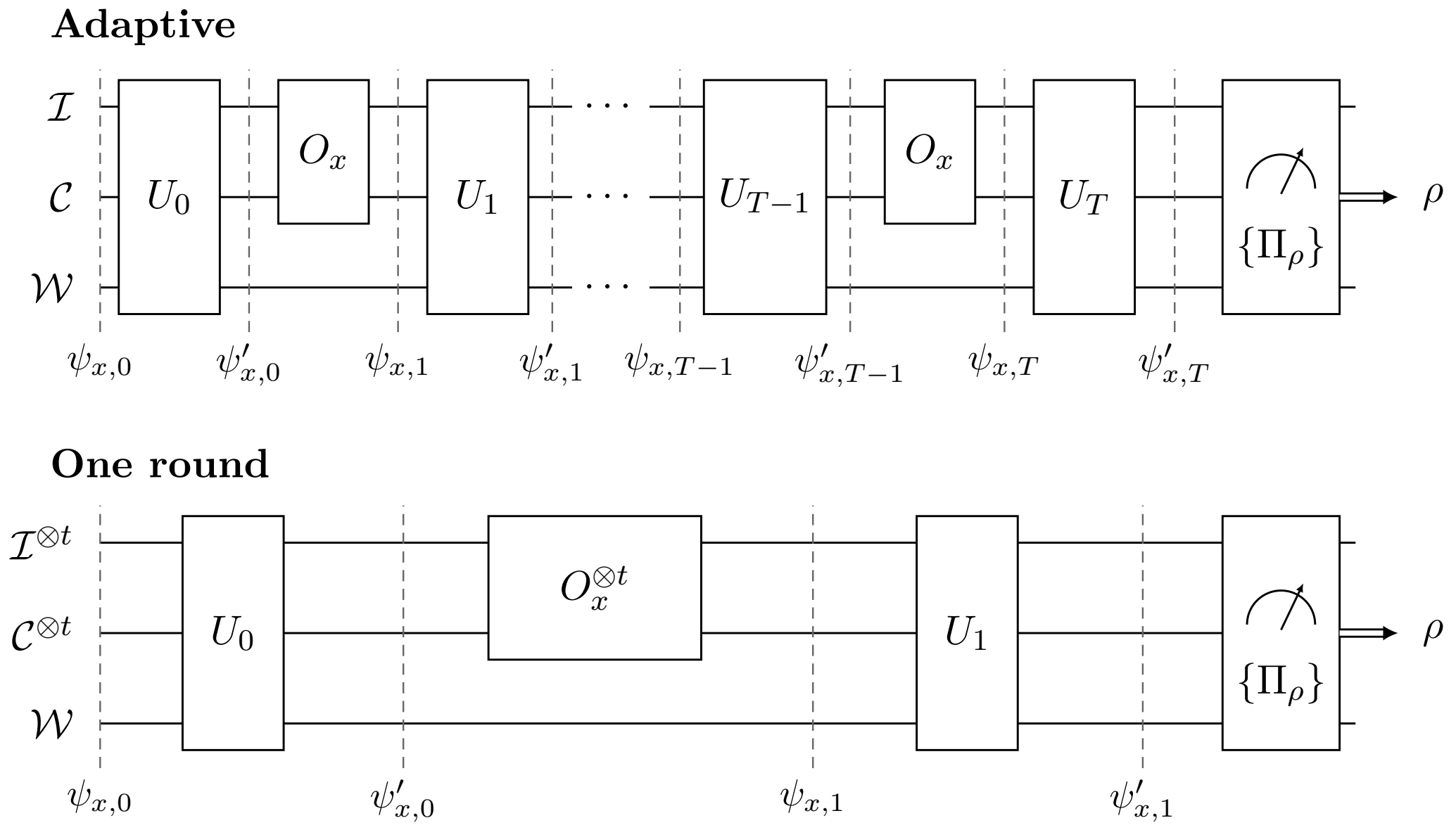
\begin{figure}[htbp]
\centering
\begin{tikzpicture}[query circuit]
 \foreach \y in {0,-.85,-1.70}
   \draw (0,\y)--(11.8,\y);
 \node[anchor=east] at (-.12,0) {$\mathcal I$};
 \node[anchor=east] at (-.12,-.85) {$\mathcal C$};
 \node[anchor=east] at (-.12,-1.70) {$\mathcal W$};
 \node[algorithm gate] at (.65,-.85) {$U_0$};
 \node[oracle gate,minimum width=.85cm,minimum height=1.35cm]
   at (2.1,-.425) {$O_x$};
 \node[algorithm gate] at (3.55,-.85) {$U_1$};
 \foreach \y in {0,-.85,-1.70}
   \node[fill=white,inner sep=3pt] at (4.8,\y) {$\cdots$};
 \node[algorithm gate,minimum width=1.15cm]
   at (6.25,-.85) {$U_{T-1}$};
 \node[oracle gate,minimum width=.85cm,minimum height=1.35cm]
   at (7.8,-.425) {$O_x$};
 \node[algorithm gate] at (9.25,-.85) {$U_T$};
 \pic at (11.1,-.85) {query measurement};
 \draw[classical output] (11.65,-.85)--(12.2,-.85);
 \node[anchor=west] at (12.3,-.85) {$\rho$};
 \node[anchor=west,font=\bfseries] at (-.6,.75) {Adaptive};
 \querycut{0}{-2.12}{.35}{\psi_{x,0}}
 \querycut{1.4}{-2.12}{.35}{\psi'_{x,0}}
 \querycut{2.8}{-2.12}{.35}{\psi_{x,1}}
 \querycut{4.25}{-2.12}{.35}{\psi'_{x,1}}
 \querycut{5.45}{-2.12}{.35}{\psi_{x,T-1}}
 \querycut{7.05}{-2.12}{.35}{\psi'_{x,T-1}}
 \querycut{8.5}{-2.12}{.35}{\psi_{x,T}}
 \querycut{10.1}{-2.12}{.35}{\psi'_{x,T}}
 \begin{scope}[yshift=-4.1cm]
 \foreach \y in {0,-.85,-1.70}
   \draw (0,\y)--(11.8,\y);
 \node[anchor=east] at (-.12,0) {$\mathcal I^{\otimes t}$};
 \node[anchor=east] at (-.12,-.85) {$\mathcal C^{\otimes t}$};
 \node[anchor=east] at (-.12,-1.70) {$\mathcal W$};
 \node[algorithm gate] at (1.25,-.85) {$U_0$};
 \node[oracle gate,minimum width=2cm,minimum height=1.35cm]
   at (4.65,-.425) {$O_x^{\otimes t}$};
 \node[algorithm gate] at (8.15,-.85) {$U_1$};
 \pic at (11.1,-.85) {query measurement};
 \draw[classical output] (11.65,-.85)--(12.2,-.85);
 \node[anchor=west] at (12.3,-.85) {$\rho$};
 \node[anchor=west,font=\bfseries] at (-.6,.75) {One round};
 \querycut{0}{-2.12}{.35}{\psi_{x,0}}
 \querycut{2.85}{-2.12}{.35}{\psi'_{x,0}}
 \querycut{6.7}{-2.12}{.35}{\psi_{x,1}}
 \querycut{9.8}{-2.12}{.35}{\psi'_{x,1}}
 \end{scope}
\end{tikzpicture}
\caption{The adaptive and non-adaptive models. Each $U_j$ may
act jointly on all accessible registers. An oracle acts on
the index and answer registers and leaves the workspace
unchanged. The non-adaptive layer contains $t$ simultaneous
calls, followed by an unrestricted joint measurement. Dashed vertical
cuts mark the states at the indicated stages, as in~\cite[Figures~3.1--3.2]{belovs2026tight}.}
\label{fig:query-models}
\end{figure}

For quantum algorithms, the relationship between query count and adaptivity
is more subtle. We distinguish the two models illustrated in
Figure~\ref{fig:query-models}. An adaptive algorithm may use information
acquired through earlier oracle calls to prepare later queries; this
dependence may remain entirely coherent and need not involve intermediate
measurements. A non-adaptive, or one-round, algorithm makes all its queries
in a single parallel layer. Its queries may be arbitrarily entangled, and
it may perform arbitrary quantum postprocessing and a joint measurement,
but it cannot use the returned information to prepare another query layer.
Thus, non-adaptivity restricts sequential oracle access, not the coherence
or computational power available within a query layer.

Known quantum collision-finding algorithms exploit precisely this sequential
dependence. Brassard, H{\o}yer, and Tapp~\cite{brassard1997quantum} gave an
adaptive quantum algorithm for $k=2$. For general $k$, Hosoyamada, Sasaki,
and Xagawa~\cite{hosoyamada2017quantum} developed a recursive algorithm that
constructs smaller collisions and extends them using quantum search, with
expected query complexity
$O_k(n^{1/2-1/(2\cdot3^{k-1})})$.
Liu and Zhandry~\cite{liu2019finding} improved this approach by constructing
lists of partial collisions and reusing each list as the target set for the
next stage.
Their algorithm uses
\[
    O_k(n^{\alpha_k})
    \qquad\text{queries, where}\qquad
    \alpha_k=\frac12-\frac{1}{2(2^k-1)}.
\]
These algorithms beat the classical birthday scale, but their successive
stages leave open how much of the improvement comes from quantum coherence
alone and how much requires adaptivity.

This raises two complementary lower-bound questions. First, \emph{is the
Liu--Zhandry query bound optimal under the exact $k$-to-one promise, even
with unrestricted quantum adaptivity?} Their matching lower bound
is proved for uniformly random functions \rev{(see \Cref{subsec:related-work})} and does not establish this
optimality for general $k$.
Resolving the adaptive complexity 
determines the largest possible quantum query advantage for this task,
rather than merely the performance of the known algorithms.

Second, \emph{can a single layer of entangled queries achieve any
asymptotic improvement over classical sampling?} Classical sampling gives a one-round upper bound of
$O_k(n^{1-1/k})$, but a quantum algorithm need not query a fixed classical
sample or explicitly observe a collision before producing its output.
A matching non-adaptive lower bound must therefore rule out all coherent
parallel strategies and joint measurements. Such a bound would show that
the sequential dependence in the known algorithms is essential, rather
than an artifact of their design, and quantify the query cost of eliminating
that dependence.

We determine both the adaptive and one-round non-adaptive quantum query complexities
under the exact $k$-to-one promise, for every feasible alphabet size.
Together, these results {identify} the optimal quantum speedup and show
that achieving it requires adaptivity. Our comparison with
\kColl further separates the cost of recovering a complete collision
from that of merely detecting the promised equality structure.

\subsection{Our Contribution}
\label{subsec:contribution}

We prove matching lower bounds for \kFind in the adaptive and
non-adaptive models (Theorems~\ref{thm:adaptive_find}
and~\ref{thm:find}, respectively). Both proofs represent information about a
random permutation by a compressed database of input--output pairs.
For adaptive algorithms, we bound how quickly queries can build larger
partial collisions. For one-round non-adaptive algorithms, we directly bound the
probability that the encoded final state contains a full collision.
This direct analysis avoids accumulating simulation error across all
parallel queries.

\begin{restatable}[Adaptive \kFind]{theorem}{adaptivefindingtheorem}
\label{thm:adaptive_find}
Let $k\ge2$ be fixed, $n=kN$, and $q\ge N$.
Every quantum
algorithm for \kFind that succeeds with constant probability
on every promised input in $\mathbb Z_q^n$ requires
\(
 \Omega_k\!\left(n^{{(2^{k-1}-1)}/{(2^k-1)}}\right)
\)
queries.
\end{restatable}

This is the exact-promise counterpart of the random-function
multicollision lower bound of Liu and Zhandry~\cite{liu2019finding}. \rev{Their lower bound does not extend to the exact $k$-to-one promise.} 
Their proof uses Zhandry's compressed-oracle method to bound the
accumulation of partial collisions and relies on the independence
of random-function values. Under the exact $k$-to-one promise,
every used value must occur exactly $k$ times, so the entries are
dependent. Their recording representation and single-query
amplitude bounds therefore do not directly apply. Moreover,
exactly $k$-to-one inputs form an exponentially small fraction of
all functions from $[n]$ to $[N]$, so conditioning on this event
does not directly transfer their average-case lower bound to our
setting. We handle these dependencies by generating balanced inputs
from a random permutation and tracking partial collisions with a
compressed permutation oracle.

With all queries restricted to one parallel layer, this gradual
construction of larger collisions is unavailable. The next theorem
gives a one-round non-adaptive lower bound that matches the cost of classical sampling. 

\begin{restatable}[Non-adaptive \kFind]{theorem}{nonadaptivefindingtheorem}
\label{thm:find}
Let $k\ge2$ be fixed, $n=kN$, and $q\ge N$. Every one-round quantum
algorithm for \kFind that succeeds with constant probability
on every promised input in $\mathbb Z_q^n$ requires
$\Omega_k(n^{1-1/k})$ parallel queries.
\end{restatable}

Theorem~\ref{thm:find} allows arbitrary entanglement across the entire
parallel layer, even at the smallest feasible alphabet size.
The proof controls the success of the final measurement,
including outputs containing positions that were not explicitly
queried in a classical sense.

\decisionrev{Our third main result shows that one-round quantum algorithms
for \kColl require as many queries as classical birthday sampling.}

\begin{restatable}[Non-adaptive \kColl]{theorem}{nonadaptivecollisiontheorem}
\label{thm:test}
Let $k\ge2$ be fixed, $k\mid n$, $q\ge n$, and
$0<\varepsilon<1/2$. Every one-round quantum algorithm for \kColl
that succeeds with probability at least $1/2+\varepsilon$ on every
promised input in $\mathbb Z_q^n$ requires $\Omega(\varepsilon\sqrt{n/k})$ parallel queries.
\end{restatable}

For $k=2$, Theorem~\ref{thm:test} answers the non-adaptive collision
question posed by Koiran, Landes, Portier, and
Yao~\cite{koiran2010adversary}. Their weighted adversary bounds treat
a parallel layer as a single query to a tuple of input positions.
Our proof compares the algorithm's reduced states under independent
and distinct block labels, allowing $q$ to be as small as $n$.

\paragraph{Tightness.}
All three lower bounds are tight for constant success probability
(constant bias for \kColl)\rev{, as summarized in} Table~\ref{tab:results}.
For Theorem~\ref{thm:adaptive_find}, the \rev{Liu--Zhandry algorithm described
above} uses $O_k(n^{\alpha_k})$ adaptive queries\rev{~\cite{liu2019finding}}.
Its cost depends on the $N$ symbols actually used, \rev{rather than the ambient
alphabet size,} so the bound holds for every $q\ge N$.

\rev{The matching one-round upper bounds are classical: sampling}
$O_k(n^{1-1/k})$ \rev{positions suffices to find a full} $k$-collision\rev{, while
sampling} $O(\sqrt{n/k})$ \rev{positions and checking for an equal pair
suffices to distinguish the two promise} cases with constant bias.
\rev{Both algorithms query uniformly chosen positions; see
Appendix~\ref{app:classical-bounds}}.

The alphabet conditions are minimal: \kFind requires $q\ge n/k$,
and \kColl requires $q\ge n$ for the injective case to exist.
Theorem~\ref{thm:test} allows variable bias, although we do not
optimize the dependence on $\varepsilon$.

\begin{table}[t!]
\centering
\small
\setlength{\tabcolsep}{4pt}
\renewcommand{\arraystretch}{1.15}
\begin{threeparttable}
\begin{tabular}{@{}lcccc@{}}
\toprule
\textbf{Problem}
& \textbf{Alphabet}
& \makecell{\textbf{Non-adaptive}\\\textbf{quantum}}
& \makecell{\textbf{Non-adaptive}\\\textbf{\& adaptive}\\\textbf{classical}}
& \makecell{\textbf{Adaptive}\\\textbf{quantum}}\\
\midrule
\parbox[c]{.27\textwidth}{\raggedright\kFind}
& $q\ge n/k$
& \makecell{$\boldsymbol{\Theta_k(n^{1-1/k})}$\\Theorem~\ref{thm:find}}
& $\Theta_k(n^{1-1/k})$
& \makecell{$\boldsymbol{\Theta_k(n^{\alpha_k})}$\nobreak\hspace{.12em}\tnote{\rev{$\dagger$}}\\Theorem~\ref{thm:adaptive_find}}\\
\midrule
\parbox[c]{.27\textwidth}{\raggedright\kColl}
& $q\ge n$
& \makecell{$\boldsymbol{\Theta(\sqrt{n/k})}$\\Theorem~\ref{thm:test}}
& $\Theta(\sqrt{n/k})$
& $\Theta_k((n/k)^{1/3})$\nobreak\hspace{.12em}\tnote{\rev{$\ddagger$}}\\
\bottomrule
\end{tabular}
\begin{tablenotes}[flushleft]
\footnotesize
\item[\rev{$\dagger$}] The adaptive upper bound for \kFind is from
\cite{liu2019finding}.
\item[\rev{$\ddagger$}] The adaptive bounds for \kColl are from
\cite{brassard1997quantum,aaronson2004quantum,kutin2005quantum,ambainis2005polynomial}.
\end{tablenotes}
\caption{Constant-error query complexity for fixed $k$.
Here $\alpha_k=\frac12-\frac1{2(2^k-1)}$.
Bold entries are the tight complexities established by this paper's
lower bounds together with the matching algorithms described above.
The classical bounds are the same with and without
adaptivity.}
\label{tab:results}
\end{threeparttable}
\end{table}

\subsection{Technical Overview}
\label{subsec:technical-overview}

We first explain the adaptive lower bound: we construct the hard
distribution, bound the growth of recorded collisions \rev{using compressed permutation oracle techniques~\cite{carolan2026compressed}}, and relate
these records to successful outputs. We then use the additional
structure of a single query round to prove the non-adaptive finding
bound, and conclude with the decision bound for \kColl.
Throughout the overview, $k\ge2$ is fixed and $n=kN$.

\subsubsection{From balanced inputs to \permBucketFind}
To prove the adaptive lower bound for \kFind, we reduce the analysis
to \permBucketFind. Intuitively, a uniformly random permutation $\pi$
generates a balanced $k$-to-one input as follows. Group the permutation's
output slots into public buckets $B_1,\ldots,B_N$ of size $k$, and
assign a distinct input symbol to each bucket. The input value
$x_i^\pi$ is the symbol of the bucket containing $\pi(i)$.
Every symbol then appears exactly $k$ times, and a full $k$-collision
consists of the $k$ input positions whose images under $\pi$ fill
one bucket. Finding these positions is precisely the task of
\permBucketFind (Definition~\ref{def:permutation-bucket-finding}).

To simulate one
input query, use a forward permutation query to compute $\pi(i)$,
apply the phase determined by its bucket label, and use a
second query to uncompute $\pi(i)$. Reversing the forward query
has the same query cost and does not require evaluating $\pi^{-1}$.
Thus a $T_0$-query algorithm for \kFind gives a
$2T_0$-query algorithm for \permBucketFind with the same average
success probability (Lemma~\ref{adp:reduction}). Access to
$\pi^{-1}$ is excluded: $k$ inverse queries would find all positions
mapping to a public bucket. For the one-round lower bound, we
analyze the original input queries directly to preserve the number
of query layers.

\subsubsection{Recording information about a random permutation}

The reduction lets us study queries to a uniform permutation using
recording techniques. Zhandry's compressed oracle~\cite{zhandry2019record}
records information about a random function in a small database.
Carolan~\cite{carolan2026compressed} constructs a unitary recorder
for permutations and bounds its difference from the genuine oracle.
We use his forward-query construction. A permutation database assigns
distinct output slots to its recorded positions, so the available
values at one position depend on the assignments elsewhere.

We apply a projection after each forward recording query to obtain
total error $O(T^2/n)$ after $T$ queries. This projection acts within
each subspace with a fixed set of recorded output slots. It can
reduce the probability mass in a bucket-occupancy class, but cannot
move mass between classes. This domain-harmonic projection is defined
below. Let $\varrho_T$ and
$\widetilde\varrho_T$ be the reduced states on the algorithm register
$\mathcal A$ for the genuine and projected recording processes. Then
\[
 \|\varrho_T-\widetilde\varrho_T\|_1
 =O\!\left(\frac{T(T+1)}n\right)
 \qquad(T+3\le n/8).
\]
The projected process may have norm less than one, and
$\operatorname{Tr}\widetilde\varrho_T$ may therefore be less than one.
We never renormalize it. Lemma~\ref{bc:simulation} proves the stronger
joint-state estimate in Eq.~\eqref{bc:simulation-bound}. 

\subsubsection{The database encoding and query comparison}

To analyze the information learned from the permutation oracle,
we introduce an auxiliary database register, initialized to the
empty database. Let $\mathcal P=\mathbb C[S_n]$ be the genuine permutation register,
and let $\mathcal I_n$ be the set of partial injections
$d:[n]\rightharpoonup[n]$. The auxiliary register is
\[
 \mathcal D=\mathbb C[\mathcal I_n]
   =\bigoplus_{s=0}^n\mathcal D_s,\qquad
 \mathcal D_s=\operatorname{Span}\{|d\rangle:|d|=s\},\qquad
 \mathcal D_{\le t}=\bigoplus_{s\le t}\mathcal D_s.
\]
Here $|d|$ is the number of assigned positions. For a residual
database $r$, an unassigned position $i\notin\operatorname{Dom}r$,
and an unused output slot $y\notin\operatorname{Im}r$, define
the two uniform extensions
\[
 |u_{i,r}\rangle
 =\frac1{\sqrt{n-|r|}}
       \sum_{z\notin\operatorname{Im}r}|r\cup\{i\mapsto z\}\rangle,
 \qquad
 |v_{y,r}\rangle
 =\frac1{\sqrt{n-|r|}}
       \sum_{z\notin\operatorname{Dom}r}|r\cup\{z\mapsto y\}\rangle.
\]
The unitary $C_i:\mathcal D\to\mathcal D$ swaps
$|r\rangle$ and $|u_{i,r}\rangle$ for each residual database $r$
not assigning $i$, and fixes the orthogonal complement of the span
of all these pairs. For a fixed phase-query charge,
$D_i:\mathcal D\to\mathcal D$ applies the query phase to the value
assigned at $i$ and fixes databases not assigning $i$. The local recorder is
$\mathcal O_i=C_iD_iC_i:\mathcal D\to\mathcal D$, as in the compressed
permutation oracles of~\cite{carolan2026compressed}
(Definitions~\ref{def:local-compression} and~\ref{def:auxiliary-query}).
It changes database size by
at most one, but may also erase or replace an assignment.

To compare this database process with genuine queries, we construct
an encoding $V:\mathcal P\to\mathcal D$ that preserves inner products.
It places permutation information of degree $s$ in databases of
size $s$. Following~\cite{rosmanis2021tight}, we define the degree-$s$
subspace $\mathcal P_s$ as the sum of representation types indexed
by partitions $\lambda\vdash n$ with first-row length $\lambda_1=n-s$
(Definition~\ref{def:permutation-degree}). Each query
increases degree by at most one. The completion map
$A_s:\mathcal D_s\to\mathcal P$ and encoding
$V_s:\mathcal P_s\to\mathcal D_s$ are
\[
 A_s|d\rangle=\frac1{\sqrt{(n-s)!}}
                   \sum_{\pi\supseteq d}|\pi\rangle,\qquad
 V_s=A_s^\dagger\Gamma_n^{-1/2}|_{\mathcal P_s},\qquad
 V=\bigoplus_{s=0}^{n-1}V_s.
\]
Here $\Gamma_n:\mathcal P\to\mathcal P$ is positive and equals
$A_sA_s^\dagger$ on $\mathcal P_s$. The normalization corrects the
overlaps between completion states for different domains.

Let $K_R,K_D:\mathcal D\to\mathcal D$ be the commuting projectors
onto the range-harmonic and domain-harmonic subspaces of
Definition~\ref{def:harmonic-spaces}. At size $s$, they retain
degree $s$ under range and domain relabelling, respectively, and
remove the lower-degree components. The number of database entries
alone can overstate the amount of information represented.
For a residual database $r$ of size $s-1$, uniformly filling one
more position averages over all its possible values; uniformly
placing a fixed unused value averages over all its possible
preimages. Both uniform extensions have exactly the same genuine completion
state as $r$ (Eq.~\eqref{eq:redundant-completions}), so their extra
entry carries no additional information. We use the domain projection
to remove part of this redundancy and reduce the simulation error.

Concretely, for $1\le s<n/2$, $i\notin\operatorname{Dom}r$, and
$y\notin\operatorname{Im}r$, the projectors annihilate
\[
 K_R|u_{i,r}\rangle=0,\qquad
 K_D|v_{y,r}\rangle=0.
\]
Their kernels at size $s$ are spanned by these uniform extensions.
Equivalently, each projection retains the vectors whose coefficients
sum to zero over the corresponding extensions. 
Lemma~\ref{bc:encoding}
gives
\[
 V^\dagger V=I_{\mathcal P},\qquad VV^\dagger=K_RK_D,\qquad
 V\left(\frac1{\sqrt{n!}}\sum_{\pi\in S_n}|\pi\rangle_{\mathcal P}\right)
       =|\emptyset\rangle_{\mathcal D}.
\]
Thus the uniform permutation state is encoded as the empty database,
and $V$ maps each degree to its corresponding database size while
satisfying both harmonic conditions. Local recording
queries can leave this encoded subspace. We project with $K_D$
to remove the error outside the domain-harmonic subspace while
preserving each set of recorded values (Lemma~\ref{bc:domain-images}).
This preserves the bucket-occupancy structure. Applying
$K_R$ as well would generally mix different image sets; instead,
we bound the remaining error outside the range-harmonic subspace.
The range zero-sum identities of the exact encoding also underpin
the one-round occupancy bound below.

Write $\mathsf O_i:\mathcal P\to\mathcal P$ for the genuine query
at the same position and charge, and let
$\mathcal P_{\le t}=\bigoplus_{s\le t}\mathcal P_s$ be the subspace
of permutation degree at most $t$.
On this subspace, the projected and unprojected comparisons satisfy,
for $t+3\le n/8$,
\[
\begin{aligned}
 \|K_D\mathcal O_iV-V\mathsf O_i\|_{\mathcal P_{\le t}}
     \le C(t+1)/n, \qquad
 \|\mathcal O_iV-V\mathsf O_i\|_{\mathcal P_{\le t}}
     \le C\sqrt{(t+1)/n}.
\end{aligned}
\]
Here and below $C$ is an absolute constant.
The projected estimate combines two error bounds: one from controlling
the component outside the range-harmonic subspace, and the other from
comparing local decompression with exact conditioning on the queried
permutation value (Lemma~\ref{bc:comparison}).

Let $\mathsf O$ and $\mathcal O$ denote the
controlled queries on $\mathcal P\otimes\mathcal A$ and
$\mathcal D\otimes\mathcal A$, respectively, and let
$U_t:\mathcal A\to\mathcal A$ be the algorithm-only unitary
preceding query $t+1$. Start the genuine
state $\psi_0$ with the uniform permutation and the auxiliary state
$\phi_0$ with the empty database, both tensored with
$|0\rangle_{\mathcal A}$. The processes in
Eqs.~\eqref{eq:genuine-process}, \eqref{eq:auxiliary-process},
and~\eqref{eq:projected-query-schedule} satisfy
\[
 \psi_{t+1}=\mathsf O U_t\psi_t,\qquad
 \phi_{t+1}=K_D\mathcal O U_t\phi_t,\qquad
 \|\phi_T-(V\otimes I_{\mathcal A})\psi_T\|
       \le C\frac{T(T+1)}n.
\]
Identity operators on other registers are suppressed. The distance
bound follows by summing the single-query errors. Each projected
operation has norm at most one, so it does not amplify previous errors.
The auxiliary state $\phi_t$ may have norm less than one.

\subsubsection{Bounding the probability of recording a full bucket}

To turn the simulation estimate into a collision lower bound,
we must bound the probability of recording a full bucket.
We adapt Liu and Zhandry's random-function multicollision
argument~\cite[Section~4]{liu2019finding} by tracking the number of
buckets with at least $j$ recorded slots, for each $1\le j\le k$.
Each bucket has exactly $k$ slots, and each slot can be recorded
only once. A recording query
may also erase an assignment or replace it by one in another
bucket. We account for these changes by bounding how much amplitude
can cross a threshold in one query. Applying $K_D$ cannot increase
this bound because it acts separately on each set of recorded
values. Define
\[
 L_j(d)=|\{y\in[N]:|\operatorname{Im}d\cap B_y|\ge j\}|
 \quad(1\le j\le k),\qquad P_{\rm full}=P_{k,\ge1},
\]
where $P_{j,\ge\ell},P_{j,=\ell},P_{j,\le\ell}:\mathcal D\to\mathcal D$
are the diagonal projectors for the corresponding conditions on
$L_j$. Thus $P_{\rm full}$ tests whether some bucket has all its
slots recorded. Lemma~\ref{adp:growth} proves, for $2\le j\le k$,
$\ell\ge1$, $R\ge0$, and $T<n$,
\[
 \|P_{j,\ge\ell}\mathcal O P_{j,=\ell-1}P_{j-1,\le R}\|
       _{\mathcal D_{\le T}\otimes\mathcal A}
 \le2\sqrt{\frac{2kR}{n-T}}.
\]
At most $R$ buckets can be extended from $j-1$ to $j$ recorded slots.
They offer at most $kR$ choices among at least $n-T$ available slots.
The amplitude increase is therefore bounded by a constant times
the square root of this fraction. One query increases $L_j$ by at most one. Since $K_D$
commutes with the projectors for every value of $L_j$, the bound
also holds after applying the projection.

Choose the cutoffs and growth parameters
\[
 R_j=\left\lceil N^{(2^{k-j}-1)/(2^k-1)}\right\rceil
 \quad(1\le j\le k-1),\qquad
 \kappa_j=2\sqrt{\frac{2kR_{j-1}}{n-T}}
 \quad(2\le j\le k).
\]
For the analysis, discard components with $L_j>R_j$ at intermediate
levels. Repeatedly applying the crossing bound controls the total
discarded amplitude by a binomial tail. Below these cutoffs, the
full-bucket amplitude increases by at most $\kappa_k$ per query.
When $T\le R_1$, Lemma~\ref{adp:hierarchy} and
Claim~\ref{bc:hierarchy-projected} therefore give
\[
 \|P_{\rm full}\phi_T\|
 \le T\kappa_k+
       \sum_{j=2}^{k-1}\binom{T}{R_j+1}\kappa_j^{R_j+1}.
\]
Set $\alpha_k=(2^{k-1}-1)/(2^k-1)$. For
$T\le cN^{\alpha_k}$, the exponents ensure
$T\kappa_j\le4\sqrt2cR_j$ at intermediate levels and
$T\kappa_k\le4\sqrt2c$ at the last level. Thus, for sufficiently
small $c>0$ and large $N$, Lemma~\ref{adp:aux-theorem}~yields
\[
 \|P_{\rm full}\phi_T\|
 \le4\sqrt2c+(k-2)2^{-N^{1/(2^k-1)}}.
\]
These cutoffs limit the number of partial collisions available for
extension. The discarded amplitude is exponentially small, and the
remaining $(k-1)$-collisions cannot produce a full bucket with
constant probability in this query range. The growth bounds apply
to both projected and unprojected recording queries; we use the
latter for the verification checks below.

\subsubsection{Verification gives a common success bound}

A correct output need not already be fully recorded in the
database: the algorithm may guess some of its positions.
A common extraction lemma accounts for these possible guesses.
Let $\psi$ be a unit state of permutation degree at most $t$,
with a proposed tuple stored in the algorithm register, and let
$p_{\rm succ}$ be its success probability for \permBucketFind.
When $t+k+3\le n/8$, Lemma~\ref{lem:bucket-extraction} gives
\[
 \sqrt{p_{\rm succ}}
 \le \|(P_{\rm full}\otimes I_{\mathcal A})(V\otimes I_{\mathcal A})\psi\|
       +C_k\sqrt{\frac{t+k+1}{n}}.
\]
The proof appends $k$ queries to check the proposed positions, solely
for the analysis. Starting from the exact encoding $V\psi$, we compare
these checks with auxiliary queries. Lemma~\ref{bc:verification}
relates their accepting component to a complete recorded bucket,
up to a small error. The one-query growth bound then controls how
much full-bucket amplitude the checks can add. The resulting
inequality refers to the state before verification and applies
to both query models. This adapts the standard output-to-database
argument for compressed oracles~\cite{zhandry2019record}; for
permutations, see~\cite[Lemmas~3.3 and~7.1]{carolan2026compressed}.

For the adaptive algorithm with $T$ permutation queries,
the projected simulation therefore gives
\[
 \sqrt{p_{\rm succ}}
 \le \|(P_{\rm full}\otimes I_{\mathcal A})\phi_T\|
       +O_k\!\left(\frac{T(T+1)}n
                    +\sqrt{\frac{T+k+1}{n}}\right).
\]
For $T\le cN^{\alpha_k}$, the hierarchy bounds the first term
by $O(c)+o(1)$, and the remaining terms vanish because
$\alpha_k<1/2$. Choosing $c$ sufficiently small in terms of the
target success probability rules out such an algorithm. The reduction
has $T=2T_0$, so $T_0=\Omega_k(N^{\alpha_k})$, proving
Theorem~\ref{thm:adaptive_find}.

\subsubsection{One-round orthogonality and a static occupancy bound}

The reduction used in the adaptive proof does not preserve the
one-round restriction. It replaces each query to $x^\pi$ by two
sequential queries to $\pi$: one computes $\pi(i)$, and the other
uncomputes it after applying the phase determined by $x_i^\pi$.
Thus a single parallel input-query layer becomes two successive
permutation-query layers.

We therefore analyze the original parallel queries directly.
A uniform $\pi$ still generates the balanced hard distribution, so
we can purify the input in $\mathcal P$ and apply the exact encoding
$V$. The queries remain the original $O_{x^\pi}$ from
Eq.~\eqref{eq:oracle}, whose phases depend on the bucket labels
$x_i^\pi$. This change of representation preserves the single query layer.

A single query round also provides additional structure.
We use the purified Fourier description of query information
underlying~\cite{zhandry2019record} and presented for general input
distributions in~\cite[Section~3]{belovs2026tight}.
Lemma~\ref{lem:one-round-phases} shows that, under any input
distribution, a single layer of $t$ queries has an expansion with
orthogonal algorithm vectors. In our permutation representation,
the joint state takes the form
\[
 \psi=\sum_a h_a\otimes u_a,\qquad
 h_a(\pi)=\prod_{i\in S_a}
          \omega_q^{c_{a,i}x_i^\pi},\qquad |S_a|\le t.
\]
Here $\omega_q=e^{2\pi\imath/q}$, $c_{a,i}\in\mathbb Z_q\setminus\{0\}$, and
a function $h$ on $S_n$ denotes the vector
$n!^{-1/2}\sum_{\pi}h(\pi)|\pi\rangle_{\mathcal P}$.
The vectors $u_a\in\mathcal A$ satisfy
$\langle u_a,u_{a'}\rangle=0$ for $a\ne a'$ and
$\sum_a\|u_a\|^2=1$. A common final unitary preserves this
orthogonality. Each $h_a$ depends on one fixed set $S_a$ of at most
$t$ positions, and database probabilities add over $a$. We use
this orthogonality before conditioning on any successful output.
A further query layer need not preserve it.

We next bound the probability that the encoded state records a
prescribed set of $k$ values. For $B\subseteq[n]$ with $|B|=k$, let
$P_B:\mathcal D\to\mathcal D$ project onto databases whose image
contains $B$. Suppose $n\ge2k$, $0\le t<n/2$, and $h$ depends on
the permutation at a fixed set of at most $t$ positions, with
$\|h\|_\infty\le1$. Lemma~\ref{lem:static} gives the static
occupancy bound
\[
 \|P_BVh\|^2\le C_k\left(\frac{t}{n-t+1}\right)^k.
\]
Here $C_k$ depends only on $k$. The bound has the same order as
in uniform sampling without replacement: each prescribed slot
contributes a factor of order $t/n$, even though $Vh$ is a quantum
superposition of databases.

To prove it, write $B=\{u_1,\ldots,u_k\}$ and let
$L_{(uv)}:\mathcal P\to\mathcal P$ exchange range labels $u,v$,
so $L_{(uv)}|\pi\rangle=|(uv)\circ\pi\rangle$.
The zero-sum identity lets us bound the amplitude of recording a
target value using differences obtained by exchanging it with
other values. Repeating this for all $k$ targets gives
Claim~\ref{lem:diff}. Since a function of at most $t$ fixed
positions has permutation degree at most $t$, it yields
\[
 \|P_BVh\|^2\le\frac1{(n-t+1)^k}
 \sum_{\substack{v_j\in[n]\setminus\{u_j,\ldots,u_k\}\\1\le j\le k}}
 \|(I_{\mathcal P}-L_{(u_kv_k)})\cdots(I_{\mathcal P}-L_{(u_1v_1)})h\|^2.
\]
The new combinatorial step, formalized in
Claim~\ref{clm:forest-differences}, organizes these mixed differences
by the graph with edges $\{u_j,v_j\}$. Each target $u_j$ is joined
to an earlier target or to a label outside $B$, so the graph is a
forest with exactly one auxiliary label in each component
(Figure~\ref{fig:difference-forest}). A pattern with $r$ components
has $O_k(n^r)$ assignments of auxiliary labels.
Its mixed difference vanishes unless the queried images meet
every component, an event of probability $O_k((t/n)^r)$ under a
uniform permutation. The difference has magnitude at most $2^k$,
and there are only $O_k(1)$ patterns. Hence, for $t\ge1$, the sum
in the numerator is~at~most
\[
 C_k\sum_{r=1}^k n^r(t/n)^r
   =C_k\sum_{r=1}^k t^r\le C'_k t^k,
\]
which proves the occupancy bound. 

Summing over the $N$ buckets and using the orthogonality of $u_a$
now gives
\[
\begin{aligned}
 \|(P_{\rm full}\otimes I_{\mathcal A})
                     (V\otimes I_{\mathcal A})\psi\|^2
 =\sum_a\|u_a\|^2\,\|P_{\rm full}Vh_a\|^2
 \le C_kN\left(\frac{t}{n-t+1}\right)^k
   =O_k\!\left(\frac{t^k}{n^{k-1}}\right).
\end{aligned}
\]

Every branch depends on at most $t$ permutation values, so
$\psi$ has permutation degree at most $t$. Coherently storing the
output tuple acts only on the algorithm register and preserves
this degree bound. The common success bound,
Lemma~\ref{lem:bucket-extraction}, therefore gives
\[
 \sqrt{p_{\rm succ}}\le
 \|(P_{\rm full}\otimes I_{\mathcal A})
                    (V\otimes I_{\mathcal A})\psi\|
       +C_k\sqrt{\frac{t+k+1}{n}},\qquad t+k+3\le n/8.
\]
Combining the estimates and squaring yields
\[
 p_{\rm succ}\le
 C_k\left(\frac{t^k}{n^{k-1}}+\frac{t+1}{n}\right).
\]
Thus constant success requires $t=\Omega_k(n^{1-1/k})$.
The complementary regime $t+k+3>n/8$ already implies a linear
query lower bound. This proves Theorem~\ref{thm:find}, matching
classical birthday sampling.

The one-round hypothesis is used only in the occupancy estimate
before verification. The added checks may depend on the output,
so the extended circuit need not be one-round. Only these $k$
checks are simulated; the original parallel layer is analyzed
through its exact encoding. The resulting error remains small
at $n^{1-1/k}$ queries, even when $k>2$ and this number exceeds $\sqrt n$.

\subsubsection{\decisionrev{The non-adaptive lower bound for \kColl}}

Fix a $t$-query one-round algorithm, and write
$\varrho_{\mathsf{S}}$ for its reduced post-query state under
input distribution $\mathsf{S}$. \decisionrev{Following the relaxed-model
viewpoint of~\cite{belovs2026tight}, we first allow labels to repeat;
we then use the same harmonic encoding as in the finding proofs to
impose distinct labels while controlling the change in the algorithm's
state.}

\paragraph{1. Comparing the relaxed distributions.}
Let $\mathsf{Fun}$ have independent uniform entries in
$\mathbb Z_q^n$, and let $\mathsf{Block}_k$ assign independent
uniform labels to the blocks of a uniformly chosen partition
of $[n]$ into sets of size $k$. Different blocks
may receive the same label. We work with $t\le\sqrt{n/(8k)}$.
The one-round decomposition has orthogonal algorithm vectors
$u_\sigma$, with $|\operatorname{Supp}\sigma|\le t$ and
$\sum_\sigma\|u_\sigma\|^2=1$.
Independent labels eliminate cross terms, giving
$\varrho_{\mathsf{Fun}}=\sum_\sigma u_\sigma u_\sigma^\dagger$.
For a hidden partition $\mu$, cross terms instead survive when
the block charges
$\tau_{\mu,\sigma}(B)=\sum_{i\in B}\sigma(i)\pmod q$ agree.

A record is good if its support meets each block at most once.
For any fixed support, the probability over the hidden partition
that the record is bad is at most
$\beta:=\binom t2(k-1)/(n-1)$.
Retaining only good records in a purification gives a
subnormalized reduced state $\varrho_{\mathsf{G}}$.
Orthogonality bounds the discarded squared norm by $\beta$, hence
\[
 \|\varrho_{\mathsf{Block}_k}-\varrho_{\mathsf{G}}\|_1
 \le2\sqrt\beta.
\]
Good records can still interfere when they assign the same charges
to different positions within the same blocks. We group pairs by
the number $m$ of charges placed at different positions.
Matching these charges within hidden blocks has probability at
most $m!(2k/n)^m$. Inclusion--exclusion and orthogonality
(Claim~\ref{lem:direct-disjointness}) bound the resulting
off-diagonal contribution by
$\sum_{m=1}^t(4kt^2/n)^m\le8kt^2/n$.
Together with diagonal loss at most $\beta$, this gives
\[
 \|\varrho_{\mathsf{G}}-\varrho_{\mathsf{Fun}}\|_1
 \le\beta+\frac{8kt^2}{n}.
\]
The triangle inequality yields Lemma~\ref{lem:direct-relaxed-decision};
using $\beta\le kt^2/n$ and the stated range of $t$, we obtain
\[
 \|\varrho_{\mathsf{Block}_k}-\varrho_{\mathsf{Fun}}\|_1
 \le2\sqrt\beta+\beta+\frac{8kt^2}{n}
 \le8t\sqrt{\frac kn}.
\]

\paragraph{2. Harmonic switching to distinct labels.}
For $q\ge n$, let $\mathsf{Inj}$ and $\mathsf{Exact}_k$ be the
uniform distributions on injective and exactly $k$-to-one inputs.
These distributions replace the independent labels on individual
positions or blocks by uniformly chosen pairwise distinct labels.
When $q$ is only linear in $n$, conditioning all labels to be distinct
can substantially change the input distribution. We instead bound
the change in the algorithm's state: for $t\le q/8$,
Lemma~\ref{lem:switch} and Corollary~\ref{cor:hidden-partition-switching}
give
\[
 \frac12\|\varrho_{\mathsf{Inj}}-\varrho_{\mathsf{Fun}}\|_1
 \le\frac{4t}{\sqrt q}, \qquad
 \frac12\|\varrho_{\mathsf{Exact}_k}-\varrho_{\mathsf{Block}_k}\|_1
 \le\frac{4t}{\sqrt q}.
\]
To prove these bounds, extend the injective label assignment to a
permutation on $q$ values. For a Fourier branch with support size
$s\le t$, compare its harmonic encoding $Vh_\sigma$ with the
random-function record $a_{0,\sigma}$. Restricting that record to
partial injections and normalizing changes it by at most
$\sqrt{s(s-1)/q}$ in Euclidean norm; harmonic projection and frame
normalization contribute at most another $3\sqrt{s/q}$
(Claims~\ref{lem:fixed} and~\ref{lem:vector}). Thus
$\|Vh_\sigma-a_{0,\sigma}\|\le4s/\sqrt q$.
Orthogonality combines these branch errors into
\[
 \left\|\sum_\sigma(Vh_\sigma-a_{0,\sigma})\otimes u_\sigma\right\|^2
 =\sum_\sigma\|Vh_\sigma-a_{0,\sigma}\|^2\|u_\sigma\|^2
 \le\frac{16t^2}{q}.
\]
Tracing out the database gives the switching bound.
For hidden blocks, aggregate the charges into $\tau_{\mu,\sigma}$,
whose support still has size at most $t$. The original vectors
$u_\sigma$ remain orthogonal even when their block-charge labels
coincide. The estimate therefore holds for each partition, and
convexity preserves it after averaging over the hidden partition.

\paragraph{3. Completion of the decision lower bound.}
Combine the relaxed comparison and the two switching bounds
by the triangle inequality:
\[
 \|\varrho_{\mathsf{Exact}_k}-\varrho_{\mathsf{Inj}}\|_1
 \le8t\sqrt{\frac kn}+\frac{16t}{\sqrt q}.
\]
Let $p_{\rm strict}$ be the success probability when the two
promised distributions are chosen with equal probability.
Binary state discrimination and $q\ge n$ give
\[
 p_{\rm strict}
 \le\frac12+\frac14\|\varrho_{\mathsf{Exact}_k}-\varrho_{\mathsf{Inj}}\|_1
 \le\frac12+2t\sqrt{\frac kn}+\frac{4t}{\sqrt q}
 \le\frac12+6t\sqrt{\frac kn}.
\]
For $1\le t\le\sqrt{n/(8k)}$, we also have $t\le q/8$;
larger $t$ already gives the lower bound, and $t=0$ gives no bias.
Thus worst-case success $1/2+\varepsilon$ requires
$t=\Omega(\varepsilon\sqrt{n/k})$, proving Theorem~\ref{thm:test}.

\subsection[Comparison with Concurrent Work]{Comparison with Concurrent Work\texorpdfstring{\protect\footnotemark}{}}
\label{subsec:concurrent-work}
\footnotetext{\rev{ChatGPT brought these concurrent works to our attention during a review of the manuscript; see \nameref{sec:aimethodology}.}}

Very recently, Carolan and Majenz~\cite{carolan2026revisited} and
Kim~\cite{kim2026soundness} concurrently improve compressed
permutation-oracle simulation, with vanishing error for
$T=o(\sqrt n)$ queries on $n$ elements. Carolan and Majenz also
support inverse queries and ideal ciphers.

\paragraph{Overlapping encoding tools.}
Our encoding tools overlap with that of Carolan and Majenz
in its uniform permutation completions, frame normalization and
spectrum, and zero-sum identities
\cite[Lemmas~4.9--4.10]{carolan2026revisited}. We include derivations
in our notation to keep the subsequent arguments self-contained.

\paragraph{Where we go beyond.}
For adaptive finding under the exact $k$-to-one promise, we adapt
Liu and Zhandry's partial-collision hierarchy~\cite{liu2019finding}
to permutation databases. Combined with witness verification,
Carolan and Majenz's simulation theorem already suffices for our
adaptive exponent; our projected forward-query process additionally
achieves $O(T^2/n)$ error, removing the $O(T/\sqrt n)$ term.
For one round, we use the encoding exactly
and prove a static occupancy bound that remains useful beyond
$\sqrt n$ queries, yielding the tight lower bound even for entangled
queries. \decisionrev{For decision, we compare hidden-block distributions
and use the same harmonic encoding to switch to distinct labels within
$O(t/\sqrt q)$ trace distance, proving the lower bound for $q\ge n$.}

\subsection{Open Problems}
\label{subsec:open-problems}

Our results leave two quantitative questions open.

First, what is the full round hierarchy for \kFind?
Theorems~\ref{thm:find} and~\ref{thm:adaptive_find} determine the
one-round and unrestricted adaptive complexities. For a fixed
number $r\ge2$ of oracle layers, what tradeoff between parallel
width and total queries interpolates between these endpoints?
One-round orthogonality need not survive an additional query layer,
while the adaptive hierarchy bounds only the total query count.

Second, can the harmonic framework relax the alphabet-size
requirement in the adaptive lower bound for \kDist?
Belovs~\cite{belovs2026tight} proves the tight bound
$\Omega_k(n^{3/4-1/(4(2^k-1))})$ under the assumption $q=\Omega(n^2)$.
The challenge is to recover this exponent for smaller alphabets,
ideally for all $q\ge n$, using exact permutation encodings and
harmonic estimates.

\subsection{Other Related Work}
\label{subsec:related-work}

\paragraph{Related collision problems.}
\rev{In random-function multicollision finding, the oracle is a uniformly
random function $F:X\to Y$. Values at distinct inputs are independent
and uniform in $Y$, so different outputs need not have the same number
of preimages. The task is to find $k$ distinct inputs with the same
output; that output may have additional preimages. This model captures
generic multicollision attacks on a hash function modeled as a random
oracle, with access limited to function evaluations
\cite{liu2019finding,zhandry2019record}.
Multicollisions also play a role in attacks that exploit the internal
structure of hash constructions. For example, Joux's attacks on
iterated and cascaded hashing use large sets of colliding messages
\cite{joux2004multicollisions}. Such attacks exploit structure beyond
access to a random-function oracle.}

We next summarize bounds for two related problems: \kColl, which distinguishes an
injective input from an exactly $k$-to-one input, and \kDist, which finds
a $k$-collision in an arbitrary input promised to contain one.

For \kColl, the Brassard--H{\o}yer--Tapp algorithm
\cite{brassard1997quantum} queries a sample, stores its values,
and uses Grover search to find another position matching the table,
using $O((n/k)^{1/3})$ queries in total.
Aaronson and Shi~\cite{aaronson2004quantum} and
Kutin~\cite{kutin2005quantum} proved a matching lower bound by
symmetrizing the acceptance polynomial and bounding its degree.
For \kDist, Belovs~\cite{belovs2012learning} gives the upper bound
$O_k(n^{3/4-1/(4(2^k-1))})$ using learning graphs that organize
the acquisition of partial collision certificates in several stages.
His subsequent work~\cite{belovs2026tight} proves this bound tight
when $q=\Omega(n^2)$.
In the non-adaptive model, the general adversary bound of Koiran,
Landes, Portier, and Yao~\cite[Theorem~6]{koiran2010adversary}
yields $\Omega_k(n)$ queries for \kDist when $q\ge n$.
Reading the entire input matches this bound.

\paragraph{Compressed-oracle techniques.}
Zhandry~\cite{zhandry2019record} introduced compressed oracles to
coherently record information learned from queries to a random function.
Liu and Zhandry~\cite{liu2019finding} used this approach to prove
tight bounds for random-function multicollisions.
For random permutations, Rosmanis~\cite{rosmanis2021tight} developed
a representation-theoretic analysis of inversion with forward queries,
and Carolan~\cite{carolan2026compressed} constructed a unitary recorder
of partial injections supporting forward and inverse queries.
Section~\ref{subsec:concurrent-work} discusses the concurrent advances
in compressed permutation oracles and their relation to our work.

\paragraph{Parallel and non-adaptive quantum queries.}
Jeffery, Magniez, and de Wolf~\cite{jeffery2017optimal} combine
parallel quantum walks and adversary bounds to obtain tight results
for \textsc{Element Distinctness} and $k$-\textsc{Sum} over sufficiently
large alphabets. Girish, Sinha, Tal, and Wu~\cite{girish2024power}
use variants of \textsc{Forrelation} to separate consecutive constant
numbers of query rounds. In the one-round model,
van Dam~\cite{van1998quantum} recovers an arbitrary $n$-bit string
with bounded error using $n/2+O(\sqrt n)$ queries, while
Montanaro~\cite{montanaro2010nonadaptive} proves an $\Omega(n)$
lower bound for every total Boolean function depending on all
$n$ variables.
Koiran, Landes, Portier, and Yao~\cite{koiran2010adversary}
develop non-adaptive adversary bounds, obtaining tight linear bounds
for unordered search and decision \textsc{Element Distinctness},
while leaving the injective-versus-two-to-one collision problem open.

\subsection{Organization of the Paper}
\label{subsec:organization}

Section~\ref{sec:pre} introduces the notation and defines the collision
problems. Section~\ref{sec:algo} describes the adaptive and
non-adaptive query models, their circuits, and the orthogonal
decomposition of a single query layer. Section~\ref{sec:representation}
fixes the representation-theoretic notation and states the lemmas
used in the later proofs.
Section~\ref{sec:harmonic-framework} constructs the database encoding,
compares genuine and recording queries, and relates successful outputs
to recorded collisions. Section~\ref{sec:adaptive} bounds the growth
of partial collisions to prove the adaptive lower bound for \kFind.
Section~\ref{sec:nonadaptive-promise} instead bounds the probability
of recording a full collision after one parallel query layer.
Section~\ref{sec:byproduct} proves the lower bound for \kColl.
Appendix~\ref{app:background} provides further background in linear
algebra and representation theory, including induction formulas and
the proofs deferred from Section~\ref{sec:representation}.
Appendix~\ref{app:representation-theory} provides concrete
irreducible models and a worked example.
\reviewrev{Appendix~\ref{app:classical-bounds} proves the tight classical
bounds for \kFind and \kColl.}

\section{Preliminaries}
\label{sec:pre}

We introduce the notation shared by the two query models and define
the collision problems studied in this paper. Section~\ref{sec:algo}
describes the query models, and Section~\ref{sec:representation}
gives the representation-theoretic notation and lemmas used below.

\subsection{Basic Notation and Asymptotics}
\label{pre:basic-notation}

Throughout the paper, \(k\ge2\) is a fixed constant. For \(k\)-to-one
inputs, we assume \(n=kN\) for an integer \(N\). For a positive integer
\(q\), the alphabet \(\Z_q=\{0,\ldots,q-1\}\) has addition modulo \(q\).
We write \([n]=\{1,\ldots,n\}\), and
\(\Z_q^A\) for the set of functions \(A\to\Z_q\); in particular,
\(\Z_q^n=\Z_q^{[n]}\). For a finite set \(S\), \(\binom Sr\)
denotes its collection of \(r\)-element subsets.
For a nonempty finite set \(X\),
\(\operatorname{Unif}(X)\) denotes the uniform distribution on \(X\).
We write \(\omega_m=e^{2\pi\imath/m}\), where \(\imath^2=-1\), and use \(\1_E\) for the
indicator of an event \(E\). The falling factorial is
\((a)_b=a(a-1)\cdots(a-b+1)\), with \((a)_0=1\).

We use standard asymptotic notation as \(n\to\infty\), with success
thresholds fixed in advance in query-complexity bounds. Apart from
these fixed thresholds, constants in \(O(\cdot)\), \(\Omega(\cdot)\), and
\(\Theta(\cdot)\) are absolute unless a subscript indicates dependence
on further fixed parameters.
For a complex number \(z\), \(\abs{z}\) is
its modulus and \(\operatorname{Re}z\) its real part.

\subsection{Linear-algebra notation}
\label{pre:linear-algebra}

For a finite set $X$, $\mathbb C[X]$ is the Hilbert space with
orthonormal basis $\{|x\rangle:x\in X\}$. A subscript on a ket names
its register; we omit it when the register is clear. We write
$I_{\mathcal X}$ for the identity on $\mathcal X$ and $H^\dagger$
for the adjoint of $H$.
The inner product is $\langle u,v\rangle=u^\dagger v$ and $\|v\|$ is
the Euclidean norm. For $H:\mathcal X\to\mathcal Y$,
\begin{equation}
\label{eq:kernel}
 \|H\|=\sup_{\|v\|=1}\|Hv\|,\qquad
 \|H\|_1=\operatorname{Tr}\sqrt{H^\dagger H},\qquad
 \ker H=\{v:Hv=0\}.
\end{equation}
We write $[H,J]=HJ-JH$. All projectors are orthogonal;
$P_{\mathcal S}$ projects onto $\mathcal S$ and
$\|H\|_{\mathcal S}:=\|H|_{\mathcal S}\|$.
An isometry satisfies $H^\dagger H=I_{\mathcal X}$, and a contraction satisfies
$\|H\|\le1$. For Hermitian $H,J$, $H\preceq J$ means that $J-H$ is
positive semidefinite.

For an orthogonal decomposition
$\mathcal H=\bigoplus_s\mathcal H_s$, put
$\mathcal H_{\le t}=\bigoplus_{s\le t}\mathcal H_s$ and similarly
define $\mathcal H_{<t}$. A controlled sum $\bigoplus_a H_a$ acts as
$\sum_a|a\rangle\langle a|\otimes H_a$.
An operator on one register is extended by the identity on the others.
For a predicate $E$ on basis labels,
\begin{equation}
\label{eq:event-projector}
 P_E:=\sum_{x:E(x)}|x\rangle\langle x|.
\end{equation}
For a count $g$, a subscript such as $g\ge\ell$ specifies the
corresponding event. A vector is subnormalized if its norm is at
most one. For such a vector, we use the squared norm of a projected
component as its probability mass, without renormalizing.

We denote density operators by $\varrho$ and measurement outcomes
by $\rho$. A density operator is positive semidefinite and has trace
one. The partial trace $\operatorname{Tr}_{\mathcal Y}$ removes
register $\mathcal Y$. The trace distance
$\frac12\|\varrho-\varrho'\|_1$ bounds the difference in probability
of any measurement event and cannot increase under partial trace.
We allow arbitrary workspace, so Naimark dilation lets us take final
measurements to be projection-valued. Appendix~\ref{app:linear-algebra}
gives the detailed conventions.

\subsection{Fourier labels and basis vectors}
\label{pre:charge-functions}

For \(\sigma\in\Z_q^A\), define
\(\Supp{\sigma}=\{a\in A:\sigma(a)\ne0\}\) and
\(|\sigma|=|\Supp{\sigma}|\). We call the positions in its support
\emph{charged} and \(\sigma(a)\) the charge at \(a\). If
\(\Supp{\sigma}=\{a_1,\ldots,a_\ell\}\) and
\(c_i=\sigma(a_i)\ne0\), we record only the nonzero entries as
\(\sigma=\{a_1\mapsto c_1,\ldots,a_\ell\mapsto c_\ell\}\).
The empty label \(\emptyset\) is the identically zero function on \(A\).
The common charged part and the part of \(\sigma\) not already present in
\(\eta\) are
\[
\begin{aligned}
 \sigma\cap\sigma'
   &=\{a\mapsto c:\sigma(a)=\sigma'(a)=c\ne0\},\\
 \sigma\setminus\eta
   &=\{a\mapsto c:\sigma(a)=c\ne0\text{ and }\eta(a)\ne c\}.
\end{aligned}
\]
In particular, for \(\eta=\sigma\cap\sigma'\), the labels
\(\sigma\setminus\eta\) and \(\sigma'\setminus\eta\) are the parts
outside the entries on which their nonzero charges agree.

For a positive integer $m$, the single-coordinate Fourier basis of
$\mathbb C[\mathbb Z_m]$ is
\begin{equation}
\label{eqn:FourierDef}
 |\widehat c\rangle
 :=\frac1{\sqrt m}\sum_{z\in\mathbb Z_m}\omega_m^{cz}|z\rangle,
 \qquad c\in\mathbb Z_m.
\end{equation}
The Fourier transform $F_m:\mathbb C[\mathbb Z_m]\to
\mathbb C[\mathbb Z_m]$ is the unitary $F_m|c\rangle=|\widehat c\rangle$.
For any finite coordinate set $A$ and charge label $\sigma\in\mathbb Z_q^A$,
we use the product Fourier basis of $\mathbb C[\mathbb Z_q^A]$:
\begin{equation}
\label{eq:product-fourier-basis}
 |\widehat\sigma\rangle
 :=\bigotimes_{a\in A}|\widehat{\sigma(a)}\rangle
 =q^{-|A|/2}\sum_{x\in\mathbb Z_q^A}
       \omega_q^{\sum_{a\in A}\sigma(a)x(a)}|x\rangle.
\end{equation}
These vectors are orthonormal.
A hat on a charge label always denotes this Fourier basis vector;
$|\sigma\rangle$ without a hat denotes a computational-basis vector.

\subsection{Partial Maps and Databases}
\label{pre:partial-maps}

\paragraph{Maps and databases.}
The notation \(f:A\hookrightarrow B\) denotes an injective map
defined on all of \(A\). We write \(\operatorname{Inj}(A,B)\) for the set
of these injections. A partial map \(d:A\rightharpoonup B\) is defined
on a subset \(\operatorname{Dom}(d)\subseteq A\), has size
\(|d|:=|\operatorname{Dom}(d)|\), and has image
\(\operatorname{Im}(d):=\{d(a):a\in\operatorname{Dom}(d)\}\subseteq B\).
It is a \emph{partial injection} if it is injective on its domain. The
empty partial map \(\emptyset\) has empty domain. We also use this
symbol for an empty charge record: omitted entries mean zero
in a charge function and unassigned in a database. The label
space determines which convention applies; the empty partial map and
the zero charge function are different objects. We use \(\bot\) for an
unassigned position, so an assigned value \(0\) is distinct from a blank.
For \(S\subseteq\operatorname{Dom}(d)\), let
\(d(S)=\{d(a):a\in S\}\), and let \(d|_S\) be its restriction to \(S\).
For \(U\subseteq B\), put
\(d^{-1}(U)=\{a\in\operatorname{Dom}(d):d(a)\in U\}\). Composition is
\((g\circ d)(a)=g(d(a))\) whenever defined. We identify a partial map
with its graph: \(d\supseteq r\) means that \(d\) extends \(r\), and
\(r\cup\{i\mapsto u\}\) adds an assignment at an unassigned position
\(i\). For a partial injection, the extension remains injective when
\(u\notin\operatorname{Im}(r)\).

Let \(S_n\) be the symmetric group on \([n]\) under composition, and put
\[
 \mathcal I_n
 :=\{d:[n]\rightharpoonup[n]:d\text{ is a partial injection}\}.
\]
A permutation \(\pi\in S_n\) is an element of \(\mathcal I_n\) of size
\(n\).

For fixed \(i,u\in[n]\), the insertion map \(\operatorname{add}_{iu}\)
maps the span of residual databases \(r\) with
\(i\notin\operatorname{Dom}(r)\) and \(u\notin\operatorname{Im}(r)\)
into \(\mathbb C[\mathcal I_n]\), by
\[
 \operatorname{add}_{iu}|r\rangle
 :=|r\cup\{i\mapsto u\}\rangle,
 \qquad i\notin\operatorname{Dom}(r),\quad
        u\notin\operatorname{Im}(r).
\]
Superscripts
\((i,u)\) on a database or permutation construction mean that input label
\(i\) and output label \(u\) have been removed; the remaining input and
output sets are identified with \([n-1]\) in increasing order. A
superscript \((n-1)\) denotes the same construction on \(n-1\) labels.

\subsection{Collision Problems}
\label{subsec:collision-problems}

For an input string \(x=(x_1,\ldots,x_n)\in \Z_q^n\), a \(k\)-collision is a
set \(S=\{i_1,\ldots,i_k\}\subseteq [n]\) of \(k\) distinct indices with
\(x_{i_1}=x_{i_2}=\cdots=x_{i_k}\).
We say that \(x\) is \emph{one-to-one} if all its entries are pairwise
distinct.  When \(n=kN\), we say that \(x\) is \emph{\(k\)-to-one} if the
indices \([n]\) can be partitioned into \(N\) disjoint blocks of size \(k\)
such that the values are equal inside each block and distinct across different
blocks.  Equivalently, every value appearing in \(x\) appears exactly \(k\)
times.

We consider the following two collision problems.

\begin{definition}[\kColl]
\label{def:CollTest}
Let \(n=kN\).  Given oracle access to a string \(x\in \Z_q^n\), promised to be
either one-to-one or \(k\)-to-one, distinguish between the two cases.
\end{definition}

For \(k=2\), this is the standard collision problem.

\begin{definition}[\kFind]
\label{def:CollFind}
Let \(n=kN\).  Given oracle access to a \(k\)-to-one string
\(x\in \Z_q^n\), output a \(k\)-collision of \(x\), i.e., output one full block
of equal-valued positions.
\end{definition}

Every promised input has $N$ valid witnesses, one for each block.
The exact multiplicities distinguish this input model from
random-function multicollision finding~\cite{liu2019finding}.

\section{Adaptive and Non-adaptive Quantum Algorithms}
\label{sec:algo}

We first define queries to an input string
$x=(x_1,\ldots,x_n)\in\mathbb Z_q^n$. We then represent a random input
by an additional quantum register and specialize the construction to
permutations. Throughout, $\mathcal A$ contains the registers accessible
to the algorithm. The register storing the unknown input is introduced
only for the analysis and is inaccessible to the algorithm.

\subsection{Value queries, phases, and query rounds}
\label{sec:query-model}

Let $\mathcal I=\mathbb C[[n]]$ be the index register,
$\mathcal C=\mathbb C[\mathbb Z_q]$ the answer register, and
$\mathcal W$ an arbitrary workspace. For a sequential query algorithm,
$\mathcal A=\mathcal I\otimes\mathcal C\otimes\mathcal W$.
The phase oracle $O_x:\mathcal I\otimes\mathcal C\to
\mathcal I\otimes\mathcal C$ is
\begin{equation}
\label{eq:oracle}
 O_x|i,c\rangle=\omega_q^{c x_i}|i,c\rangle,
 \qquad i\in[n],\quad c\in\mathbb Z_q.
\end{equation}
It is equivalent to the value query $|i,z\rangle\mapsto|i,z+x_i\rangle$
by conjugating the answer register with $F_q$ and $F_q^\dagger$.
This input-independent change of basis preserves both query cost and
query rounds. We use phase queries in the analysis and value queries
for witness verification. Conjugating a value query by answer-register
negation $|z\rangle\mapsto|-z\rangle$ implements its inverse at the
same query cost.

Let $U_j:\mathcal A\to\mathcal A$ be input-independent unitaries.
Write $|0\rangle_{\mathcal A}$ for a fixed normalized initial state
that is independent of the input, and set $\psi_{x,0}=|0\rangle_{\mathcal A}$ and
\[
 \psi'_{x,j}=U_j\psi_{x,j}\quad(0\le j\le T),\qquad
 \psi_{x,j+1}=O_x\psi'_{x,j}\quad(0\le j<T).
\]
An unrestricted $T$-query algorithm has final state
\[
 \psi'_{x,T}
 =U_T O_x U_{T-1}O_x\cdots U_1O_xU_0|0\rangle_{\mathcal A}.
\]
The unitaries $U_j$ are fixed in advance, but the state on which a
later query acts can depend on earlier answers. This is how the model
allows adaptivity, even without intermediate classical observations.
Classical randomness and intermediate measurements may be purified
into $\mathcal W$. We take the final measurement to be a PVM
$\{\Pi_\rho\}_{\rho\in R}$ on $\mathcal A$, where $R$ is the response
set. \mbox{The response probability is $\|\Pi_\rho\psi'_{x,T}\|^2$}.

More generally, an $r$-round algorithm uses $t_j$ parallel queries
in layer $j$, with arbitrary input-independent unitaries between
layers and total query cost $T=\sum_{j=1}^r t_j$.
A \emph{one-round non-adaptive} algorithm has the form
\begin{equation}
\label{eq:non-adaptiveALG}
 U_1(O_x^{\otimes t}\otimes I_{\mathcal W})U_0,
 \qquad
 \mathcal A=\mathcal I^{\otimes t}\otimes
             \mathcal C^{\otimes t}\otimes\mathcal W.
\end{equation}
Its $t$ index and answer registers may be arbitrarily entangled.
The definition restricts the number of oracle layers, not the
initial state or the final measurement.
Here subscripts $0,1$ count~query~layers:
\[
 \psi_{x,0}\xrightarrow{\,U_0\,}\psi'_{x,0}
 \xrightarrow{\,O_x^{\otimes t}\,}\psi_{x,1}
 \xrightarrow{\,U_1\,}\psi'_{x,1}.
\]
A \emph{basis query record} is
$a=(i_1,\ldots,i_t,c_1,\ldots,c_t,w)$, with basis vector
\begin{equation}
\label{eq:basis-query-record}
 |a\rangle_{\mathcal A}
 =|i_1,\ldots,i_t\rangle_{\mathcal I^{\otimes t}}
  |c_1,\ldots,c_t\rangle_{\mathcal C^{\otimes t}}
  |w\rangle_{\mathcal W}.
\end{equation}
These records index quantum basis components; they do not represent
measurement outcomes.

\FloatBarrier
\subsection{Purifying the input distribution}
\label{sec:input-purification}

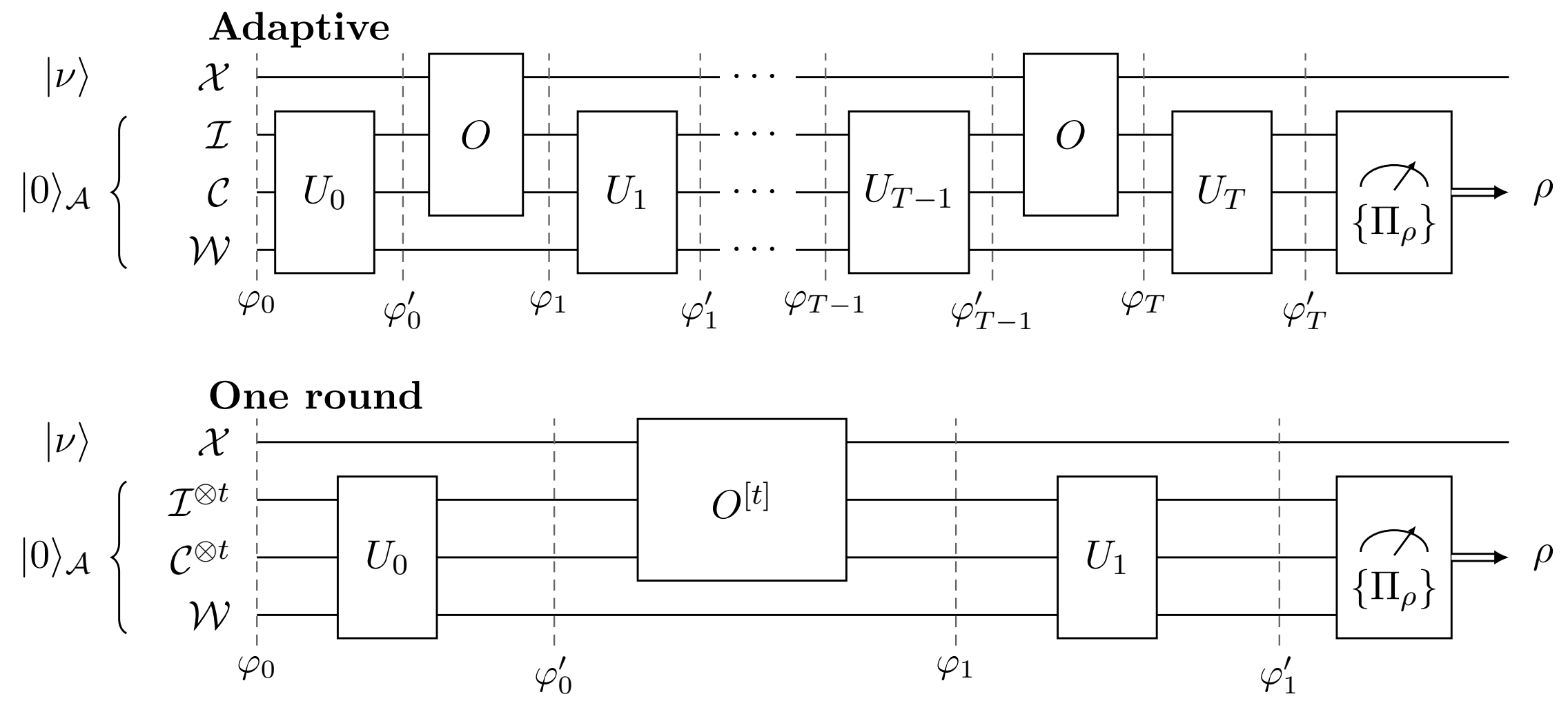
\begin{figure}[!b]
\centering
\begin{tikzpicture}[query circuit,y=.65cm,
  algorithm gate/.append style={minimum height=1.55cm},
  oracle gate/.append style={minimum height=1.55cm}]
 \draw (0,0)--(12.0,0);
 \foreach \y in {-.85,-1.70,-2.55}
   \draw (0,\y)--(11.45,\y);
 \node[anchor=east] at (-.12,0) {$\mathcal X$};
 \node[anchor=east] at (-.12,-.85) {$\mathcal I$};
 \node[anchor=east] at (-.12,-1.70) {$\mathcal C$};
 \node[anchor=east] at (-.12,-2.55) {$\mathcal W$};
 \node[anchor=east] at (-1.45,0) {$|\nu\rangle$};
 \draw[decorate,decoration={brace,amplitude=4pt}]
   (-1.25,-2.82)--(-1.25,-.58);
 \node[anchor=east] at (-1.45,-1.70) {$|0\rangle_{\mathcal A}$};
 \node[algorithm gate] at (.65,-1.70) {$U_0$};
 \node[oracle gate,minimum width=.9cm] at (2.1,-.85) {$O$};
 \node[algorithm gate] at (3.55,-1.70) {$U_1$};
 \foreach \y in {0,-.85,-1.70,-2.55}
   \node[fill=white,inner sep=3pt] at (4.8,\y) {$\cdots$};
 \node[algorithm gate,minimum width=1.15cm]
   at (6.25,-1.70) {$U_{T-1}$};
 \node[oracle gate,minimum width=.9cm] at (7.8,-.85) {$O$};
 \node[algorithm gate] at (9.25,-1.70) {$U_T$};
 \pic at (10.9,-1.70) {query measurement};
 \draw[classical output] (11.45,-1.70)--(12.0,-1.70);
 \node[anchor=west] at (12.1,-1.70) {$\rho$};
 \node[anchor=west,font=\bfseries] at (-.6,.7) {Adaptive};
 \querycut{0}{-3.0}{.35}{\varphi_0}
 \querycut{1.4}{-3.0}{.35}{\varphi'_0}
 \querycut{2.8}{-3.0}{.35}{\varphi_1}
 \querycut{4.25}{-3.0}{.35}{\varphi'_1}
 \querycut{5.45}{-3.0}{.35}{\varphi_{T-1}}
 \querycut{7.05}{-3.0}{.35}{\varphi'_{T-1}}
 \querycut{8.5}{-3.0}{.35}{\varphi_T}
 \querycut{10.05}{-3.0}{.35}{\varphi'_T}
 \begin{scope}[yshift=-3.5cm]
 \draw (0,0)--(12.0,0);
 \foreach \y in {-.85,-1.70,-2.55}
   \draw (0,\y)--(11.45,\y);
 \node[anchor=east] at (-.12,0) {$\mathcal X$};
 \node[anchor=east] at (-.12,-.85) {$\mathcal I^{\otimes t}$};
 \node[anchor=east] at (-.12,-1.70) {$\mathcal C^{\otimes t}$};
 \node[anchor=east] at (-.12,-2.55) {$\mathcal W$};
 \node[anchor=east] at (-1.45,0) {$|\nu\rangle$};
 \draw[decorate,decoration={brace,amplitude=4pt}]
   (-1.25,-2.82)--(-1.25,-.58);
 \node[anchor=east] at (-1.45,-1.70) {$|0\rangle_{\mathcal A}$};
 \node[algorithm gate] at (1.25,-1.70) {$U_0$};
 \node[oracle gate,minimum width=2cm] at (4.65,-.85) {$O^{[t]}$};
 \node[algorithm gate] at (8.15,-1.70) {$U_1$};
 \pic at (10.9,-1.70) {query measurement};
 \draw[classical output] (11.45,-1.70)--(12.0,-1.70);
 \node[anchor=west] at (12.1,-1.70) {$\rho$};
 \node[anchor=west,font=\bfseries] at (-.6,.7) {One round};
 \querycut{0}{-3.0}{.35}{\varphi_0}
 \querycut{2.85}{-3.0}{.35}{\varphi'_0}
 \querycut{6.7}{-3.0}{.35}{\varphi_1}
 \querycut{9.8}{-3.0}{.35}{\varphi'_1}
 \end{scope}
\end{tikzpicture}
\caption{Purified adaptive (top) and one-round (bottom) algorithms.
The inaccessible input register starts in $|\nu\rangle_{\mathcal X}$,
and the accessible register starts in $|0\rangle_{\mathcal A}$.
Every $U_j$ and the final measurement act only on $\mathcal A$.
The $t$ parallel query pairs share the same register $\mathcal X$.
Dashed vertical cuts mark the joint states $\varphi_j$ and $\varphi'_j$.}
\label{fig:QAAvg}
\end{figure}

For a distribution $\nu$ on $\mathbb Z_q^n$, introduce the inaccessible
input register
$\mathcal X=\mathcal X_1\otimes\cdots\otimes\mathcal X_n=\mathbb C[\mathbb Z_q^n]$, with each $\mathcal X_i\cong\mathbb C^q$.
Its initial state is
\begin{equation}
\label{eq:input-purification}
 |\nu\rangle_{\mathcal X}
   =\sum_{x\in\mathbb Z_q^n}\sqrt{\nu(x)}|x\rangle.
\end{equation}
The controlled queries $O$ and $O^{[t]}$ act on
$\mathcal X\otimes\mathcal I\otimes\mathcal C$ and
$\mathcal X\otimes\mathcal I^{\otimes t}\otimes\mathcal C^{\otimes t}$,
respectively, and are~defined~by
\begin{equation}
\label{eq:oracleAvg}
 O:=\sum_{x\in\mathbb Z_q^n}|x\rangle\langle x|\otimes O_x,
 \qquad
 O^{[t]}:=\sum_{x\in\mathbb Z_q^n}|x\rangle\langle x|
                         \otimes O_x^{\otimes t}.
\end{equation}
All $t$ query pairs share this single input register.
We write $O^{[t]}$ to make this explicit: $O^{\otimes t}$ as a
literal tensor product would introduce $t$ input registers.
For either query model, write the joint states as
\begin{equation}
\label{eq:purified-process-states}
 \varphi_j:=\sum_x\sqrt{\nu(x)}|x\rangle\otimes\psi_{x,j},
 \qquad
 \varphi'_j:=\sum_x\sqrt{\nu(x)}|x\rangle\otimes\psi'_{x,j}.
\end{equation}
Here $j$ counts sequential queries or parallel layers according to the
model. Both models start in
$\varphi_0=|\nu\rangle_{\mathcal X}|0\rangle_{\mathcal A}$.
For an adaptive $T$-query algorithm, the evolution satisfies
\[
 \varphi'_j=U_j\varphi_j\quad(0\le j\le T),\qquad
 \varphi_{j+1}=O\varphi'_j\quad(0\le j<T).
\]
For a one-round algorithm, it is
\[
 \varphi_0\xrightarrow{\,U_0\,}\varphi'_0
 \xrightarrow{\,O^{[t]}\,}\varphi_1
 \xrightarrow{\,U_1\,}\varphi'_1,
\]
where identities on untouched registers are suppressed.
Figure~\ref{fig:QAAvg} depicts both purified circuits.
Equivalently, the oracle in a fixed index/charge sector is the
operator $O_{i,c}:\mathcal X\to\mathcal X$ given by
\begin{equation}
\label{eqn:zhandryOic}
 O_{i,c}|x\rangle=\omega_q^{c x_i}|x\rangle.
\end{equation}
All algorithm-only unitaries act as the identity on $\mathcal X$.
The final joint state in the one-round case is $\varphi'_1$; tracing out
$\mathcal X$ produces
$\sum_x\nu(x)|\psi'_{x,1}\rangle\langle\psi'_{x,1}|$.
Thus the purified process has exactly the same output distribution
as the original algorithm averaged over $x\sim\nu$. Applying an
isometry only to the inaccessible register preserves this distribution.

\FloatBarrier
\subsection{The orthogonal decomposition of a single layer}
\label{sec:one-round-decomposition}

A parallel query layer multiplies each basis query record by an
input-dependent phase and leaves the record itself unchanged.
The resulting components are therefore orthogonal in the algorithm
register. After encoding the input register, probabilities of events
measured only in the database register add over these components.
The following lemma makes this decomposition explicit.

\begin{lemma}[One-round orthogonality]
\label{lem:one-round-phases}
For every $t$-query one-round algorithm and every input distribution
$\nu$ on $\mathbb Z_q^n$, its joint post-query state
$\varphi_1\in\mathcal X\otimes\mathcal A$ has the form
\[
\begin{gathered}
 \varphi_1=\sum_a|\nu,h_a\rangle_{\mathcal X}\otimes u_a,\qquad
 |\nu,h_a\rangle_{\mathcal X}
       :=\sum_x\sqrt{\nu(x)}\,h_a(x)|x\rangle_{\mathcal X},\\
 \langle u_a,u_b\rangle=0\quad(a\ne b),\qquad
 \sum_a\|u_a\|^2=1.
\end{gathered}
\]
Here $u_a\in\mathcal A$ and there are fixed sets $S_a\subseteq[n]$ and
nonzero charges $c_{a,i}\in\mathbb Z_q$ such that
\[
 h_a(x)=\prod_{i\in S_a}\omega_q^{c_{a,i}x_i},
 \qquad |S_a|\le t.
\]
The vectors $u_a$ depend only on the algorithm, not on $\nu$.
The same conclusion holds for $\varphi'_1$, with each $u_a$ replaced by
$U_1u_a$.
\end{lemma}
\begin{proof}
Expand $U_0|0\rangle_{\mathcal A}=\sum_a\alpha_a|a\rangle_{\mathcal A}$
in the basis query records of Eq.~\eqref{eq:basis-query-record}.
For each record $a$, define its aggregate charge label and algorithm
vector by
\[
 \sigma_a(i):=\sum_{\substack{1\le j\le t\\i_j=i}}c_j\pmod q,
 \qquad
 u_a:=\alpha_a|a\rangle_{\mathcal A}.
\]
The parallel query multiplies this basis component by
$\prod_{j=1}^t\omega_q^{c_jx_{i_j}}
=\omega_q^{\sum_i\sigma_a(i)x_i}$, and $|\sigma_a|\le t$.
Set $S_a=\Supp{\sigma_a}$ and $c_{a,i}=\sigma_a(i)$ for $i\in S_a$.
Applying the query to $|\nu\rangle_{\mathcal X}\otimes U_0|0\rangle_{\mathcal A}$
therefore gives
\[
 \varphi_1=\sum_{x,a}\sqrt{\nu(x)}\,h_a(x)\alpha_a
                    |x\rangle_{\mathcal X}\otimes|a\rangle_{\mathcal A}
      =\sum_a|\nu,h_a\rangle_{\mathcal X}\otimes u_a.
\]
Distinct basis records are orthogonal, so
$\langle u_a,u_b\rangle=0$ for $a\ne b$ and
$\sum_a\|u_a\|^2=\sum_a|\alpha_a|^2=1$.
The common final unitary $U_1$ preserves their inner products and norms.
\end{proof}

We may also group records with the same aggregate charge label.
The resulting vectors $u_\sigma:=\sum_{a:\sigma_a=\sigma}u_a$
are pairwise orthogonal and satisfy $\sum_\sigma\|u_\sigma\|^2=1$.
The input vectors $|\nu,h_a\rangle_{\mathcal X}$ need not be orthogonal.
Orthogonality of the algorithm vectors is specific to one round
and can fail after further queries.

\FloatBarrier
\subsection{Permutation registers and forward access}
\label{sec:permutation-model}
For a uniformly random permutation, the inaccessible register is
$\mathcal P=\mathbb C[S_n]$, initialized to
\[
 |\widehat\emptyset\rangle_{\mathcal P}
   =\frac1{\sqrt{n!}}\sum_{\pi\in S_n}|\pi\rangle.
\]

We use $\mathcal C_\pi=\mathbb C[\mathbb Z_{n+1}]$ as the answer
register. The values $1,\ldots,n$ denote valid permutation slots,
and $0$ denotes a blank.
Write $\mathcal W_\pi$ for the remaining workspace, so that the
accessible register is
\begin{equation}
\label{eq:permutation-registers}
 \mathcal A=\mathcal I\otimes\mathcal C_\pi\otimes\mathcal W_\pi.
\end{equation}
The subscript $\pi$ specifies the permutation-query model; these
register spaces do not depend on the unknown permutation.

\begin{definition}[Forward permutation queries]
\label{def:permutation-query}
The value query for a fixed $\pi$ maps
$|i,z\rangle\mapsto|i,z+\pi(i)\bmod(n+1)\rangle$.
Its phase form $O_\pi$ acts on $\mathcal I\otimes\mathcal C_\pi$,
its purified blocks $\mathsf O_{i,c}$ act on $\mathcal P$, and the
full purified oracle $\mathsf O$ acts on
$\mathcal P\otimes\mathcal I\otimes\mathcal C_\pi$. They are~defined~by
\[
\begin{gathered}
 O_\pi|i,c\rangle=\omega_{n+1}^{c\pi(i)}|i,c\rangle,
 \qquad
 \mathsf O_{i,c}|\pi\rangle
       =\omega_{n+1}^{c\pi(i)}|\pi\rangle,\\
 \mathsf O
 =\sum_{\pi\in S_n}|\pi\rangle\langle\pi|\otimes O_\pi
 =\bigoplus_{i\in[n],\,c\in\mathbb Z_{n+1}}\mathsf O_{i,c}.
\end{gathered}
\]
For fixed $c$ we write $\mathsf O_i=\mathsf O_{i,c}$.
\end{definition}

The accessible register is still denoted by $\mathcal A$;
when a permutation query is applied, the original function
answer register can be retained as part of its workspace.
No oracle evaluating~$\pi^{-1}$~is~available.

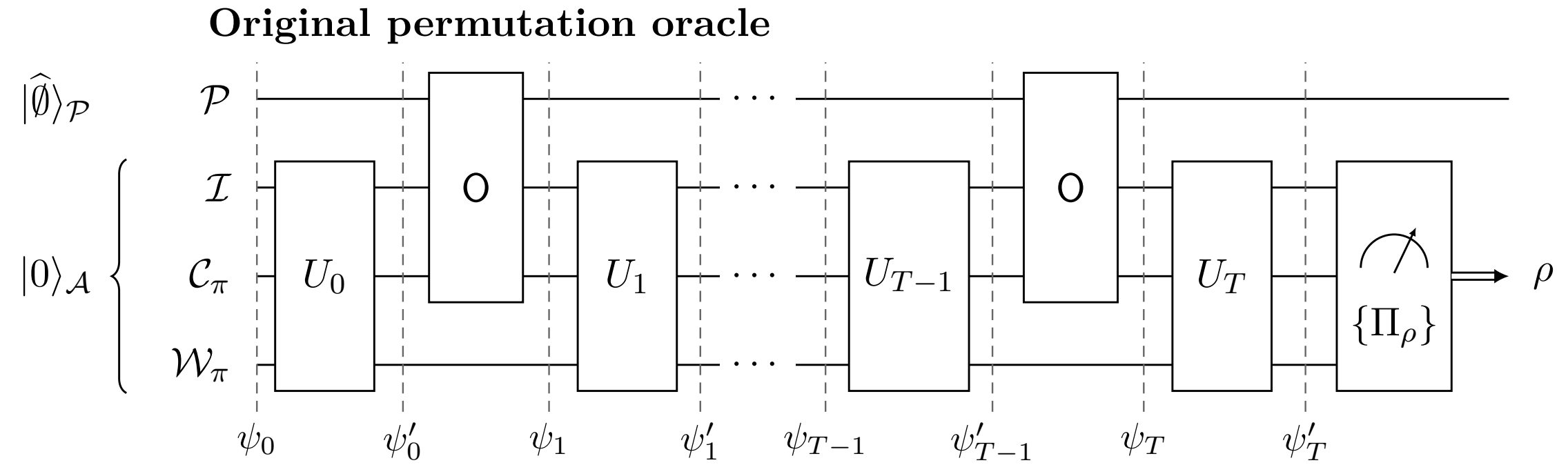
\begin{figure}[htbp]
\centering
\begin{tikzpicture}[query circuit]
 \draw (0,0)--(12.0,0);
 \foreach \y in {-.85,-1.70,-2.55}
   \draw (0,\y)--(11.45,\y);
 \node[anchor=east] at (-.12,0) {$\mathcal P$};
 \node[anchor=east] at (-.12,-.85) {$\mathcal I$};
 \node[anchor=east] at (-.12,-1.70) {$\mathcal C_\pi$};
 \node[anchor=east] at (-.12,-2.55) {$\mathcal W_\pi$};
 \node[anchor=east] at (-1.45,0)
   {$|\widehat\emptyset\rangle_{\mathcal P}$};
 \draw[decorate,decoration={brace,amplitude=4pt}]
   (-1.25,-2.82)--(-1.25,-.58);
 \node[anchor=east] at (-1.45,-1.70) {$|0\rangle_{\mathcal A}$};
 \node[algorithm gate] at (.65,-1.70) {$U_0$};
 \node[oracle gate,minimum width=.9cm] at (2.1,-.85) {$\mathsf O$};
 \node[algorithm gate] at (3.55,-1.70) {$U_1$};
 \foreach \y in {0,-.85,-1.70,-2.55}
   \node[fill=white,inner sep=3pt] at (4.8,\y) {$\cdots$};
 \node[algorithm gate,minimum width=1.15cm]
   at (6.25,-1.70) {$U_{T-1}$};
 \node[oracle gate,minimum width=.9cm] at (7.8,-.85) {$\mathsf O$};
 \node[algorithm gate] at (9.25,-1.70) {$U_T$};
 \pic at (10.9,-1.70) {query measurement};
 \draw[classical output] (11.45,-1.70)--(12.0,-1.70);
 \node[anchor=west] at (12.1,-1.70) {$\rho$};
 \node[anchor=west,font=\bfseries] at (-.6,.7)
   {Original permutation oracle};
 \querycut{0}{-3.0}{.35}{\psi_0}
 \querycut{1.4}{-3.0}{.35}{\psi'_0}
 \querycut{2.8}{-3.0}{.35}{\psi_1}
 \querycut{4.25}{-3.0}{.35}{\psi'_1}
 \querycut{5.45}{-3.0}{.35}{\psi_{T-1}}
 \querycut{7.05}{-3.0}{.35}{\psi'_{T-1}}
 \querycut{8.5}{-3.0}{.35}{\psi_T}
 \querycut{10.05}{-3.0}{.35}{\psi'_T}
 \end{tikzpicture}
\caption{The original forward-permutation query circuit. The inaccessible
register starts in $|\widehat\emptyset\rangle_{\mathcal P}$, and the
accessible register starts in $|0\rangle_{\mathcal A}$.
Algorithm-only gates and the final PVM act on
$\mathcal A=\mathcal I\otimes\mathcal C_\pi\otimes\mathcal W_\pi$.
Dashed vertical cuts mark the joint states before and after each~\mbox{algorithm-only}~unitary.}
\label{fig:QAPerm}
\end{figure}

For permutation inputs, write the joint states as
$\psi_j,\psi'_j\in\mathcal P\otimes\mathcal A$:
\begin{equation}
\label{eq:genuine-process}
 \psi_0=|\widehat\emptyset\rangle_{\mathcal P}\otimes|0\rangle_{\mathcal A},
 \qquad \psi'_j=U_j\psi_j,\qquad
 \psi_{j+1}=\mathsf O\psi'_j.
\end{equation}
Figure~\ref{fig:QAPerm} depicts this process, with the final PVM
acting only on $\mathcal A$.
For a layer of $t$ parallel queries, the shared-register oracle
$\mathsf O^{[t]}$ acts on
$\mathcal P\otimes\mathcal I^{\otimes t}\otimes\mathcal C_\pi^{\otimes t}$:
\begin{equation}
\label{eq:parallel-permutation-query}
 \mathsf O^{[t]}:=\sum_{\pi\in S_n}|\pi\rangle\langle\pi|
                         \otimes O_\pi^{\otimes t}.
\end{equation}
All query pairs share the same inaccessible permutation register.

\FloatBarrier

\section{Representations of the Symmetric Group}
\label{pre:representation-notation}
\label{sec:representation}

We fix the representation-theoretic notation and state the lemmas
used below to analyze permutation queries. All representations are finite-dimensional, complex,
and unitary. Appendix~\ref{app:background} collects definitions and
supporting arguments; Appendix~\ref{app:representation-theory} gives
irreducible models and a worked example.

\subsection{Notation}

For a finite group $G$, $(W,\chi)$ denotes a representation, with
$\chi$ extended linearly to $\mathbb C[G]$. An intertwiner
$T:W\to\widetilde W$ satisfies $T\chi(g)=\widetilde\chi(g)T$.
The $\lambda$-isotypic subspace is the sum of all irreducible copies
of type $\lambda$; their number is its multiplicity.
For $H\le G$, write $\operatorname{Res}_H^G W$ for restriction to $H$
and $W^H=\{w:\chi(h)w=w\text{ for all }h\in H\}$ for its fixed subspace.

For a partition $\lambda\vdash m$, $\lambda_i$ and $\lambda'_j$ are
its row and column lengths. The irreducible types of $S_m$ have
representatives $(V_\lambda,\chi_\lambda)$ with
\begin{equation}
\label{eq:irreducible-types}
 V_\lambda\cong V_\mu\ \Longleftrightarrow\ \lambda=\mu,
 \qquad \dim V_\lambda=f^\lambda.
\end{equation}
Here $f^{\lambda/\mu}$ counts standard Young tableaux of the skew
diagram $\lambda/\mu$, and $f^\lambda=f^{\lambda/\emptyset}$.
We use $f^\emptyset=1$, $f^{\lambda/\mu}=0$ when
$\mu\not\subseteq\lambda$, and $(0)=\emptyset$.
A \textit{horizontal strip} has at most one box in each column;
$\nu\nearrow\lambda$ means that $\lambda$ adds one box to $\nu$.
The notation $(m-s,\nu)$ places $\nu$ below the first row.
The content sum and hook lengths are
\[
 c(\lambda)=\sum_{(i,j)\in\lambda}(j-i),\qquad
 h_\lambda(i,j)=\lambda_i-j+\lambda'_j-i+1.
\]

The commuting left and right regular actions on $\mathbb C[G]$ are
\[
 L_a|g\rangle=|ag\rangle,\qquad R_b|g\rangle=|gb^{-1}\rangle.
\]
On permutations and partial injections, their joint action is
$|\pi\rangle\mapsto|\tau\circ\pi\circ\sigma^{-1}\rangle$ and
$|d\rangle\mapsto|\tau\circ d\circ\sigma^{-1}\rangle$,
respectively. The first factor relabels the range and the second
the domain. Write $\mathcal P_\lambda$ for the left-regular
$\lambda$-isotypic subspace. For a product group, joint type
$(\lambda,\mu)$ means $V_\lambda\otimes V_\mu$.

We use the transposition class sum and point-multiplication operators
\begin{equation}
\label{eq:transposition-class-sum}
 Z_m:=\sum_{1\le a<b\le m}(a\ b)\in\mathbb C[S_m],
\end{equation}
where $(a\ b)$ exchanges $a$ and $b$, and, for $i\in[m]$ and
$g:[m]\to\mathbb C$,
\begin{equation}
\label{eq:point-multiplier}
 M_{i,g}:\mathbb C[S_m]\to\mathbb C[S_m]:\quad
 |\pi\rangle\longmapsto g(\pi(i))|\pi\rangle.
\end{equation}

\subsection{Representation-theoretic lemmas}

The regular decomposition matches range and domain representation
types. This matching will let us encode permutation information in a
database.

\begin{lemma}[{\cite[Section~1.10]{sagan2001symmetric},
\cite[Section~7, Eqs.~(14)--(15)]{rosmanis2021tight}} Regular representation]
\label{lem:pre-regular}
For the commuting range and domain actions of \(S_m\times S_m\),
\begin{equation}
\label{pre:regular-decomposition}
 \mathbb C[S_m]=\bigoplus_{\lambda\vdash m}\mathcal P_\lambda,
 \qquad
 \mathcal P_\lambda\cong V_\lambda\otimes V_\lambda.
\end{equation}
The summands are orthogonal and irreducible for the joint action.
Under either action alone, $\mathcal P_\lambda$ consists of $f^\lambda$
copies of $V_\lambda$, so $\dim\mathcal P_\lambda=(f^\lambda)^2$.
It is the $\lambda$-isotypic subspace for both actions.
\end{lemma}

Schur's lemma lets us analyze intertwiners separately on each
isotypic subspace.

\begin{lemma}[{\cite[Sections~1.5--1.7]{sagan2001symmetric}}
Schur's lemma]
\label{lem:pre-schur}
An intertwiner \(T:W\to\widetilde W\) between representations of a
finite group maps each isotypic subspace of \(W\) into the corresponding
subspace of \(\widetilde W\). If \(W\) is irreducible, every intertwiner
\(W\to W\) is a scalar multiple of the identity.
\end{lemma}

Averaging a group action enforces invariance under that group.
The following projector formulas turn the constraints defining
permutation completions into explicit frame operators.

\begin{lemma}[{\cite[Sections~1.5--1.7]{sagan2001symmetric}} Subgroup averages]
\label{lem:pre-averages}
Let \((W,\chi)\) be a unitary representation of a finite group \(G\),
and let \(H\le G\). Then
\begin{equation}
 P_{W^H}=\frac1{|H|}\sum_{h\in H}\chi(h).
\end{equation}
For \(H\le S_m\) and an irreducible representation \(V_\lambda\),
\begin{equation}
\label{pre:conjugate-average}
 \frac1{m!}\sum_{\sigma\in S_m}
 \chi_\lambda(\sigma)P_{V_\lambda^H}\chi_\lambda(\sigma)^\dagger
 =\frac{\dim(V_\lambda^H)}{f^\lambda}I_{V_\lambda}.
\end{equation}
\end{lemma}

The branching rule tracks representation types when labels are fixed.
We use it both to control how a query changes first-row length and to
compute dimensions of invariant subspaces arising from prescribed
assignments.

\begin{lemma}[{\cite[Section~2.8]{sagan2001symmetric},
\cite[Section~6.2]{rosmanis2021tight}} Branching rule and fixed subspaces]
\label{lem:pre-branching}
For \(m\ge1\), let \(S_{m-1}\le S_m\) fix any chosen label. Then
\begin{equation}
 \operatorname{Res}_{S_{m-1}}^{S_m}V_\lambda
 \cong\bigoplus_{\nu\nearrow\lambda}V_\nu.
\end{equation}
More generally, for \(0\le a\le m\), let \(S_a\le S_m\) permute
any \(a\) labels and fix the others. Then
\begin{equation}
 \dim(V_\lambda^{S_a})=f^{\lambda/(a)}.
\end{equation}
\end{lemma}

The Pieri rule decomposes the space of injections under relabellings
of values and positions. We use it to distinguish information requiring
$s$ recorded assignments from information already present with fewer
assignments.

\begin{lemma}[{\cite[Section~4.9]{sagan2001symmetric},
\cite[Section~6.2, Eq.~(12), and Appendix~A.3, proof of Claim~21]{rosmanis2021tight}}
Pieri rule and the representation on injections]
\label{lem:pre-pieri}
For $0\le s\le m$, the space of injections carries the value action
of $S_m$ and the position action of $S_s$, with decomposition
\begin{equation}
\label{pre:injection-decomposition}
 \mathbb C[\operatorname{Inj}([s],[m])]
 \cong
 \bigoplus_{\substack{\lambda\vdash m,\ \mu\vdash s\\
                      \lambda/\mu\ \text{a horizontal strip}}}
       V_\lambda\otimes V_\mu.
\end{equation}
Here $\lambda/\mu$ is a horizontal strip exactly when
$\lambda_i\ge\mu_i\ge\lambda_{i+1}$ for every $i\ge1$.
For $\lambda=(m-r,\nu)$, this gives $r\le s$ and
$\nu\subseteq\mu$, with $\nu=\mu$ when $r=s$.
\end{lemma}

The transposition class sum acts by a scalar on each irreducible
representation. The following formula will make the spectrum of the
compression Laplacian explicit.

\begin{lemma}[{\cite[Problem~5.16.2]{etingof2011representation}}
Transposition class sum]
\label{lem:pre-content}
For \(\lambda\vdash m\), the transposition class sum from
Eq.~\eqref{eq:transposition-class-sum} acts on \(V_\lambda\) by
\begin{equation}
 \chi_\lambda(Z_m)=c(\lambda)I_{V_\lambda}.
\end{equation}
\end{lemma}

We also need ratios of representation dimensions to normalize the
completion frames. The hook-length formula expresses these dimensions
through the boxes of the Young diagram.

\begin{lemma}[{\cite[Section~3.10]{sagan2001symmetric}} Hook-length formula]
\label{lem:pre-hook}
For \(\lambda\vdash m\),
\begin{equation}
 f^\lambda=\frac{m!}{\prod_{(i,j)\in\lambda}h_\lambda(i,j)}.
\end{equation}
\end{lemma}

The next lemma identifies the database subspace representing permutation
degree $s$ by matching first-row lengths under the range and domain
actions. Its proof is in Appendix~\ref{app:partial-bijections-proof}.

\begin{lemma}[Partial bijections and simultaneous first-row types]
\label{lem:pre-partial-bijections}
For $0\le s<n$, the sum of joint isotypic subspaces of
$\mathbb C[\{d\in\mathcal I_n:|d|=s\}]$ whose range and domain types
both have first row $n-s$ is isomorphic to
\begin{equation}
\label{pre:simultaneous-first-row}
 \bigoplus_{\substack{\eta\vdash s\\\eta_1\le n-s}}
 V_{(n-s,\eta)}\otimes V_{(n-s,\eta)}.
\end{equation}
\end{lemma}

A phase query at $i$ multiplies each permutation basis vector by a
function of $\pi(i)$. The next lemma bounds its change in first-row
length by one; see Appendix~\ref{app:point-multiplier} for the proof.

\begin{lemma}[First-row support of a multiplication operator]
\label{lem:pre-point-multiplier}
For every $i\in[m]$, $g:[m]\to\mathbb C$,~and~$\lambda\vdash m$,
\begin{equation}
\label{pre:point-multiplier-support}
 M_{i,g}\mathcal P_\lambda
 \subseteq\bigoplus_{\substack{\mu\vdash m\\|\lambda_1-\mu_1|\le1}}
                   \mathcal P_\mu.
\end{equation}
\end{lemma}

\section{A Harmonic Framework for Collision Lower Bounds}
\label{sec:harmonic-framework}

\label{sec:setting}
This section develops the tools shared by the adaptive and non-adaptive
lower bounds. We represent information about a uniform permutation of
$[n]$ in a database of partial assignments. There are two ways to
describe a query in this database space: an exact change of basis for
the genuine oracle, and a simpler local operation that adds, erases,
or changes an assignment. We construct the exact encoding, compare
the two query operations, and show how a projection reduces their
difference. These constructions do not depend on the algorithm's
final output.

We then relate a correct output to the database. An algorithm may
guess positions that it has not recorded, so we append queries that
check its proposed witness. The resulting bound expresses success
in terms of the probability that the exact encoded database already
contains a full bucket, together with the error from these checks.
Sections~\ref{sec:adaptive} and~\ref{sec:nonadaptive-promise} bound
that probability in different ways. The decision lower bound uses
the same encoding to compare input distributions in
Lemma~\ref{lem:switch}.

\subsection{Partial injections and local recording queries}
\label{adp:auxiliary}

The local compression below follows the forward-query rule of
Carolan~\cite{carolan2026compressed}.
Its definition gives an auxiliary unitary on databases. We will bound its
difference from a genuine permutation query.
\begin{definition}[Database space and size decomposition]
\label{def:database-space}
The \emph{database space} is $\mathcal D=\mathbb C[\mathcal I_n]$.
For $0\le s,t\le n$, define
\begin{equation}
\label{adp:database-grading}
 \mathcal D_s:=\operatorname{Span}\{|d\rangle:d\in\mathcal I_n,\ |d|=s\},
 \qquad \mathcal D:=\bigoplus_{s=0}^n\mathcal D_s,
 \qquad \mathcal D_{\le t}:=\bigoplus_{s=0}^{t}\mathcal D_s.
\end{equation}
Set $\mathcal D_s=\{0\}$ for $s<0$ or $s>n$.
\end{definition}

Thus $\mathcal D_{\le t}$ contains superpositions of databases with at
most $t$ assignments, and
$\mathcal D_0=\operatorname{Span}\{|\emptyset\rangle_{\mathcal D}\}$.
Here $|\emptyset\rangle_{\mathcal D}$ is the empty-database basis vector,
while $\ket{\widehat{\emptyset}}_{\mathcal P}$ is a uniform superposition
of permutations.
For a database vector $w$, we write $w(d):=\langle d|w\rangle$ for its
coefficient at $d$; thus $w=\sum_d w(d)|d\rangle$.

\begin{definition}[Local compression]
\label{def:local-compression}
For $i\in[n]$ and $r\in\mathcal I_n$ with $i\notin\operatorname{Dom}r$,
and for $y\notin\operatorname{Im}r$, put
\[
\begin{gathered}
 \mathcal D_{i,r}:=\operatorname{Span}
   \bigl(\{|r\rangle\}\cup
     \{|r\cup\{i\mapsto z\}\rangle:z\notin\operatorname{Im}r\}\bigr),\\
 \ket{u_{i,r}}:=\frac1{\sqrt{n-|r|}}
       \sum_{z\notin\operatorname{Im}r}|r\cup\{i\mapsto z\}\rangle,
 \qquad
 \ket{v_{y,r}}:=\frac1{\sqrt{n-|r|}}
       \sum_{z\notin\operatorname{Dom}r}|r\cup\{z\mapsto y\}\rangle.
\end{gathered}
\]
The local and full \emph{compression operators}
$C_{i,r}:\mathcal D_{i,r}\to\mathcal D_{i,r}$ and
$C_i:\mathcal D\to\mathcal D$ are
\begin{equation}
\label{adp:compressor}
\begin{aligned}
 C_{i,r}:=I_{\mathcal D_{i,r}}-(|r\rangle-\ket{u_{i,r}})
          (\langle r|-\langle u_{i,r}|), \qquad
 C_i:=\sum_{\substack{r\in\mathcal I_n\\i\notin\operatorname{Dom}r}}
                       C_{i,r}P_{\mathcal D_{i,r}}.
\end{aligned}
\end{equation}
\end{definition}

The vector $\ket{u_{i,r}}$ is the normalized uniform superposition
of the $n-|r|$ available assignments at $i$. It is orthogonal to
$|r\rangle$, which leaves $i$ unassigned.

Deleting the assignment at $i$ gives a unique residual database $r$,
so $\mathcal D=\bigoplus_{r:\,i\notin\operatorname{Dom}r}\mathcal D_{i,r}$
orthogonally. The operator $C_{i,r}$ swaps
$|r\rangle$ and $\ket{u_{i,r}}$ and fixes their orthogonal
complement. Thus each $C_{i,r}$, and hence $C_i$, is a unitary reflection.

Use the answer alphabet and query registers from
Definition~\ref{def:permutation-query}, with $0$ reserved for a blank.

\begin{definition}[Auxiliary database queries]
\label{def:auxiliary-query}
For $i\in[n]$ and $c\in\mathbb Z_{n+1}$, define
\begin{equation}
 D_{i,c}:\mathcal D\to\mathcal D:\quad
 |d\rangle\longmapsto\begin{cases}
   \omega_{n+1}^{c d(i)}|d\rangle,&i\in\operatorname{Dom}d,\\
   |d\rangle,&i\notin\operatorname{Dom}d.
 \end{cases}
\end{equation}
The \emph{auxiliary phase queries} $\mathcal O_{i,c}$ act on
$\mathcal D$, and their controlled sum $\mathcal O$ acts on
$\mathcal D\otimes\mathcal I\otimes\mathcal C_\pi$:
\begin{equation}
\label{adp:aux-query}
 \mathcal O_{i,c}:\mathcal D\to\mathcal D:\quad
 w\longmapsto C_iD_{i,c}C_iw,
 \qquad
 \mathcal O:=\bigoplus_{i\in[n],\,c\in\mathbb Z_{n+1}}\mathcal O_{i,c}.
\end{equation}
\end{definition}

Here the direct sum acts on
$\mathcal D\otimes\mathcal I\otimes\mathcal C_\pi$,
with $\mathcal O_{i,c}$ acting on the database factor of the sector
$\mathcal D\otimes\operatorname{Span}\{|i\rangle|c\rangle\}$.
The unitary $D_{i,c}$ is the phase form of database evaluation, using the
Fourier convention of~\eqref{eq:oracle}. It has the same role as
$O_{i,c}$ in~\eqref{eqn:zhandryOic}, with the database replacing the
genuine input register. In the local recorder $C_iD_{i,c}C_i$, the first
compression exchanges a blank at $i$ with the uniform superposition of
available assignments, $D_{i,c}$ applies the phase, and the second
compression reverses the change of representation.
The full query $\mathcal O$ is controlled by
the algorithm's index and Fourier-answer registers, so both labels may
be in coherent superposition. For fixed $c$, write $D_i=D_{i,c}$ and
$\mathcal O_i=\mathcal O_{i,c}$. Auxiliary value queries refer to the
Fourier-equivalent additive form of the same operation.

The estimates below also allow an arbitrary phase of modulus one
for each output slot and for a blank position, provided the phases
do not depend on the residual database $r$. In particular, the phase
may depend only on the bucket containing $d(i)$.

We retain the genuine states $\psi_t,\psi'_t$ and the algorithm-only
unitaries $U_t:\mathcal A\to\mathcal A$ from
Eq.~\eqref{eq:genuine-process}. For auxiliary query operators
$\mathcal Q_{t+1}:
\mathcal D\otimes\mathcal A\to\mathcal D\otimes\mathcal A$,
define the \emph{auxiliary states}
$\phi_t,\phi'_t\in\mathcal D\otimes\mathcal A$ by
\begin{equation}
\label{eq:auxiliary-process}
 \phi_0:=|\emptyset\rangle_{\mathcal D}\otimes|0\rangle_{\mathcal A},
 \qquad \phi'_t:=U_t\phi_t,\qquad
 \phi_{t+1}:=\mathcal Q_{t+1}\phi'_t.
\end{equation}
Here $\phi_t$ is the joint state immediately after $t$ queries;
algorithm-only unitaries act as the identity on $\mathcal D$.
Unless a projection is specified, $\mathcal Q_{t+1}=\mathcal O$.
A final algorithm-only unitary can be absorbed into the final measurement;
we then use $\psi_T$ and $\phi_T$ for the final states.
The genuine process is depicted in Figure~\ref{fig:QAPerm}, and the
auxiliary process with $\mathcal Q_t=\mathcal O$ is depicted in
Figure~\ref{fig:QAAux}.

\begin{center}
\begin{minipage}{\linewidth}
    \centering
    \captionsetup{type=figure}
    \begin{tikzpicture}[query circuit]

      \draw (0,0) -- (12.45,0);
      \foreach \y in {-.85,-1.70,-2.55}
        \draw (0,\y) -- (11.55,\y);

      \node[anchor=east] at (-.12,0) {$\mathcal D$};
      \node[anchor=east] at (-.12,-.85) {$\mathcal I$};
      \node[anchor=east] at (-.12,-1.70) {$\mathcal C_\pi$};
      \node[anchor=east] at (-.12,-2.55) {$\mathcal W_\pi$};
      \node[anchor=east] at (-1.50,0) {$\ket{\emptyset}_{\mathcal D}$};
      \draw[decorate,decoration={brace,amplitude=4pt}]
        (-1.30,-2.82) -- (-1.30,-.58);
      \node[anchor=east] at (-1.50,-1.70) {$\ket{0}_{\mathcal A}$};

      \node[algorithm gate] at (.85,-1.70) {$U_0$};
      \node[oracle gate] at (2.30,-.85) {$\mathcal O$};
      \node[algorithm gate] at (3.75,-1.70) {$U_1$};
      \foreach \y in {0,-.85,-1.70,-2.55}
        \node[fill=white,inner xsep=5pt,inner ysep=2pt]
          at (5.00,\y) {$\cdots$};
      \node[algorithm gate,minimum width=1.10cm] at (6.50,-1.70) {$U_{T-1}$};
      \node[oracle gate] at (8.05,-.85) {$\mathcal O$};
      \node[algorithm gate] at (9.55,-1.70) {$U_T$};

      \pic at (11.00,-1.70) {query measurement};
      \draw[classical output] (11.55,-1.70) -- (12.45,-1.70);
      \node[anchor=west] at (12.55,-1.70) {$\rho$};
     \querycut{0}{-3.0}{.35}{\phi_0}
 \querycut{1.48}{-3.0}{.35}{\phi'_0}
 \querycut{3.1}{-3.0}{.35}{\phi_1}
 \querycut{4.4}{-3.0}{.35}{\phi'_1}
 \querycut{5.7}{-3.0}{.35}{\phi_{T-1}}
 \querycut{7.25}{-3.0}{.35}{\phi'_{T-1}}
 \querycut{8.8875}{-3.0}{.35}{\phi_T}
 \querycut{10.25}{-3.0}{.35}{\phi'_T}
\end{tikzpicture}
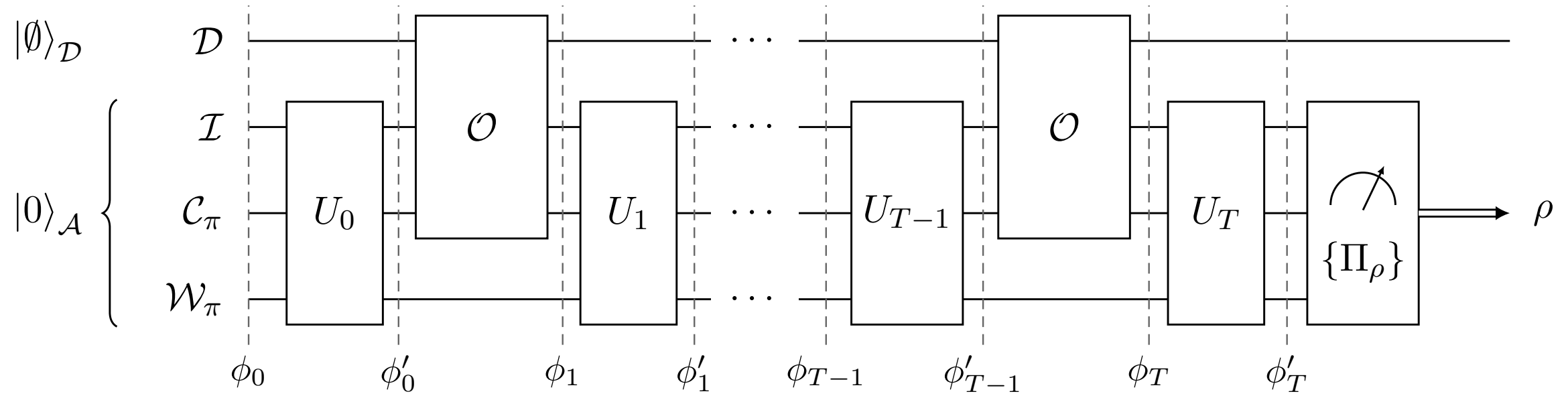
    \caption{The auxiliary quantum algorithm with query operator
    $\mathcal O$. The database register starts in
    $\ket{\emptyset}_{\mathcal D}$, and each auxiliary query acts on
    $\mathcal D\otimes\mathcal I\otimes\mathcal C_\pi$.
    The final PVM $\{\Pi_\rho\}$ acts on $\mathcal A$ and outputs
    a candidate set $\rho\subseteq[n]$.}
    \label{fig:QAAux}
\end{minipage}
\end{center}

Both $C_i$ and $D_{i,c}$ preserve every subspace $\mathcal D_{i,r}$.
Consequently, so does $\mathcal O_{i,c}$: a query leaves the residual database
$r$ unchanged and may add, erase, or replace the assignment at $i$.
Since $\mathcal D_{i,r}\subseteq\mathcal D_{|r|}\oplus\mathcal D_{|r|+1}$,
and a size-$s$ basis database belongs to a block with $|r|=s$ or $s-1$,
we obtain
\begin{equation}
\label{adp:database-query-support}
 \mathcal O_{i,c}\mathcal D_s
 \subseteq\mathcal D_{s-1}\oplus\mathcal D_s\oplus\mathcal D_{s+1},
 \qquad
 \mathcal O(\mathcal D_{\le t}\otimes\mathcal A)
 \subseteq\mathcal D_{\le t+1}\otimes\mathcal A.
\end{equation}
Algorithm-only unitaries preserve $\mathcal D_{\le t}\otimes\mathcal A$.
Starting from $\phi_0\in \mathcal D_{\le 0}\otimes\mathcal A$, induction therefore shows that the post-query state
after $t$ queries satisfies $\phi_t\in\mathcal D_{\le t}\otimes\mathcal A$ and
\begin{equation}
 \phi_t=\sum_{d\in\mathcal I_n:\,|d|\le t}|d\rangle\otimes a_{t,d},
 \quad a_{t,d}\in\mathcal A.
\end{equation}
In the adaptive setting, the vectors $a_{t,d}$ need not be mutually orthogonal.

\subsection{Harmonic database spaces and exact permutation frames}
\label{bc:harmonic-spaces}

Recall the database decomposition
$\mathcal D=\bigoplus_{s=0}^n\mathcal D_s$
from~\eqref{adp:database-grading}.
A vector supported on size-$s$ databases can still represent
information obtainable from fewer than $s$ assignments. For example, the uniform extension
$|u_{i,r}\rangle$ of a size-$(s-1)$ database $r$ names one more
position but averages over all its remaining values. Its genuine
completion state is exactly that of $r$; the same holds when a
fixed unused value is placed uniformly at the remaining positions
(Eq.~\eqref{eq:redundant-completions}). We introduce harmonic
spaces to remove these redundant directions
by imposing zero-sum conditions. This will let the exact encoding represent
each permutation degree at the corresponding database size.

\begin{definition}[Permutation degree]
\label{def:permutation-degree}
For $0\le s\le n-1$, put
\[
 \mathcal P_s:=\bigoplus_{\substack{\lambda\vdash n\\\lambda_1=n-s}}
                 \mathcal P_\lambda,
 \qquad \mathcal P_{\le t}:=\bigoplus_{s\le t}\mathcal P_s,
\]
and set $\mathcal P_s=\{0\}$ outside this range.
A joint state has \emph{permutation degree at most $t$} if it lies~in~$\mathcal P_{\le t}\otimes\mathcal A$.
\end{definition}

Grouping the regular decomposition of Lemma~\ref{lem:pre-regular}
by first-row length gives the~orthogonal~decomposition
$\mathcal P=\bigoplus_{s=0}^{n-1}\mathcal P_s$.

The number of queries bounds the permutation degree of the algorithm's state.

\begin{lemma}[Permutation degree after forward queries]
\label{bc:query-degree}
For a $T$-query algorithm, the states $\psi_t$ in
Eq.~\eqref{eq:genuine-process} satisfy
\[
 \psi_t\in\mathcal P_{\le t}\otimes\mathcal A
 \qquad(0\le t\le T).
\]
\end{lemma}
\begin{proof}
For fixed $i,c$, write $\mathsf O_{i,c}=M_{i,g_c}$ with
$g_c(y)=\omega_{n+1}^{cy}$. Lemma~\ref{lem:pre-point-multiplier},
applied with $m=n$, gives
\[
 \mathsf O_{i,c}\mathcal P_s
 \subseteq\mathcal P_{s-1}\oplus\mathcal P_s\oplus\mathcal P_{s+1}.
\]
Summing over $s\le t$ shows that each $\mathsf O_{i,c}$ maps
$\mathcal P_{\le t}$ into $\mathcal P_{\le t+1}$.
The same degree bound holds for the coherently controlled query
by taking the direct sum over $i,c$ and tensoring with the workspace.

The initial uniform permutation vector belongs to $\mathcal P_0$,
and algorithm-only operations preserve permutation degree.
Induction using Eq.~\eqref{eq:genuine-process} proves the claim.
\end{proof}

For $0\le s\le n$ and $\tau,\sigma\in S_n$, use the range and domain
representations on $\mathcal D_s$ given by
\[
 \chi_R^{(s)}(\tau)|d\rangle:=|\tau\circ d\rangle,\qquad
 \chi_D^{(s)}(\sigma)|d\rangle:=|d\circ\sigma^{-1}\rangle.
\]
Write $\mathcal D_{s,\lambda}^R$ and $\mathcal D_{s,\lambda}^D$ for
their respective $\lambda$-isotypic subspaces.

\begin{definition}[Range- and domain-harmonic database spaces]
\label{def:harmonic-spaces}
For $0\le s\le n$, define the \emph{range-harmonic} and
\emph{domain-harmonic spaces} below. Their projectors
$K_R^{(s)},K_D^{(s)}:\mathcal D_s\to\mathcal D_s$ and the full
projectors $K_R,K_D:\mathcal D\to\mathcal D$ are
\begin{equation}
\begin{gathered}
 \mathcal D_R^{(s)}:=\bigoplus_{\substack{\lambda\vdash n\\\lambda_1=n-s}}
                       \mathcal D_{s,\lambda}^R,
 \qquad
 \mathcal D_D^{(s)}:=\bigoplus_{\substack{\lambda\vdash n\\\lambda_1=n-s}}
                       \mathcal D_{s,\lambda}^D,\\
 K_R^{(s)}:=P_{\mathcal D_R^{(s)}},\qquad
 K_D^{(s)}:=P_{\mathcal D_D^{(s)}},\qquad
 K_R:=\bigoplus_{s=0}^nK_R^{(s)},\qquad
 K_D:=\bigoplus_{s=0}^nK_D^{(s)}.
\end{gathered}
\end{equation}
\end{definition}

At database size $s$, these spaces keep only representation types with
first row $n-s$. The range condition concerns output values, and the
domain condition concerns input positions. The decompositions into
types give
\[
 \mathcal D_s=\bigoplus_{\lambda\vdash n}\mathcal D_{s,\lambda}^R
            =\bigoplus_{\lambda\vdash n}\mathcal D_{s,\lambda}^D,
 \qquad
 K_R^{(s)}=\sum_{\lambda:\,\lambda_1=n-s}P_{\mathcal D_{s,\lambda}^R},
 \quad
 K_D^{(s)}=\sum_{\lambda:\,\lambda_1=n-s}P_{\mathcal D_{s,\lambda}^D}.
\]
An empty sum is zero.

Apply the injection decomposition~\eqref{pre:injection-decomposition}
in Lemma~\ref{lem:pre-pieri}, with $m=n$, to each fixed domain set for
the range action, and to each fixed image set for the domain action.
It follows that only types with $\lambda_1\ge n-s$ can occur in these
decompositions.
Consequently, $K_R^{(s)}$ annihilates precisely the components of
range degree $n-\lambda_1<s$, and $K_D^{(s)}$ annihilates those of
domain degree less than $s$. For $1\le s<n/2$, the
uniform-extension description of these removed components is given in
Eq.~\eqref{eq:harmonic-annihilation} below.
The range and domain actions commute, so each range-isotypic projector
commutes with each domain-isotypic projector. In particular,
$K_RK_D=K_DK_R$.
Thus $K_R^{(s)}K_D^{(s)}$ is the orthogonal projector onto
$\mathcal D_R^{(s)}\cap\mathcal D_D^{(s)}$.

\begin{definition}[Extension contractions and the compression Laplacian]
\label{def:compression-laplacian}
For $i\in[n]$, define the projectors
$P_{i=\bot},P_{i\ne\bot}:\mathcal D\to\mathcal D$ by
\[
 P_{i=\bot}:=\sum_{\substack{d\in\mathcal I_n\\i\notin\operatorname{Dom}d}}
                         |d\rangle\langle d|,\qquad
 P_{i\ne\bot}:=I_{\mathcal D}-P_{i=\bot}.
\]
The \emph{extension contraction} $\mathcal L_i:\mathcal D\to\mathcal D$
and, for $0\le s\le n$, the \emph{compression Laplacian}
$\Delta_s:\mathcal D_s\to\mathcal D_s$ are~defined~by
\[
 \mathcal L_i:=P_{i=\bot}C_i,\qquad
 \Delta_s:=\left.\sum_{i=1}^n\mathcal L_i^\dagger\mathcal L_i
                                                    \right|_{\mathcal D_s}.
\]
\end{definition}

The operator $\mathcal L_i$ takes a normalized sum over the possible
values at position $i$. Thus $\langle w,\Delta_s w\rangle$ measures the
total squared size of these sums. To compute its eigenvalues, introduce
the following sums of transpositions.
For $1\le s<n/2$ and a domain set $S\subseteq[n]$ of size $s$, define
operators on $\mathcal D_s$ and scalars for $\lambda\vdash n$, $\mu\vdash s$ by
\begin{equation}
\label{eq:database-transposition-sums}
\begin{gathered}
 Z_n^R:=\sum_{1\le a<b\le n}\chi_R^{(s)}((a\ b)),\qquad
 Z_S^D:=\sum_{\substack{i,j\in S\\i<j}}\chi_D^{(s)}((i\ j)),\\
 e_{\lambda,\mu}:=
 \frac{s+c(\lambda)-c(\mu)-\binom{n-s}{2}}{n-s+1}.
\end{gathered}
\end{equation}

\begin{claim}[Compression Laplacian and its eigenvalues]
\label{bc:laplacian-spectrum}
For $1\le s<n/2$ and every domain set $S\subseteq[n]$ of size $s$,
the operators from Eq.~\eqref{eq:database-transposition-sums} satisfy,
on the subspace spanned by databases with~domain~$S$,
\begin{equation}
\label{bc:laplacian}
 \Delta_s=\sum_{i\in S}\mathcal L_i^\dagger\mathcal L_i
 =\frac{sI_{\mathcal D_s}+Z_n^R-Z_S^D-\binom{n-s}{2}I_{\mathcal D_s}}{n-s+1}.
\end{equation}
Identify $S$ with $[s]$. In the injection decomposition
\eqref{pre:injection-decomposition} with $m=n$, the operator $\Delta_s$
acts on each constituent $V_\lambda\otimes V_\mu$ by the scalar
$e_{\lambda,\mu}$ from Eq.~\eqref{eq:database-transposition-sums}.
Writing $\lambda=(n-r,\nu)$, one has $\nu\subseteq\mu$ and $0\le r\le s$.
With $u:=s-r$, the same eigenvalue is
\[
 e_{\lambda,\mu}
 =\frac{u(n-s+1)+\binom u2-\bigl(c(\mu)-c(\nu)\bigr)}{n-s+1}.
\]
\end{claim}
\begin{proof}
For $i\notin S$, a basis database with domain $S$ is blank at $i$.
The compression $C_i$ sends it to states assigning $i$, so
$\mathcal L_i=P_{i=\bot}C_i$ annihilates it.
For $i\in S$, removing and restoring the assignment at $i$ preserves $S$.
Thus $\Delta_s$ preserves this fixed-domain subspace and acts on it as
$\sum_{i\in S}\mathcal L_i^\dagger\mathcal L_i$.

Here $Z_n^R$ sums over all exchanges of two output values, whereas $Z_S^D$
sums over all exchanges of two input positions in $S$. The individual operators act by
\[
 \chi_R^{(s)}((a\,b))|d\rangle=|(a\,b)\circ d\rangle,
 \qquad
 \chi_D^{(s)}((i\,j))|d\rangle=|d\circ(i\,j)\rangle.
\]
Each sum commutes with its respective group action, since conjugation
permutes the transpositions. These are the central sums of transpositions
in the value and position actions.

To verify~\eqref{bc:laplacian}, fix a basis database $d:S\hookrightarrow[n]$,
so $|\operatorname{Im}d|=s$.
For $i\in S$, put $d_i^\circ:=d|_{S\setminus\{i\}}$.
After removing the assignment at $i$, the available values are
\[
 [n]\setminus\operatorname{Im}d_i^\circ
 =\{d(i)\}\cup([n]\setminus \operatorname{Im}d),
\]
so there are $n-s+1$ of them, including the original value $d(i)$.
We compute the two actions directly from the compression formula.
In the block $\mathcal D_{i,d_i^\circ}$, the blank state is
$|d_i^\circ\rangle$, and the normalized vector over the
available values is
\[
 \ket{u_{i,d_i^\circ}}
 =\frac1{\sqrt{n-s+1}}
      \sum_{y\notin\operatorname{Im}d_i^\circ}
          |d_i^\circ\cup\{i\mapsto y\}\rangle.
\]
Since $|d\rangle=|d_i^\circ\cup\{i\mapsto d(i)\}\rangle$ is one of the
basis states in this sum, while $|d_i^\circ\rangle$ is orthogonal to all~of~them,
\[
 \langle d_i^\circ|d\rangle=0,\qquad
 \langle u_{i,d_i^\circ}|d\rangle=\frac1{\sqrt{n-s+1}},\qquad
 \langle u_{i,d_i^\circ}|d_i^\circ\rangle=0.
\]
Substituting these inner products into~\eqref{adp:compressor} gives
\[
\begin{aligned}
 C_i|d\rangle
 &=|d\rangle-
   \bigl(|d_i^\circ\rangle-\ket{u_{i,d_i^\circ}}\bigr)
   \bigl(\langle d_i^\circ|d\rangle-
                     \langle u_{i,d_i^\circ}|d\rangle\bigr)\\
 &=|d\rangle-\frac{\ket{u_{i,d_i^\circ}}}{\sqrt{n-s+1}}
             +\frac{|d_i^\circ\rangle}{\sqrt{n-s+1}}.
\end{aligned}
\]
Both $|d\rangle$ and $\ket{u_{i,d_i^\circ}}$ are supported on
databases that assign $i$, whereas $d_i^\circ$ leaves $i$ unassigned. Applying $P_{i=\bot}$ therefore yields
\[
 \mathcal L_i|d\rangle
 =P_{i=\bot}C_i|d\rangle
 =\frac{P_{i=\bot}\bigl(|d_i^\circ\rangle-
                       \ket{u_{i,d_i^\circ}}\bigr)}
        {\sqrt{n-s+1}}
 =\frac{|d_i^\circ\rangle}{\sqrt{n-s+1}}.
\]

For the adjoint, $C_i^\dagger=C_i$ and
$P_{i=\bot}^\dagger=P_{i=\bot}$ give
$\mathcal L_i^\dagger=(P_{i=\bot}C_i)^\dagger=C_iP_{i=\bot}$.
Now $P_{i=\bot}|d_i^\circ\rangle=|d_i^\circ\rangle$,
$\langle d_i^\circ|d_i^\circ\rangle=1$, and
$\langle u_{i,d_i^\circ}|d_i^\circ\rangle=0$. Thus
\[
\begin{aligned}
 \mathcal L_i^\dagger|d_i^\circ\rangle
 =C_iP_{i=\bot}|d_i^\circ\rangle=C_i|d_i^\circ\rangle
 =\ket{u_{i,d_i^\circ}}
 =\frac1{\sqrt{n-s+1}}
      \sum_{y\notin\operatorname{Im}d_i^\circ}
          |d_i^\circ\cup\{i\mapsto y\}\rangle.
\end{aligned}
\]
Substituting the adjoint formula gives
\[
\begin{aligned}
 \mathcal L_i^\dagger\mathcal L_i|d\rangle
 =\frac1{\sqrt{n-s+1}}\mathcal L_i^\dagger|d_i^\circ\rangle
 =\frac1{n-s+1}\left(
       |d\rangle+\sum_{y\notin \operatorname{Im}d}|d_i^\circ\cup\{i\mapsto y\}\rangle\right).
\end{aligned}
\]
The factor $1/(n-s+1)$ comes from the two normalization factors in
$\mathcal L_i$ and $\mathcal L_i^\dagger$. When we sum over $i\in S$,
the unchanged term $|d\rangle$ therefore occurs $s$ times.

Divide the transpositions in $Z_n^R$ into three cases according to
which values lie in $\operatorname{Im}d$.
\begin{enumerate}
\item If neither value belongs to $\operatorname{Im}d$, the transposition leaves $d$ unchanged.
There are $\binom{n-s}{2}$ such transpositions, contributing
$\binom{n-s}{2}|d\rangle$.
\item If both values belong to $\operatorname{Im}d$, write them as $d(i)$ and $d(j)$ for
the unique distinct positions $i,j\in S$. Then
$(d(i)\,d(j))\circ d=d\circ(i\,j)$.
The unordered pairs of used values correspond bijectively to the unordered
pairs of positions in $S$, so these terms contribute exactly $Z_S^D|d\rangle$.
\item If one value is $d(i)\in \operatorname{Im}d$ and the other is $y\notin \operatorname{Im}d$, the
transposition changes only the assignment at $i$:
$(d(i)\,y)\circ d=d_i^\circ\cup\{i\mapsto y\}$.
Each pair $(i,y)$ with $i\in S$ and $y\notin \operatorname{Im}d$ occurs~exactly~once.
\end{enumerate}
Combining these three contributions gives
\[
 Z_n^R|d\rangle
 =\binom{n-s}{2}|d\rangle+Z_S^D|d\rangle
   +\sum_{i\in S}\sum_{y\notin \operatorname{Im}d}|d_i^\circ\cup\{i\mapsto y\}\rangle.
\]
Comparing this decomposition with the formula for $\mathcal L_i^\dagger\mathcal L_i|d\rangle$ and
summing that formula over $i\in S$ proves~\eqref{bc:laplacian} on
$|d\rangle$. These vectors span the subspace with domain $S$, proving
the operator identity.

Identify $S$ with $[s]$ and apply the injection
decomposition~\eqref{pre:injection-decomposition} in
Lemma~\ref{lem:pre-pieri} with $m=n$. Its constituents are
$V_\lambda\otimes V_\mu$, where $\lambda/\mu$ is a horizontal strip of
size $n-s$. Lemma~\ref{lem:pre-content}, applied with $m=n$ and $m=s$,
says that $Z_n^R$ and $Z_S^D$ act on the two factors by $c(\lambda)$ and
$c(\mu)$, respectively. Substituting these scalars into~\eqref{bc:laplacian}
gives the first eigenvalue formula.

For $\lambda=(n-r,\nu)$, the interlacing inequalities in
Lemma~\ref{lem:pre-pieri} give $\mu_i\ge\lambda_{i+1}=\nu_i$ for every $i$.
Hence $\nu\subseteq\mu$ and $r=|\nu|\le|\mu|=s$.
The first row contributes $\binom{n-r}{2}$ to $c(\lambda)$, and shifting
$\nu$ down by one row reduces its content sum by $r$, so
$c(\lambda)=\binom{n-r}{2}+c(\nu)-r$.
With $u=s-r$, this gives
\[
 s+c(\lambda)-c(\mu)-\binom{n-s}{2}
 =u(n-s+1)+\binom u2-\bigl(c(\mu)-c(\nu)\bigr),
\]
which proves the second eigenvalue formula.
\end{proof}

The eigenvalues above identify the kernel of $\Delta_s$. A vector
belongs to this kernel exactly when its coefficients sum to zero over
every possible extension at a fixed position. The next lemma shows
that this is precisely the range-harmonic condition.

\begin{lemma}[Range-harmonic zero-sum characterization]
\label{bc:row-zero-sums}
For $1\le s<n/2$, let
$a=\sum_{|d|=s}a_d|d\rangle\in\mathcal D_s$. The following conditions are equivalent:
\[
 a\in\mathcal D_R^{(s)}
 \quad\Longleftrightarrow\quad
 \mathcal L_i a=0\quad\text{for every }i\in[n].
\]
In coefficients, $a\in\mathcal D_R^{(s)}$ if and only if, for every
$i\in[n]$ and every database $r$ with $|r|=s-1$ and
$i\notin\operatorname{Dom}r$,
\begin{equation}
\label{bc:row-sum}
 \langle r|\mathcal L_i a\rangle
 =\frac1{\sqrt{n-s+1}}
       \sum_{y\notin\operatorname{Im}r}a_{r\cup\{i\mapsto y\}}
 =0.
\end{equation}
\end{lemma}
\begin{proof}
For $|r|=s-1$, $i\notin\operatorname{Dom}r$, and
$y\notin\operatorname{Im}r$, the compression~\eqref{adp:compressor} has
matrix element
\[\langle r|C_i|r\cup\{i\mapsto y\}\rangle=\frac{1}{\sqrt{n-s+1}}.\]
A size-$s$ basis state not assigning $i$ is sent by $C_i$ to states assigning
$i$, which are removed by $P_{i=\bot}$. This proves the coefficient formula
for $\mathcal L_i$. Moreover,
$\mathcal L_i a\in\operatorname{span}\{|r\rangle:|r|=s-1,\
i\notin\operatorname{Dom}r\}$, so the vanishing of all the coefficients
in~\eqref{bc:row-sum} is equivalent to $\mathcal L_i a=0$.
It remains to identify the common kernel of these operators.
For every $a\in\mathcal D_s$, the definition of $\Delta_s$ gives
\[
 \ker\Delta_s=\{a\in\mathcal D_s:\Delta_s a=0\},\qquad
 \langle a,\Delta_s a\rangle
 =\sum_{i=1}^n\|\mathcal L_i a\|^2.
\]
If $\Delta_s a=0$, the sum on the right is zero. Each summand is
nonnegative, so every $\mathcal L_i a$ is zero. Conversely, if every
$\mathcal L_i a$ is zero, then the definition gives $\Delta_s a=0$.
Consequently,
\[
 \ker\Delta_s
 =\{a\in\mathcal D_s:\mathcal L_i a=0
                     \text{ for every }i\in[n]\}.
\]
Apply Claim~\ref{bc:laplacian-spectrum} on each fixed-domain subspace.
For a constituent $V_\lambda\otimes V_\mu$, write $\lambda=(n-r,\nu)$
and $u=s-r$. The claim gives $\nu\subseteq\mu$ and $u\ge0$.
If $u=0$, then $\mu=\nu$, so the claim gives $e_{\lambda,\mu}=0$.
If $u\ge1$, the diagram $\mu/\nu$ has $u$ boxes. Since $|\mu|=s$,
each of these boxes has content at most $s-1<n-s+1$. Thus
$c(\mu)-c(\nu)<u(n-s+1)$, so the eigenvalue formula in the claim gives
$e_{\lambda,\mu}>0$.
We therefore have
\[
 e_{\lambda,\mu}=0
 \quad\Longleftrightarrow\quad
 u=0
 \quad\Longleftrightarrow\quad
 r=s
 \quad\Longleftrightarrow\quad
 \lambda_1=n-s.
\]
To relate this eigenvalue calculation to an arbitrary $a\in\mathcal D_s$,
let $a_S$ be its component supported on databases with~domain~$S$, so
$a=\sum_{S\subseteq[n],\,|S|=s}a_S$.
Within that subspace, use the orthogonal constituent decomposition above:
\[
 a_S=\sum_{\lambda,\mu}a_{S,\lambda,\mu},
 \qquad
 \Delta_s a_S=\sum_{\lambda,\mu}
                    e_{\lambda,\mu}a_{S,\lambda,\mu}.
\]
Here $a_{S,\lambda,\mu}$ is the component in the constituent
$V_\lambda\otimes V_\mu$; the sums range over the pairs occurring
in~\eqref{pre:injection-decomposition}, with $m=n$.
Since these components are mutually orthogonal, $\Delta_s a_S=0$
if and only if $e_{\lambda,\mu}a_{S,\lambda,\mu}=0$ for every pair.
For $\lambda_1=n-s$, the eigenvalue is zero, so any component of that
type is annihilated. Every other occurring type has $\lambda_1>n-s$
and a strictly positive eigenvalue, so its component must vanish if
$\Delta_s a_S=0$. Thus the kernel on this fixed-domain subspace consists
exactly of the components with $\lambda_1=n-s$.

Different domain sets $S$ give orthogonal invariant subspaces.
Since $\Delta_s$ preserves these subspaces, $\Delta_s a=0$ if and only if
$\Delta_s a_S=0$ for every $S$. Taking their direct sum therefore gives
\[
 \ker\Delta_s
 =\bigoplus_{\substack{\lambda\vdash n\\\lambda_1=n-s}}
                  \mathcal D_{s,\lambda}^R
 =\mathcal D_R^{(s)}.
\]
Combining this identity with the common-kernel characterization above yields
\[
 a\in\mathcal D_R^{(s)}
 \quad\Longleftrightarrow\quad
 \Delta_s a=0
 \quad\Longleftrightarrow\quad
 \mathcal L_i a=0\ \text{for every }i\in[n].
\]
Finally, the coefficient formula proved at the start says that
$\mathcal L_i a=0$ exactly when all its coefficients
$\langle r|\mathcal L_i a\rangle$ vanish. These are the normalized sums
in~\eqref{bc:row-sum}, giving the stated zero-sum condition.
\end{proof}

The coefficient test describes exactly what the harmonic projectors
remove. Fix $1\le s<n/2$ and a residual database $r$ of size $s-1$.
For $i\notin\operatorname{Dom}r$ and $y\notin\operatorname{Im}r$,
Lemma~\ref{bc:row-zero-sums} and its version with domains and images
interchanged imply
\begin{equation}
\label{eq:harmonic-annihilation}
 K_R^{(s)}|u_{i,r}\rangle=0,\qquad
 K_D^{(s)}|v_{y,r}\rangle=0.
\end{equation}
Indeed, Eq.~\eqref{bc:row-sum} says that the range-harmonic space
is the orthogonal complement of the span of the vectors
$|u_{i,r}\rangle$ from Definition~\ref{def:local-compression}.
Thus these uniform extensions span the entire kernel of
$K_R^{(s)}$. Transposing each partial injection exchanges range
and domain actions, giving the analogous statement for $K_D^{(s)}$.
The first projection removes directions that average over the
possible values at an additional position; the second removes
directions that average over the possible positions of an~additional~value.

The next lemma makes the zero-sum test quantitative: if the extension
sums have small total squared norm, then the vector is close to the
range-harmonic subspace. This estimate will bound the error introduced
by a recording query.

\begin{lemma}[Range-harmonic spectral gap]
\label{bc:row-gap}
For $1\le s<n/2$, every $w\in\mathcal D_s$ satisfies
\begin{equation}
\label{bc:gap}
 \|(I_{\mathcal D_s}-K_R^{(s)})w\|^2
 \le \frac{n-s+1}{n-2s+2}\sum_{i=1}^n\|\mathcal L_iw\|^2.
\end{equation}
\end{lemma}
\begin{proof}
Lemma~\ref{bc:row-zero-sums} identifies $\ker\Delta_s$ as
$\mathcal D_R^{(s)}$. To bound the nonzero eigenvalues, apply
Claim~\ref{bc:laplacian-spectrum} on each subspace with a fixed domain
set $S$. For a constituent $V_\lambda\otimes V_\mu$, write
$\lambda=(n-r,\nu)$ and $u=s-r$, with $\nu\subseteq\mu$ and $|\mu|=s$.
If $u=0$, then $\mu=\nu$ and the claim gives a zero eigenvalue.
Thus a nonzero eigenvalue requires $u\ge1$.
The $u$ boxes of $\mu/\nu$ each have content at most $s-1$, so
$c(\mu)-c(\nu)\le u(s-1)$. Substituting this into the second eigenvalue
formula of Claim~\ref{bc:laplacian-spectrum} gives
\[
 e_{\lambda,\mu}
 \ge\frac{u(n-2s+2)+\binom u2}{n-s+1}
 \ge\frac{n-2s+2}{n-s+1}.
\]
This bound holds for every domain set $S$, hence for every nonzero
eigenvalue of $\Delta_s$ on $\mathcal D_s$.

The projector $K_R^{(s)}$ is the orthogonal projector onto
$\ker\Delta_s$. Decompose
$w=K_R^{(s)}w+(I_{\mathcal D_s}-K_R^{(s)})w$.
The first component is annihilated by $\Delta_s$, and every eigenvalue
on the second component is at least the bound above. Therefore
\[
 \sum_i\|\mathcal L_iw\|^2
 =\langle w,\Delta_s w\rangle
 \ge\frac{n-2s+2}{n-s+1}\|(I_{\mathcal D_s}-K_R^{(s)})w\|^2,
\]
which is the claimed inequality.
\end{proof}

We next examine how the domain projection affects a database's image set.
It can redistribute assignments among positions but preserves the
set of occupied values. Consequently, it commutes with the
bucket-occupancy projectors used in the adaptive lower bound.

\begin{lemma}[Image-set preservation by the domain projection]
\label{bc:domain-images}
At every database size,~$K_D$~satisfies
\[
 \langle d'|K_D|d\rangle=0
 \quad\text{whenever }\operatorname{Im}d'\ne\operatorname{Im}d,
 \qquad d,d'\in\mathcal I_n.
\]
Consequently, $K_D$ commutes with every database-size projector and every
diagonal projector whose event depends only on the image set of a database.
\end{lemma}
\begin{proof}
The domain action satisfies
$\operatorname{Im}(d\circ\sigma^{-1})=\operatorname{Im}d$.
It therefore preserves every summand of the~orthogonal~decomposition
\[
 \mathcal D_s=
 \bigoplus_{\substack{Y\subseteq[n]\\|Y|=s}}
       \operatorname{span}\{|d\rangle:\operatorname{Im}d=Y\}.
\]
Decompose each summand into irreducible domain representations.
For every $\lambda$, the global $\lambda$-isotypic subspace is the
orthogonal sum of the corresponding isotypic subspaces for the individual
image sets. Hence each domain-isotypic projector, and therefore
$K_D^{(s)}$, acts separately on each image set. The asserted matrix
elements are zero. A projector depending only on the image set acts as a
scalar on each summand, and thus commutes with $K_D$. The same applies to
database size, \mbox{since $|d|=|\operatorname{Im}d|$}.
\end{proof}

For example, if $\operatorname{Im}d=\{2,5\}$, every basis database in
$K_D|d\rangle$ still records exactly the values $2$ and $5$. Only their
positions and the matching between positions and values may change.
This is the property needed for the progress bounds below.

We now introduce the completion maps used in the frame normalization.
\begin{definition}[Completion and synthesis maps]
\label{def:completion-maps}
For $0\le s\le n$, the completion maps $A_s:\mathcal D_s\to\mathcal P$
and full synthesis map $A:\mathcal D\to\mathcal P$ are~defined~by
\begin{equation}
\label{bc:completion}
 A_s|d\rangle:=|v_d\rangle
       :=\frac1{\sqrt{(n-s)!}}\sum_{\pi\supseteq d}|\pi\rangle,
 \qquad A:=\sum_{s=0}^n A_sP_{\mathcal D_s}.
\end{equation}
Here the sum over $\pi$ ranges over $S_n$, and $A_s$ is extended linearly.
\end{definition}

The outputs of the different size components are added in $\mathcal P$.
In particular, $|v_{\emptyset}\rangle=\ket{\widehat{\emptyset}}_{\mathcal P}$.
The completion vectors $|v_d\rangle$ can have overlapping supports and
need not be orthogonal, so they do not form a Fourier basis of $\mathcal P$.
The map $A$ is not an isometry on the entire database space. All uses of a
norm bound for $A$ below explicitly specify the appropriate harmonic or
isotypic subspace.

The completion maps make the redundancy of uniform extensions
explicit. For $1\le s\le n$, $|r|=s-1$,
$i\notin\operatorname{Dom}r$, and $y\notin\operatorname{Im}r$,
\begin{equation}
\label{eq:redundant-completions}
 A_s|u_{i,r}\rangle
 =A_s|v_{y,r}\rangle
 =A_{s-1}|r\rangle.
\end{equation}
Each permutation extending $r$ occurs exactly once in either
extension sum: at its value $\pi(i)$ or at the preimage of $y$.
The normalization factors give the normalized completion of $r$.
Thus these size-$s$ vectors represent information already
expressible at size $s-1$. The harmonic projections remove these
redundant directions at size $s$. The encoding below represents
the information they carry in the smaller database subspaces.

\begin{definition}[Frame normalization and the encoding map]
\label{def:encoding-map}
For $\lambda=(n-s,\nu)\vdash n$ with $|\nu|=s$, put
$\gamma_{n,\lambda}:=\binom ns f^\nu/f^\lambda$.
The frame normalization $\Gamma_n:\mathcal P\to\mathcal P$, component encodings
$V_s:\mathcal P_s\to\mathcal D_s$, and full encoding
$V:\mathcal P\to\mathcal D$ are
\begin{equation}
\label{bc:isometry}
\begin{gathered}
 \Gamma_n:=\sum_{\lambda\vdash n}\gamma_{n,\lambda}P_{\mathcal P_\lambda},\\
 V_s:=A_s^\dagger\Gamma_n^{-1/2}|_{\mathcal P_s}
 \quad(0\le s\le n-1),\qquad
 V:=\sum_{s=0}^{n-1}V_sP_{\mathcal P_s}.
\end{gathered}
\end{equation}
\end{definition}

Here $s=n-\lambda_1$ varies with $\lambda$, and all $\gamma_{n,\lambda}$
are positive, so $\Gamma_n^{-1/2}$ is well defined. The summands
defining $V$ lie in mutually orthogonal database-size subspaces
$\mathcal D_s$.

The next lemma gives the normalization that makes completion an
isometric encoding. The eigenvalues of $\Gamma_n$ are close to one at
low degree, which will also control the query comparison.

\begin{lemma}[Exact frame normalization]
\label{bc:frame}
For $\lambda=(n-s,\nu)\vdash n$ with $|\nu|=s$,
\begin{equation}
\label{bc:gamma}
\begin{gathered}
 \left.(A_sA_s^\dagger)\right|_{\mathcal P_\lambda}
 =\gamma_{n,\lambda}I_{\mathcal P_\lambda},\qquad
 \text{where}\quad
 \gamma_{n,\lambda}
 =\binom ns\frac{f^\nu}{f^\lambda}
 =\prod_{j=1}^{\nu_1}
       \left(1+\frac{\nu'_j}{n-s-j+1}\right).
\end{gathered}
\end{equation}
In particular, for $s<n/2$,
\begin{equation}
\label{bc:gamma-bounds}
 1\le\gamma_{n,\lambda}
 \le\exp\!\left(\frac{s}{n-2s+1}\right).
\end{equation}
\end{lemma}
\begin{proof}
For a fixed size-$s$ domain set, the completion vectors for its assignments
are orthonormal. Let $H_S\cong S_{n-s}$ be the domain subgroup fixing
every point of that set $S$. Its orbits are exactly the permutations
having the same assignments on $S$.
Write $R_\sigma|\pi\rangle:=|\pi\circ\sigma^{-1}\rangle$ for the domain
action. By Lemma~\ref{lem:pre-averages}, the projector onto the span of
the completion vectors with domain $S$ is the average over $H_S$.
Summing these projectors gives
\[
 A_sA_s^\dagger
 =\sum_{\substack{S\subseteq[n]\\|S|=s}}
       \sum_{\operatorname{Dom}(d)=S}|v_d\rangle\langle v_d|
 =\sum_{\substack{S\subseteq[n]\\|S|=s}}
       \frac1{(n-s)!}\sum_{\sigma\in H_S}R_\sigma.
\]
The domain action permutes the sets $S$ transitively, so this sum is
$\binom ns$ times the conjugate average of the projector for any one
fixed $S$.
Under the domain action, $\mathcal P_\lambda$ is a direct sum of
$f^\lambda$ copies of $V_\lambda$, by Lemma~\ref{lem:pre-regular}.
Apply the conjugate-average formula~\eqref{pre:conjugate-average} of
Lemma~\ref{lem:pre-averages} on each copy, with $m=n$ and
$H=H_S\cong S_{n-s}$.
It gives the same scalar on every copy and hence on all of
$\mathcal P_\lambda$.
Lemma~\ref{lem:pre-branching}, with $(m,a)=(n,n-s)$, then identifies
the fixed-space dimension:
\[
 \left.(A_sA_s^\dagger)\right|_{\mathcal P_\lambda}
 =\binom ns\frac{\dim(V_\lambda^{H_S})}{f^\lambda}
          I_{\mathcal P_\lambda}
 =\binom ns\frac{f^{\lambda/(n-s)}}{f^\lambda}
          I_{\mathcal P_\lambda}.
\]
For $\lambda_1=n-s$, the skew tableau is just the lower diagram $\nu$, giving
the operator identity in~\eqref{bc:gamma} with $\gamma_{n,\lambda}$ as
defined above. Grouping these identities over $\lambda_1=n-s$ gives
$\left.(A_sA_s^\dagger)\right|_{\mathcal P_s}
 =\left.\Gamma_n\right|_{\mathcal P_s}$.
Applying Lemma~\ref{lem:pre-hook} to $\lambda$ and $\nu$ gives the
product expression: the hooks below the first row are the hooks of $\nu$,
and the first-row hook in column $j\le\nu_1$ is
$n-s-j+1+\nu'_j$. Taking logarithms and using
$\sum_j\nu'_j=s$ proves~\eqref{bc:gamma-bounds}.
\end{proof}

\reviewrev{With this normalization, $V$ preserves inner products and maps the
permutation register onto the database subspace selected by both
harmonic projections. At low degree, this is exactly the subspace
satisfying both zero-sum conditions.}

\begin{lemma}[\reviewrev{Properties of the encoding}]
\label{bc:encoding}
The completion maps and the~encoding~satisfy
\begin{equation}
\label{bc:physical}
\begin{gathered}
 V^\dagger V=I_{\mathcal P},\qquad
 AV=\Gamma_n^{1/2},\qquad VV^\dagger=K_RK_D,\\
 A_sK_R^{(s)}=A_sK_D^{(s)}=P_{\mathcal P_s}A_s.
\end{gathered}
\end{equation}
\end{lemma}
\begin{proof}
We first verify directly that completion respects both relabelling actions.
For domain and range permutations $\sigma,\tau$, the definition
\eqref{bc:completion} gives
\[
\begin{aligned}
 A_s|\tau\circ d\circ\sigma^{-1}\rangle
 =\frac1{\sqrt{(n-s)!}}
       \sum_{\pi'\supseteq\tau\circ d\circ\sigma^{-1}}|\pi'\rangle
 =\frac1{\sqrt{(n-s)!}}
       \sum_{\pi\supseteq d}|\tau\circ\pi\circ\sigma^{-1}\rangle.
\end{aligned}
\]
The second equality uses the bijection
$\pi'=\tau\circ\pi\circ\sigma^{-1}$ between the two sets of completions.
Thus $A_s$ intertwines both relabelling actions and preserves both
representation types by Lemma~\ref{lem:pre-schur}.
In the genuine register, the domain and range
isotypic subspaces coincide by Lemma~\ref{lem:pre-regular}. Therefore
$A_sK_R^{(s)}=A_sK_D^{(s)}=P_{\mathcal P_s}A_s$.
Taking adjoints shows that
\[
 K_R^{(s)}V_s=V_s,\qquad K_D^{(s)}V_s=V_s,
 \qquad
 V_s\mathcal P_s
 \subseteq\mathcal D_R^{(s)}\cap\mathcal D_D^{(s)}.
\]
On each $\mathcal P_s$, Lemma~\ref{bc:frame}, Eq.~\eqref{bc:gamma}, gives
$\left.(A_sA_s^\dagger)\right|_{\mathcal P_s}
 =\left.\Gamma_n\right|_{\mathcal P_s}$.
Together with the definition~\eqref{bc:isometry}, this implies
$V_s^\dagger V_s=I_{\mathcal P_s}$ and
$A_sV_s=\left.\Gamma_n^{1/2}\right|_{\mathcal P_s}$.
Different $s$ use orthogonal database-size registers, so $V$ is an
isometry. Summing the identities for $A_sV_s$ gives
$AV=\Gamma_n^{1/2}$.

To complete the proof of~\eqref{bc:physical}, it remains to show that
the range inclusion above is an equality. By Definition~\ref{def:database-space},
$\mathcal D_s=\mathbb C[\{d\in\mathcal I_n:|d|=s\}]$.
Its range and domain actions are the relabellings used in
Lemma~\ref{lem:pre-partial-bijections}.
The projectors $K_R^{(s)}$ and $K_D^{(s)}$ retain the types whose
respective first rows are $n-s$. For $0\le s<n$, the simultaneous
first-row conclusion~\eqref{pre:simultaneous-first-row} of that lemma gives
\[
 \mathcal D_R^{(s)}\cap\mathcal D_D^{(s)}
 \cong\bigoplus_{\substack{\eta\vdash s\\\eta_1\le n-s}}
             V_{(n-s,\eta)}\otimes V_{(n-s,\eta)}.
\]
By Lemma~\ref{lem:pre-regular}, Eq.~\eqref{pre:regular-decomposition},
these are precisely the constituents of $\mathcal P_s$, each occurring
once. Hence
$\dim(\mathcal D_R^{(s)}\cap\mathcal D_D^{(s)})=\dim\mathcal P_s$.
The isometry $V_s$ maps $\mathcal P_s$ into this space, so the inclusion
proved above is an equality. Consequently
\[
 V_sV_s^\dagger=K_R^{(s)}K_D^{(s)}
 \quad\text{on }\mathcal D_s
 \qquad(0\le s\le n-1).
\]
For $s=n$, no partition $\lambda\vdash n$ has $\lambda_1=0$, so
$K_R^{(n)}=K_D^{(n)}=0$.
Also $V\mathcal P\subseteq\bigoplus_{s=0}^{n-1}\mathcal D_s$,
so $V^\dagger$ annihilates $\mathcal D_n$.
Thus both $VV^\dagger$ and $K_RK_D$ vanish on $\mathcal D_n$.
Summing over all database sizes proves~\eqref{bc:physical} on
$\mathcal D$.
\end{proof}

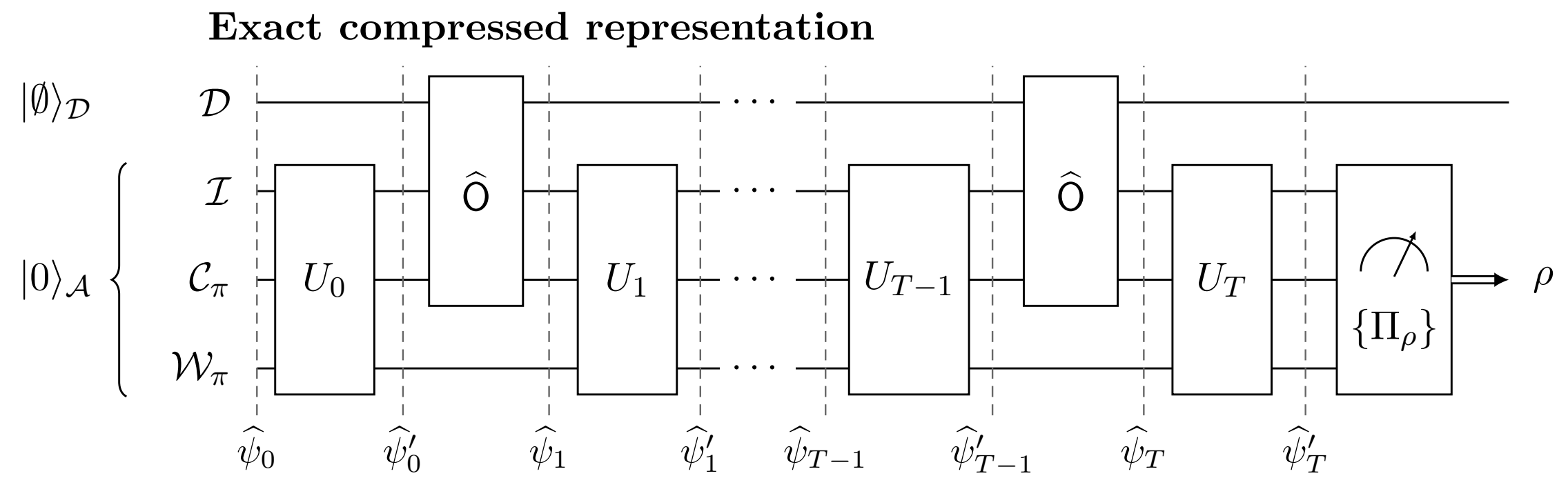
\begin{figure}[htbp]
\centering
\begin{tikzpicture}[query circuit]
 \draw (0,0)--(12.0,0);
 \foreach \y in {-.85,-1.70,-2.55}
   \draw (0,\y)--(11.45,\y);
 \node[anchor=east] at (-.12,0) {$\mathcal D$};
 \node[anchor=east] at (-.12,-.85) {$\mathcal I$};
 \node[anchor=east] at (-.12,-1.70) {$\mathcal C_\pi$};
 \node[anchor=east] at (-.12,-2.55) {$\mathcal W_\pi$};
 \node[anchor=east] at (-1.45,0)
   {$|\emptyset\rangle_{\mathcal D}$};
 \draw[decorate,decoration={brace,amplitude=4pt}]
   (-1.25,-2.82)--(-1.25,-.58);
 \node[anchor=east] at (-1.45,-1.70) {$|0\rangle_{\mathcal A}$};
 \node[algorithm gate] at (.65,-1.70) {$U_0$};
 \node[oracle gate,minimum width=.9cm] at (2.1,-.85)
   {$\widehat{\mathsf O}$};
 \node[algorithm gate] at (3.55,-1.70) {$U_1$};
 \foreach \y in {0,-.85,-1.70,-2.55}
   \node[fill=white,inner sep=3pt] at (4.8,\y) {$\cdots$};
 \node[algorithm gate,minimum width=1.15cm]
   at (6.25,-1.70) {$U_{T-1}$};
 \node[oracle gate,minimum width=.9cm] at (7.8,-.85)
   {$\widehat{\mathsf O}$};
 \node[algorithm gate] at (9.25,-1.70) {$U_T$};
 \pic at (10.9,-1.70) {query measurement};
 \draw[classical output] (11.45,-1.70)--(12.0,-1.70);
 \node[anchor=west] at (12.1,-1.70) {$\rho$};
 \node[anchor=west,font=\bfseries] at (-.6,.7)
   {Exact compressed representation};
 \querycut{0}{-3.0}{.35}{\widehat\psi_0}
 \querycut{1.4}{-3.0}{.35}{\widehat\psi'_0}
 \querycut{2.8}{-3.0}{.35}{\widehat\psi_1}
 \querycut{4.25}{-3.0}{.35}{\widehat\psi'_1}
 \querycut{5.45}{-3.0}{.35}{\widehat\psi_{T-1}}
 \querycut{7.05}{-3.0}{.35}{\widehat\psi'_{T-1}}
 \querycut{8.5}{-3.0}{.35}{\widehat\psi_T}
 \querycut{10.05}{-3.0}{.35}{\widehat\psi'_T}
\end{tikzpicture}
\caption{The exact database encoding of the circuit in
Figure~\ref{fig:QAPerm}. The registers start in
$|\emptyset\rangle_{\mathcal D}\otimes|0\rangle_{\mathcal A}$.
Each $\widehat{\mathsf O}$ implements a genuine permutation query on
the encoded subspace. The dashed cuts mark the encoded joint states
$\widehat\psi_j$ and $\widehat\psi'_j$.}
\label{fig:QACompressed}
\end{figure}

\paragraph{The exact encoded circuit.}
\label{sec:compressed-circuits}
Lemma~\ref{bc:encoding} shows that $V:\mathcal P\to\mathcal D$ is an
isometry. Moreover, $\gamma_{n,(n)}=1$, so the size-zero completion
map gives
\[
 V|\widehat\emptyset\rangle_{\mathcal P}=|\emptyset\rangle_{\mathcal D}.
\]
We can therefore express the genuine process in the database space,
starting from the empty database. Define its \emph{exact encoded oracle}
$\widehat{\mathsf O}:\mathcal D\otimes\mathcal A
\to\mathcal D\otimes\mathcal A$ by
\[
 \widehat{\mathsf O}
   =(V\otimes I_{\mathcal A})\mathsf O
                        (V^\dagger\otimes I_{\mathcal A})
     +(I_{\mathcal D}-VV^\dagger)\otimes I_{\mathcal A}.
\]
The first term implements the genuine query on
$V\mathcal P\otimes\mathcal A$; the second acts as the identity on
its orthogonal complement. Thus $\widehat{\mathsf O}$ is unitary on
all of $\mathcal D\otimes\mathcal A$.
Because $V$ acts only on the inaccessible register, the encoded
process has joint states
\[
 \widehat\psi_j=(V\otimes I_{\mathcal A})\psi_j,\qquad
 \widehat\psi'_j=(V\otimes I_{\mathcal A})\psi'_j
              =U_j\widehat\psi_j.
\]
They obey $\widehat\psi_{j+1}=\widehat{\mathsf O}\widehat\psi'_j$.
Figure~\ref{fig:QACompressed} shows the resulting circuit. It uses the
same algorithm-only unitaries and final measurement as
Figure~\ref{fig:QAPerm}, and has exactly the same output distribution.

For a parallel layer, replace $\mathsf O$ by the shared-register
query $\mathsf O^{[t]}$ from Eq.~\eqref{eq:parallel-permutation-query} in
the same construction, with $\mathcal A$ containing all query pairs.
This encodes the single shared permutation register as a single
database register. More generally, $V$ can be applied directly to
the joint state of a one-round algorithm whose input is determined
by the permutation. Section~\ref{sec:nonadaptive-promise} uses this
exact encoding to analyze the original parallel layer.

The next two subsections compare the recording query in
Figure~\ref{fig:QAAux} with the exact encoded query in
Figure~\ref{fig:QACompressed}. We first show that compressions at
different positions almost commute. This bounds the failure of the
range zero-sum condition after a query. We then compare local
decompression with exact conditioning on the queried value. Applying
the domain projection gives the smaller query error used in our
adaptive lower bound.

\FloatBarrier

\subsection{The error caused by a local recording query}
\label{bc:commutators}

The next lemma bounds the component that a local recording query sends
outside the range-harmonic subspace. We use this bound to compare
recording and genuine queries.

\begin{lemma}[Query error outside the range-harmonic subspace]
\label{bc:row-error}
For $t+1\le n/8$, $i\in[n]$, and $v\in\bigoplus_{s=0}^t\mathcal D_R^{(s)}$,
\begin{equation}
\label{bc:row-error-bound}
 \|(I_{\mathcal D}-K_R)\mathcal O_iv\|
 \le \frac{4\sqrt{2t}}{n-t-1}\|v\|.
\end{equation}
\end{lemma}

The bound scales as $\sqrt t/n$ when $t\le n/8$; the factor $n$
is outside the square root. To prove it, we first bound the
commutator of local compressions at distinct positions.

\begin{claim}[Two-coordinate compression commutator]
\label{bc:commutator}
Fix distinct positions $i,z$, remove both assignments from the database, and
suppose $m\ge2$ output slots remain unused by the residual database. On the
resulting invariant subspace,
\begin{equation}
\label{bc:commutator-exact}
 \|[C_i,C_z]\|
 =\frac{2}{m-1}\sqrt{1-\frac1{4(m-1)^2}}
 \le\frac2{m-1}.
\end{equation}
Consequently, on databases of size at most $t<n-1$,
\[
 \|[C_z,\mathcal O_i]\|\le\frac4{n-t-1}
 \qquad (i\ne z).
\]
\end{claim}
\begin{proof}
Suppress the residual database. The basis is the empty pair, states $|i,a\rangle$
and $|z,a\rangle$, and states $|i,a;z,b\rangle$ with $a\ne b$ in the $m$
available slots. On the four-dimensional invariant space spanned by the empty
pair and the uniform single- and double-assignment vectors, $C_i$ and $C_z$
are~commuting~swaps.

Choose an orthonormal basis of zero-sum vectors $h\in\mathbb C^m$. For each
such $h$, define
\[
\begin{aligned}
 i_h&=\sum_a h_a|i,a\rangle,
 &z_h&=\sum_b h_b|z,b\rangle,\\
 R_h&=\frac1{\sqrt{m-1}}\sum_{a\ne b}h_a|i,a;z,b\rangle,
 &S_h&=\frac1{\sqrt{m-1}}\sum_{a\ne b}h_b|i,a;z,b\rangle.
\end{aligned}
\]
The latter two vectors have norm one and
$\langle R_h,S_g\rangle=-\langle h,g\rangle/(m-1)$.
On their span together with $i_h,z_h$, the operators are the reflections with
unit normal vectors
\[
 a_h=(z_h-S_h)/\sqrt2,\qquad b_h=(i_h-R_h)/\sqrt2,
 \qquad \langle a_h,b_h\rangle=-\frac1{2(m-1)}.
\]
For two reflections $I_{\mathcal H}-2aa^\dagger,I_{\mathcal H}-2bb^\dagger$
on $\mathcal H$ with unit normal vectors $a,b$, put $u=|\langle a,b\rangle|$.
A direct two-dimensional calculation gives
commutator norm $4u\sqrt{1-u^2}$. The remaining double-assignment subspace
has zero row and column sums and is fixed by both compressions. This proves
\eqref{bc:commutator-exact}; the same calculation covers the degeneracy at
$m=2$.

For $z\ne i$, $C_z$ commutes with the diagonal phase operation $D_i$.
Expanding $[C_z,C_iD_iC_i]$ gives two terms, each bounded by
$\|[C_z,C_i]\|$. The residual after deleting $i,z$ has size at most $t$,
so $m\ge n-t$. Taking direct sums proves the remaining claims.
\end{proof}

\begin{proof}[Proof of Lemma~\ref{bc:row-error}]
Fix the phase query, so that $\mathcal O_i=C_iD_iC_i$ and $C_i^2=I_{\mathcal D}$.
For each $z\in[n]$, use $P_{z\ne\bot}$ and $P_{z=\bot}$ from
Definition~\ref{def:compression-laplacian}.
If $z\notin\operatorname{Dom}d$, the compression formula
\eqref{adp:compressor} sends $|d\rangle$ to the uniform superposition
of extensions assigning $z$. Projecting back onto databases with $z$
blank therefore gives zero. Consequently,
\[
 \mathcal L_zP_{z=\bot} = P_{z=\bot}C_zP_{z=\bot}=0,\qquad
 \mathcal L_z=\mathcal L_zP_{z\ne\bot}.
\]
Write $v=\sum_{s=0}^t v_s$ with $v_s\in\mathcal D_R^{(s)}$.
For $1\le s\le t<n/2$, Lemma~\ref{bc:row-zero-sums} gives
$\mathcal L_zv_s=0$. The vector $v_0$ is a multiple of the empty
database and is annihilated by $\mathcal L_zP_{z=\bot}=0$ as well.
By linearity, $\mathcal L_zP_{z\ne\bot}v=\mathcal L_zv=0$
for every $z\in[n]$.

\emph{The queried position $z=i$.}
The phase operator $D_i$ is diagonal in the database basis and hence
commutes with $P_{i=\bot}$. Using this commutation and $C_i^2=I_{\mathcal D}$ gives
\[
\begin{aligned}
 \mathcal L_i\mathcal O_iv
 =P_{i=\bot}C_i(C_iD_iC_i)v
 =P_{i=\bot}D_iC_iv
 =D_iP_{i=\bot}C_iv
 =D_i\mathcal L_iv=0.
\end{aligned}
\]

\emph{A different position $z\ne i$.}
The query $\mathcal O_i$ changes only the assignment at $i$, leaving
the residual database outside $i$ unchanged. In particular, it preserves
whether $z$ is assigned. Thus
\[
 P_{z=\bot}\mathcal O_i=\mathcal O_iP_{z=\bot},\qquad
 P_{z\ne\bot}\mathcal O_i=\mathcal O_iP_{z\ne\bot}.
\]
First insert the assigned projector using
$\mathcal L_z=\mathcal L_zP_{z\ne\bot}$, then move it through the query:
\[
\begin{aligned}
 \mathcal L_z\mathcal O_iv
 &=P_{z=\bot}C_zP_{z\ne\bot}\mathcal O_iv
  =P_{z=\bot}C_z\mathcal O_iP_{z\ne\bot}v\\
 &=P_{z=\bot}[C_z,\mathcal O_i]P_{z\ne\bot}v
   +P_{z=\bot}\mathcal O_iC_zP_{z\ne\bot}v.
\end{aligned}
\]
Here $[C_z,\mathcal O_i]=C_z\mathcal O_i-\mathcal O_iC_z$.
The second term vanishes because
\[
\begin{aligned}
 P_{z=\bot}\mathcal O_iC_zP_{z\ne\bot}v
 =\mathcal O_iP_{z=\bot}C_zP_{z\ne\bot}v
 =\mathcal O_i\mathcal L_zP_{z\ne\bot}v
 =\mathcal O_i\mathcal L_zv=0.
\end{aligned}
\]
We have therefore proved the exact identity
\[
 \mathcal L_z\mathcal O_iv
 =P_{z=\bot}[C_z,\mathcal O_i]P_{z\ne\bot}v
 \qquad(z\ne i).
\]
The vector $P_{z\ne\bot}v$ still belongs to $\mathcal D_{\le t}$,
since this diagonal projector preserves database size.
Applying the commutator bound of Claim~\ref{bc:commutator} to this
vector, and using $\|P_{z=\bot}\|\le1$, gives
\[
 \|\mathcal L_z\mathcal O_iv\|
 \le\|[C_z,\mathcal O_i]P_{z\ne\bot}v\|
 \le\frac4{n-t-1}\|P_{z\ne\bot}v\|.
\]

\emph{Sum over the positions.}
For a basis database $d$, exactly $|d|$ of its input positions are assigned. Summing the assignment projectors therefore gives
\[
 \left(\sum_{z=1}^nP_{z\ne\bot}\right)|d\rangle
 =|d|\,|d\rangle,\qquad
 \sum_{z=1}^nP_{z\ne\bot}
 =\sum_{s=0}^n sP_{\mathcal D_s}.
\]
Since each $P_{z\ne\bot}$ is an orthogonal projector and the components
$v_s$ are mutually orthogonal, this operator identity implies
\[
\begin{aligned}
 \sum_{z=1}^n\|P_{z\ne\bot}v\|^2
 =\sum_{z=1}^n\langle v,P_{z\ne\bot}v\rangle
 =\sum_{s=0}^t s\|v_s\|^2
 \le t\sum_{s=0}^t\|v_s\|^2=t\|v\|^2.
\end{aligned}
\]
Combining this bound with the preceding commutator estimate and
$\mathcal L_i\mathcal O_iv=0$ yields
\[
\begin{aligned}
 \sum_{z=1}^n\|\mathcal L_z\mathcal O_iv\|^2
 =\sum_{z\ne i}\|\mathcal L_z\mathcal O_iv\|^2
 \le\frac{16}{(n-t-1)^2}\sum_{z\ne i}\|P_{z\ne\bot}v\|^2
 \le\frac{16t}{(n-t-1)^2}\|v\|^2.
\end{aligned}
\]

\emph{Apply the spectral gap to each output size.}
By~\eqref{adp:database-query-support}, the output has size at most
$t+1$. Write
\[
 w:=\mathcal O_iv=\sum_{s=0}^{t+1}w_s,\qquad
 w_s:=P_{\mathcal D_s}\mathcal O_iv.
\]
The size-zero component is harmonic, so
$(I_{\mathcal D_0}-K_R^{(0)})w_0=0$.
For $1\le s\le t+1\le n/8<n/2$, apply
Lemma~\ref{bc:row-gap}, Eq.~\eqref{bc:gap}, to $w_s\in\mathcal D_s$:
\[
 \|(I_{\mathcal D_s}-K_R^{(s)})w_s\|^2
 \le\frac{n-s+1}{n-2s+2}\sum_{z=1}^n\|\mathcal L_zw_s\|^2
 \le2\sum_{z=1}^n\|\mathcal L_zw_s\|^2.
\]
The last inequality uses $n-s+1\le n$ and
$n-2s+2\ge3n/4+2\ge n/2$.
For each fixed $z$, the vectors $\mathcal L_zw_s\in\mathcal D_{s-1}$
for different $s\ge1$ are mutually orthogonal, while
$\mathcal L_zw_0=0$. Therefore
\[
 \sum_{s=1}^{t+1}\|\mathcal L_zw_s\|^2
 =\left\|\sum_{s=1}^{t+1}\mathcal L_zw_s\right\|^2
 =\|\mathcal L_z\mathcal O_iv\|^2.
\]
Also $K_R$ acts separately on the mutually orthogonal database-size
subspaces. Summing the spectral-gap inequalities and then using the
bound on the extension sums gives
\[
\begin{aligned}
 \|(I_{\mathcal D}-K_R)\mathcal O_iv\|^2
 &=\sum_{s=1}^{t+1}\|(I_{\mathcal D_s}-K_R^{(s)})w_s\|^2
 \le2\sum_{s=1}^{t+1}\sum_{z=1}^n\|\mathcal L_zw_s\|^2\\
 &=2\sum_{z=1}^n\|\mathcal L_z\mathcal O_iv\|^2
 \le\frac{32t}{(n-t-1)^2}\|v\|^2.
\end{aligned}
\]
Taking square roots, with $\sqrt{32t}=4\sqrt{2t}$ and $n-t-1>0$,
proves~\eqref{bc:row-error-bound}.
All calculations remain valid after tensoring with an identity on
ancillary registers. For coherent control, apply the same bound in each
query-index and phase-label sector and sum the squared norms over these
orthogonal sectors. The constant is independent of both labels, so the
same estimate holds for the controlled query.
\end{proof}

\subsection{Comparison with an exact conditional decompression}
\label{bc:conditional}

\reviewrev{We now compare a recording query with a genuine permutation query
under the encoding $V$. Table~\ref{tab:query-error-scales} distinguishes
the error outside the range-harmonic subspace from the full comparison
errors with and without the domain projection. The projected bound is
used for the adaptive simulation, and the unprojected bound for verification.}

\begin{table}[htbp]
\centering
\small
\setlength{\tabcolsep}{5pt}
\renewcommand{\arraystretch}{1.3}
\begin{tabular}{@{}lccc@{}}
\toprule
\textbf{Error}
& \textbf{Operator}
& \textbf{Norm bound}
& \textbf{Reference}\\
\midrule
\makecell[l]{Outside the\\range-harmonic subspace}
& $(I_{\mathcal D}-K_R)\mathcal O_iV$
& $O\!\left(\frac{\sqrt t}{n}\right)$
& Lemma~\ref{bc:row-error}\\
After domain projection
& $K_D\mathcal O_iV-V\mathsf O_i$
& $O\!\left(\frac{t+1}{n}\right)$
& Eq.~\eqref{bc:projected}\\
Without projection
& $\mathcal O_iV-V\mathsf O_i$
& $O\!\left(\sqrt{\frac{t+1}{n}}\right)$
& Eq.~\eqref{bc:raw}\\
\bottomrule
\end{tabular}
\caption{Three single-query error bounds for $t+3\le n/8$.
All operator norms are restricted to $\mathcal P_{\le t}$, and the
implicit constants are absolute. The first row controls one error
component; the other two compare with the genuine query.}
\label{tab:query-error-scales}
\end{table}

\begin{lemma}[Projected forward-query comparison]
\label{bc:comparison}
For $t+3\le n/8$, the following two bounds hold with a universal~constant~$C$:
\begin{align}
 \|K_D\mathcal O_iV-V\mathsf O_i\|_{\mathcal P_{\le t}}
 &\le C(t+1)/n,\label{bc:projected}\\
 \|\mathcal O_iV-V\mathsf O_i\|_{\mathcal P_{\le t}}
 &\le C\sqrt{(t+1)/n}.\label{bc:raw}
\end{align}
The estimates hold uniformly for arbitrary phase vectors, ancillary registers,
and coherent control of the query position and answer label.
\end{lemma}

To prove Lemma~\ref{bc:comparison}, we compare the local decompression
$C_iV$ with an exact isometry conditioned on the queried value $\pi(i)$.

For fixed $i,y\in[n]$, identify each of $[n]\setminus\{i\}$ and
$[n]\setminus\{y\}$ with $[n-1]$ in increasing order.
\reviewrev{Then $\pi^\circ\in S_{n-1}$ denotes a bijection from
$[n]\setminus\{i\}$ to $[n]\setminus\{y\}$.
Write $A_r^{(i,y)}$ and $V^{(i,y)}$ for the size-$r$ completion map
and the encoding on these residual label sets, obtained from
Eqs.~\eqref{bc:completion} and~\eqref{bc:isometry} with $n$ replaced by $n-1$.}

Insert the deleted assignment using the map
\[
 J_{iy}:\mathbb C[S_{n-1}]\to\mathcal P:\quad
 |\pi^\circ\rangle\longmapsto|\pi^\circ\cup\{i\mapsto y\}\rangle.
\]
The map $W_i$ first conditions on the exact value $\pi(i)$, encodes
the remaining permutation, and records $i\mapsto\pi(i)$. Define this
\emph{exact conditional decompression}
$W_i:\mathcal P\to\mathcal D$ and its overlap operator
$\mathcal B_i:\mathcal P\to\mathcal P$ by
\begin{equation}
\label{eq:conditional-decompression}
 W_i:=\sum_{y=1}^n\operatorname{add}_{iy}V^{(i,y)}J_{iy}^\dagger,
 \qquad \mathcal B_i:=W_i^\dagger C_iV.
\end{equation}

The images of $J_{iy}$, for fixed $i$ and varying $y$, are mutually
orthogonal and sum to $\mathcal P$.
By Lemma~\ref{bc:encoding}, Eq.~\eqref{bc:physical}, each residual
encoding is an isometry. Adding $i\mapsto y$ preserves norms and
keeps different $y$ sectors orthogonal, so $W_i$ is also an isometry.

Its image consists of databases assigning $i$ whose residual belongs to
the image of the residual encoding. For any phase vector, it obeys
\begin{equation}
\label{eq:conditional-query-intertwining}
 D_iW_i=W_i\mathsf O_i.
\end{equation}
In particular, the image of $W_i$ and its orthogonal complement are invariant
under $D_i$.

For $s\ge1$, deleting the queried assignment from a degree-$s$ sector leaves
residual degree $s-1$ or $s$. The residual encoding matches these
with database sizes $s-1$ and $s$, respectively; keeping both cases
is necessary to compute the overlap $\mathcal B_i$.
For the query comparison, the residual
$F=C_iV-W_i\mathcal B_i$ is orthogonal to the image of $W_i$.
The invariance under $D_i$ makes the mixed terms vanish, so $F$
appears at both ends of the remaining term, $F^\dagger D_iF$.
We bound one factor on $\mathcal P_{\le t}$ and the other on
$\mathcal P_{\le t+1}$, since a query increases degree by at most
one. Each factor has norm of order $\sqrt{(t+1)/n}$, so their
product contributes only $O((t+1)/n)$ to the query comparison.

\begin{claim}[Overlap with exact conditional decompression]
\label{bc:overlap}
For $t+2\le n/8$, the overlap error satisfies the following bound
with a universal~constant~$C_0$:
\begin{equation}
\label{bc:overlap-bound}
 \|\mathcal B_i-I_{\mathcal P}\|_{\mathcal P_{\le t}}
 \le C_0\frac{t+1}{n}.
\end{equation}
Moreover, $\mathcal B_i$ is a contraction and preserves every
permutation-degree subspace.
\end{claim}
\begin{proof}
Define the residual frame operator $\Gamma_i^{\rm res}:\mathcal P\to\mathcal P$ by
\[
 \Gamma_i^{\rm res}
 =\sum_y J_{iy}\Gamma_{n-1}^{(i,y)}J_{iy}^{\dagger}.
\]
Let $G_i\le S_n$ be the domain subgroup fixing $i$. In each sector
$\pi(i)=y$, the residual operator is the same linear combination of
the operators $R_\sigma$ for $\sigma\in G_i$, with coefficients
constant on conjugacy classes of $G_i$. Thus
$\Gamma_i^{\rm res}$ is a central operator for this subgroup. The
operator $\Gamma_n$ and all full permutation-degree projectors commute
with every $R_\sigma$, so they commute with
$\Gamma_i^{\rm res}$ as well.

\reviewrev{Let $P_\mu^{G_i}$ denote the $\mu$-isotypic projector for the domain
action of $G_i$ on $\mathcal P$. For $1\le s\le t$, denote the projectors
onto residual degrees $s-1$ and $s$ within $\mathcal P_s$ by
$\Pi_{i,s}^-$ and $\Pi_{i,s}^+$, respectively. Explicitly,
\[
 \Pi_{i,s}^-:=P_{\mathcal P_s}
     \sum_{\substack{\mu\vdash n-1\\\mu_1=n-s}}P_\mu^{G_i},\qquad
 \Pi_{i,s}^+:=P_{\mathcal P_s}
     \sum_{\substack{\mu\vdash n-1\\\mu_1=n-s-1}}P_\mu^{G_i}.
\]
The full degree projector $P_{\mathcal P_s}$ commutes with the
$G_i$-action, so these are orthogonal projectors. They select residual
degrees $s-1$ and $s$, respectively, since the residual degree of type
$\mu$ is $(n-1)-\mu_1$. The branching rule in
Lemma~\ref{lem:pre-branching} gives
\[
 \Pi_{i,s}^-+\Pi_{i,s}^+=P_{\mathcal P_s},\qquad
 \Pi_{i,s}^-\Pi_{i,s}^+=\Pi_{i,s}^+\Pi_{i,s}^-=0.
\]
The centrality properties above give the commutation relations
\[
\begin{gathered}
 [\Gamma_n,\Gamma_i^{\rm res}]=0,\qquad
 [P_{\mathcal P_s},\Gamma_n]
   =[P_{\mathcal P_s},\Gamma_i^{\rm res}]=0,\\
 [\Pi_{i,s}^{\pm},\Gamma_n]
   =[\Pi_{i,s}^{\pm},\Gamma_i^{\rm res}]=0.
\end{gathered}
\]
For $h\in\mathcal P_s$, put $h_-:=\Pi_{i,s}^-h$ and
$h_+:=\Pi_{i,s}^+h$, so $h=h_-+h_+$ orthogonally.}
All maps used below respect domain relabellings fixing $i$, so they
preserve these residual domain types.

For $v=V_sh$, Lemmas~\ref{bc:encoding} and
\ref{bc:row-zero-sums} give $P_{i=\bot}C_iv=\mathcal L_iv=0$.
Thus $C_iv$ is supported entirely on databases that assign $i$. Its size-$s$ part is
$P_{i\ne\bot}v$, and its other part has size $s+1$.
After deleting the assignment at $i$, these parts have residual sizes
$s-1$ and $s$, respectively. We also use the exact identity
\begin{equation}
\label{bc:AC}
 AC_i=A.
\end{equation}
Indeed, partitioning the completions of a blank residual $r$ according
to their value at $i$ gives
$A\sum_y|r\cup\{i\mapsto y\}\rangle
=\sqrt{n-|r|}\,A|r\rangle$; on an assigned basis vector the two
correction terms in $C_i$ cancel after applying $A$.

We compute the overlap separately on $h_-$ and $h_+$.
The adjoint of the residual encoding at residual size $r$ annihilates
every domain type except degree $r$. This follows from
Lemma~\ref{bc:encoding}, Eq.~\eqref{bc:physical}, applied on $n-1$
labels. In particular, the size-$(s+1)$ part of $C_iV_sh_-$ is killed by
$W_i^\dagger$: its residual size is $s$, but its residual domain degree
is $s-1$.
For its size-$s$ part, the definition of the~encoding~gives
\begin{equation}
\label{bc:assigned-identity}
\begin{aligned}
 P_{i\ne\bot}V_sh_-
 =\sum_y\operatorname{add}_{iy}
   \bigl(A_{s-1}^{(i,y)}\bigr)^\dagger
   J_{iy}^\dagger\Gamma_n^{-1/2}h_-
 =W_i(\Gamma_i^{\rm res})^{1/2}\Gamma_n^{-1/2}h_-.
\end{aligned}
\end{equation}
The first equality holds coefficientwise: deleting $i\mapsto y$ from
a size-$s$ completion vector gives the residual size-$(s-1)$
completion vector with the same normalization. For the second,
$J_{iy}^\dagger\Gamma_n^{-1/2}h_-$ has residual degree $s-1$, so
the residual encoding formula applies. Since $W_i$ is an isometry,
\[
 \mathcal B_i h_-
 =(\Gamma_i^{\rm res})^{1/2}\Gamma_n^{-1/2}h_-.
\]

For $h_+$, the size-$s$ part of $C_iV_sh_+$ vanishes. After deleting
$i$, this part would have residual size $s-1$ and residual domain degree
$s$. The injection decomposition in Lemma~\ref{lem:pre-pieri}, applied
with domain and range interchanged, shows that a size-$r$ database
space has domain degree at most $r$.
The remaining part has residual size and residual domain degree both
equal to $s$. On this subspace, the adjoint encoding identity in
Lemma~\ref{bc:encoding}, Eq.~\eqref{bc:physical}, yields
\[
 W_i^\dagger z=(\Gamma_i^{\rm res})^{-1/2}Az.
\]
Here completion on databases assigning $i$ agrees exactly with
residual completion followed by $J_{iy}$.
\reviewrev{Combining this identity with~\eqref{bc:AC} and
$AV_sh_+=\Gamma_n^{1/2}h_+$ gives the block-diagonal identity below,
with respect to the orthogonal decomposition
$\mathcal P_s=\operatorname{Im}\Pi_{i,s}^-\oplus
\operatorname{Im}\Pi_{i,s}^+$:
\begin{equation}
\label{bc:B-formula}
 \left.\mathcal B_i\right|_{\mathcal P_s}
 =\begin{pmatrix}
  \left.(\Gamma_i^{\rm res})^{1/2}\Gamma_n^{-1/2}
    \right|_{\operatorname{Im}\Pi_{i,s}^-} & 0\\[4pt]
  0 & \left.(\Gamma_i^{\rm res})^{-1/2}\Gamma_n^{1/2}
    \right|_{\operatorname{Im}\Pi_{i,s}^+}
 \end{pmatrix}.
\end{equation}
In particular,
$[\mathcal B_i,\Pi_{i,s}^{\pm}]|_{\mathcal P_s}=0$.}

\reviewrev{On these two residual-degree subspaces, Lemma~\ref{bc:frame},
Eq.~\eqref{bc:gamma-bounds}, with label counts $n$ and $n-1$, gives,
with $a_s=s/(n-2s)$,}
\[
 I_{\mathcal P_s}\preceq\Gamma_n|_{\mathcal P_s}
                    \preceq e^{a_s}I_{\mathcal P_s},\qquad
 I_{\mathcal P_s}\preceq\Gamma_i^{\rm res}|_{\mathcal P_s}
                    \preceq e^{a_s}I_{\mathcal P_s}.
\]
\reviewrev{By the commutation relations above, each diagonal block
in~\eqref{bc:B-formula} has spectrum in
$[e^{-a_s/2},e^{a_s/2}]$. Orthogonality of the two ranges gives}
\[
 \|(\mathcal B_i-I_{\mathcal P})h\|
 \le (e^{a_s/2}-1)\|h\|
 =O(s/n)\|h\|.
\]
For $s=0$, the two decompressions agree exactly.

Finally, $C_i$, $V$, and $W_i$ intertwine the full range-relabelling
action: a range relabelling merely permutes the assigned-value sectors
and relabels their residual encodings. Hence $\mathcal B_i$ commutes
with that action. Lemmas~\ref{lem:pre-regular} and
\ref{lem:pre-schur} imply that it preserves each $\mathcal P_s$.
The degreewise bounds therefore combine orthogonally to
give~\eqref{bc:overlap-bound}. The contraction property follows from
$\mathcal B_i=W_i^\dagger C_iV$, since $W_i$ and $C_iV$ are isometries.
\end{proof}

\begin{proof}[Proof of Lemma~\ref{bc:comparison}]
Let $F:\mathcal P\to\mathcal D$ be $F:=C_iV-W_i\mathcal B_i$.
By Eq.~\eqref{eq:conditional-decompression},
$\mathcal B_i=W_i^\dagger C_iV$.
By Lemma~\ref{bc:encoding}, Eq.~\eqref{bc:physical}, and the construction of
$W_i$, both $C_iV$ and $W_i$ are isometries.
Then $W_i^\dagger F=0$. By Eq.~\eqref{eq:conditional-query-intertwining},
$D_i$ preserves the image of $W_i$ and acts there as $\mathsf O_i$, so
\begin{equation}
\label{bc:sandwich-decomp}
 V^\dagger\mathcal O_iV
 =(C_iV)^\dagger D_i(C_iV)
 =\mathcal B_i^\dagger \mathsf O_i\mathcal B_i+F^\dagger D_iF.
\end{equation}
Since $\mathcal B_i$ preserves each degree subspace, so does
$F^\dagger F=I_{\mathcal P}-\mathcal B_i^\dagger\mathcal B_i$.
Writing $\delta_u=O((u+1)/n)$ for Claim~\ref{bc:overlap}, we therefore have
\[
 \|F\|_{\mathcal P_{\le u}}^2
 =\|I_{\mathcal P}-\mathcal B_i^\dagger\mathcal B_i\|_{\mathcal P_{\le u}}
 \le 2\delta_u.
\]
By the one-query bound in the proof of Lemma~\ref{bc:query-degree},
the genuine query maps
$\mathcal P_{\le t}$ into $\mathcal P_{\le t+1}$.
The same holds for $V^\dagger\mathcal O_iV$: the encoding $V$ maps
permutation degree $s$ to database size $s$, and a database query changes
size by at most one. Since $\mathcal B_i$ preserves degree, both terms in
Eq.~\eqref{bc:sandwich-decomp} restrict to maps
$\mathcal P_{\le t}\to\mathcal P_{\le t+1}$. The first term differs from $\mathsf O_i$ by at most
$\delta_{t+1}+\delta_t$, since $\|\mathcal B_i\|\le1$. For the second
term, the output support just proved allows us to insert
$P_{\mathcal P_{\le t+1}}$ on the left. Hence
\[
 \|F^\dagger D_iF\|_{\mathcal P_{\le t}}
 \le \|FP_{\mathcal P_{\le t+1}}\|\,
       \|FP_{\mathcal P_{\le t}}\|
 \le 2\sqrt{\delta_{t+1}\delta_t}.
\]
Together these estimates give the intermediate bound
\begin{equation}
\label{bc:sandwich}
 \|V^\dagger\mathcal O_iV-\mathsf O_i\|_{\mathcal P_{\le t}}
 \le C(t+1)/n.
\end{equation}

By Lemma~\ref{bc:encoding}, Eq.~\eqref{bc:physical},
$VV^\dagger=K_RK_D$ on $\mathcal D$. Then
\begin{equation}
\label{eq:projected-error-decomposition}
 K_D\mathcal O_iV-V\mathsf O_i
 =V(V^\dagger\mathcal O_iV-\mathsf O_i)
    +K_D(I_{\mathcal D}-K_R)\mathcal O_iV.
\end{equation}
For $h\in\mathcal P_{\le t}$, Lemma~\ref{bc:encoding},
Eq.~\eqref{bc:physical}, gives
\[
 Vh=\sum_{s=0}^t V_sP_{\mathcal P_s}h
 \in\bigoplus_{s=0}^t\mathcal D_R^{(s)}.
\]
Applying Lemma~\ref{bc:row-error} with $v=Vh$, and using
$\|K_D\|\le1$ and $\|Vh\|=\|h\|$, bounds the second term applied to $h$
by $O(\sqrt t/n)\|h\|$. Together with~\eqref{bc:sandwich}, this
proves~\eqref{bc:projected}.
Finally, for a unit vector $h$ of degree at most $t$,
both $\mathcal O_iVh$ and $V\mathsf O_ih$ have norm one by
Lemma~\ref{bc:encoding}, Eq.~\eqref{bc:physical}.
Equation~\eqref{bc:sandwich} therefore gives
\[
 \|\mathcal O_iVh-V\mathsf O_ih\|^2
 =2-2\operatorname{Re}\langle \mathsf O_ih,
                         V^\dagger\mathcal O_iVh\rangle
 \le 2C(t+1)/n.
\]
This proves~\eqref{bc:raw}. All estimates were operator-norm estimates uniform
in $i$ and the diagonal unitary $D_i$; direct sums and tensoring with identity
prove the controlled versions.
\end{proof}

\subsection{Simulation by projected recording queries}
\label{bc:simulation-section}

We now apply the domain projection after each recording query.
This reduces the error relative to the exact encoding and preserves
each subspace with a fixed set of recorded values. By Eq.~\eqref{bc:physical}, every
exactly encoded vector is fixed by both $K_R$ and $K_D$.
A local query $\mathcal O_i$ can leave this intersection. Applying
$K_D$ removes its domain-nonharmonic component
$(I_{\mathcal D}-K_D)\mathcal O_iV$. The remaining error splits as in
Eq.~\eqref{eq:projected-error-decomposition}: the part inside the
encoded subspace has norm $O((t+1)/n)$ on degree at most $t$,
and the remaining range-nonharmonic part has norm $O(\sqrt t/n)$
by Lemma~\ref{bc:row-error}. Adding the two bounds gives
Eq.~\eqref{bc:projected}.

The domain projection also preserves the conditions used to track
partial collisions. Lemma~\ref{bc:domain-images} shows that $K_D$ acts
within each fixed image set, so it commutes with every
bucket-occupancy projector. It can decrease the mass in an
occupancy sector but cannot transfer mass into another sector.
The range projection generally mixes output slots and does not have
this property.
We therefore apply $K_D$ and bound the remaining error with
Lemma~\ref{bc:row-error}. The projected process is used only for the
analysis. Lemma~\ref{bc:simulation} bounds its accumulated difference
from the genuine process.

The \emph{projected auxiliary query} at fixed $i,c$ is the contraction
$K_D\mathcal O_{i,c}:\mathcal D\to\mathcal D$.
The projected evolution is the auxiliary process
in Eq.~\eqref{eq:auxiliary-process}, with the same algorithm-only
unitaries as the genuine process and
\begin{equation}
\label{eq:projected-query-schedule}
 \mathcal Q_t=K_D\mathcal O\qquad(1\le t\le T).
\end{equation}
No renormalization is applied, so the states $\phi_t$ may be subnormalized.

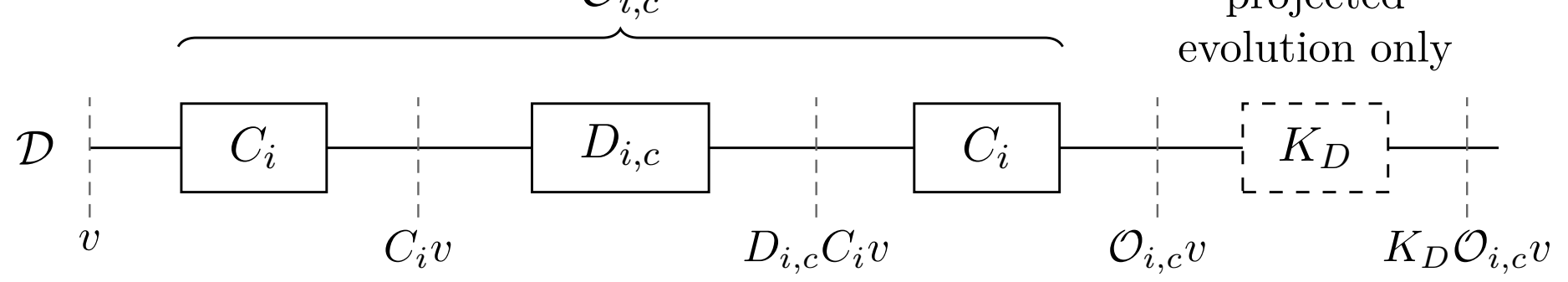
\begin{figure}[htbp]
\centering
\begin{tikzpicture}[query circuit]
 \draw (0,0)--(11.15,0);
 \node[anchor=east] at (-.15,0) {$\mathcal D$};
 \node[query gate,minimum width=1.15cm,minimum height=.7cm]
   at (1.3,0) {$C_i$};
 \node[query gate,minimum width=1.4cm,minimum height=.7cm]
   at (4.2,0) {$D_{i,c}$};
 \node[query gate,minimum width=1.15cm,minimum height=.7cm]
   at (7.1,0) {$C_i$};
 \node[query gate,dashed,minimum width=1.15cm,minimum height=.7cm]
   at (9.7,0) {$K_D$};
 \node[font=\small,align=center] at (9.7,.98) {projected\\evolution only};
 \draw[decorate,decoration={brace,amplitude=4pt}]
   (.7,.8)--(7.7,.8);
 \node at (4.2,1.2) {$\mathcal O_{i,c}$};
 \querycut{0}{-.55}{.4}{v}
 \querycut{2.6}{-.55}{.4}{C_iv}
 \querycut{5.75}{-.55}{.4}{D_{i,c}C_iv}
 \querycut{8.45}{-.55}{.4}{\mathcal O_{i,c}v}
 \querycut{10.9}{-.55}{.4}{K_D\mathcal O_{i,c}v}
\end{tikzpicture}
\caption{The auxiliary compressed query for fixed index $i$ and
charge $c$, acting on a vector $v\in\mathcal D$. The first three
gates decompress, apply the phase, and recompress, forming the unitary recorder.
The dashed projection is appended only in the projected
evolution used for the adaptive comparison. General queries
coherently control this circuit by $\mathcal I$ and $\mathcal C_\pi$.}
\label{fig:local-compression}
\end{figure}

Figure~\ref{fig:local-compression} shows a fixed index/charge sector:
the three unitary gates implement the local recorder, and the final
projection gives the query used in the projected evolution.

\FloatBarrier

The single-query errors accumulate without amplification because
the projected evolution consists of contractions. Summing them gives
an $O(T(T+1)/n)$ bound on the distance between the final states.
The adaptive proof uses this estimate together with
Lemma~\ref{lem:bucket-extraction}.

\begin{lemma}[Simulation error]
\label{bc:simulation}
Let $\psi_t$ and $\phi_t$ be the states in
Eqs.~\eqref{eq:genuine-process} and~\eqref{eq:auxiliary-process} respectively.
There is a universal constant $C$ such that every $T$-query algorithm
with $T+3\le n/8$ satisfies
\begin{equation}
\label{bc:simulation-bound}
 \|\phi_T-(V\otimes I_{\mathcal A})\psi_T\|\le C T(T+1)/n.
\end{equation}
\end{lemma}
\begin{proof}
By Lemma~\ref{bc:encoding}, Eq.~\eqref{bc:physical}, $V$ is an isometry.
Lemma~\ref{bc:frame}, Eq.~\eqref{bc:gamma}, gives $\gamma_{n,(n)}=1$;
the definitions~\eqref{bc:completion} and~\eqref{bc:isometry} therefore give
$V\ket{\widehat{\emptyset}}_{\mathcal P}=|\emptyset\rangle_{\mathcal D}$ for the initial uniform permutation
vector.
Algorithm-only unitaries commute with $V$ and the database
projections. Lemma~\ref{bc:query-degree} gives
$(P_{\mathcal P_{\le t}}\otimes I_{\mathcal A})\psi_t=\psi_t$.
Thus the controlled version of~\eqref{bc:projected} applies to $\psi_t$,
with the right-hand projection leaving the state unchanged.
Since each projected query is a contraction,
\[
 \|\phi_{t+1}-(V\otimes I_{\mathcal A})\psi_{t+1}\|
 \le \|\phi_t-(V\otimes I_{\mathcal A})\psi_t\|+C(t+1)/n.
\]
The initial distance is zero, so summing over $0\le t<T$
proves~\eqref{bc:simulation-bound}.

Coherent intermediate measurements and variable
query choices are already covered by the arbitrary workspace and controlled
query version of Lemma~\ref{bc:comparison}.
\end{proof}

\subsection{Verification and recorded targets}
\label{sec:generic-verification}

A correct output need not be fully recorded: the algorithm may guess
some positions. We therefore relate success to the database statistics
used by the progress arguments.

Both finding lower bounds use the same connection between a
successful output and recorded bucket occupancy. The progress
arguments will bound the database before any witness checks:
the adaptive argument uses a projected recording process,
whereas the one-round argument uses the exact encoding.
Verification is used only to prove the common success bound
in Lemma~\ref{lem:bucket-extraction}.

We first establish consistency between an auxiliary verifier's
returned values and its database. We then bound the additional
bucket amplitude created by the checks and compare them with
genuine queries. The resulting success bound applies to every
state of bounded permutation degree; it imposes no restriction
on the number of query rounds.

The checks are appended solely for the analysis and are not required
of the original algorithm. After measuring the candidate, they could
be performed using at most $k$ classical forward queries without
changing its success probability. We use coherent checks to compare
the genuine and auxiliary states. This follows the output-to-database
principle of compressed-oracle arguments~\cite{zhandry2019record};
for permutations, see~\cite[Lemmas~3.3 and~7.1]{carolan2026compressed}.

\begin{definition}[Target families and verification projectors]
\label{def:target-family}
Let $1\le k\le n$ and let $\mathscr B\subseteq\binom{[n]}k$ be a
publicly known family of target range sets. Suppose $\mathcal A$
stores proposed positions $i_1,\ldots,i_k$ and returned slots
$y_1,\ldots,y_k$. Using the event-projector notation of
Eq.~\eqref{eq:event-projector}, define
$P_{\rm full}:\mathcal D\to\mathcal D$ and
$P_{\rm ver}:\mathcal A\to\mathcal A$ by
\[
\begin{aligned}
 P_{\rm full}:=
 \sum_{\substack{d\in\mathcal I_n\\
                  \exists B\in\mathscr B:\ B\subseteq\operatorname{Im}d}}
       |d\rangle\langle d|,\qquad 
 P_{\rm ver}:=
 P_{\{i_1,\ldots,i_k\}\in\binom{[n]}k\;\wedge\;
     \{y_1,\ldots,y_k\}\in\mathscr B}.
\end{aligned}
\]
\end{definition}

The projector $P_{\rm full}$ detects a complete recorded target.
It depends only on the image set and therefore commutes with $K_D$
by Lemma~\ref{bc:domain-images}. The projector $P_{\rm ver}$ requires
distinct valid proposed positions and returned slots forming a target.
Since every target is a $k$-element subset of $[n]$, the returned
slots must also be valid and distinct.

Let $\eta_0\in\mathcal D_{\le t}\otimes\mathcal A$ satisfy
$\|\eta_0\|\le1$, with a proposed ordered $k$-tuple stored in
$\mathcal A$. Coherently flag invalid or repeated positions
as rejecting branches and retain these branches. On valid
branches, query the proposed positions into $k$ fresh answer
registers using unprojected auxiliary value queries; use
dummy queries on rejecting branches. Write $\eta_j$ for the
state after $j$ checks; we never renormalize.

The proposed tuple may be in coherent superposition, and $\eta_0$
need not have been prepared by auxiliary queries. The parameter $t$
bounds its database size. As in Definition~\ref{def:permutation-query},
valid slots are $1,\ldots,n$, and the blank symbol is $0$.

\begin{lemma}[Verification and recorded targets]
\label{bc:verification}
If $t+k<n$, then
\begin{equation}
\label{bc:extraction}
 \|(I_{\mathcal D}\otimes P_{\rm ver})\eta_k\|
 \le
 \|(P_{\rm full}\otimes I_{\mathcal A})\eta_k\|
 +k\sqrt{\frac{2}{n-t-k}}.
\end{equation}
\end{lemma}
\begin{proof}
Immediately before the final recompression in a value query to position
$i$, its fresh answer register equals the database value at $i$. For a
fixed assigned value $y$ and residual database $r$, put $a=n-|r|$. Then
\[
 (C_i-I_{\mathcal D})|r\cup\{i\mapsto y\}\rangle
 =\frac1{\sqrt a}|r\rangle
      -\frac1a\sum_{z\notin\operatorname{Im}r}
                       |r\cup\{i\mapsto z\}\rangle,
\]
whose norm is $\sqrt{2/a}$. The answer register keeps different valid
values $y$ orthogonal, and different residual databases occupy orthogonal
compression blocks. Therefore this estimate holds for an arbitrary
entangled state after restricting the answer to a valid slot. At every
verification step, $a\ge n-t-k$.

For $1\le j\le k$, let $Q_j$ be the diagonal projector on
$\mathcal D\otimes\mathcal A$ asserting that the database assignment at
position $i_j$ equals the $j$th returned slot. On each valid proposed-tuple
branch the positions are distinct, so every later verification query
preserves $Q_j$: it modifies only a different position. Thus, at the end,
the component with a valid $j$th answer but failing $Q_j$ has norm at most
$\sqrt{2/(n-t-k)}$, using that the starting state has norm at most one.
The projectors $Q_j$ commute. A telescoping decomposition of the complement
of $\prod_jQ_j$ bounds the accepting amplitude failing at least one
consistency condition by $k\sqrt{2/(n-t-k)}$.
If acceptance and every $Q_j$ hold, the database contains all returned
slots, and these form a target in $\mathscr B$. That component lies in
$P_{\rm full}$. The triangle inequality proves~\eqref{bc:extraction}.
All operations in this argument are block diagonal in the proposed tuple,
so the same bound holds for a coherent superposition of tuples.
\end{proof}

Lemma~\ref{bc:verification} controls the database after the checks.
For bucket finding, the next subsection bounds the bucket amplitude
they can add and derives a common success bound in terms of the
state before verification.

\subsection{Bucket occupancies and a common success bound}
\label{sec:bucket-progress}

We now specialize the target sets to a public partition into buckets.
The resulting search problem and one-query progress estimate will be
used in both the adaptive and non-adaptive finding arguments.

\begin{definition}[\permBucketFind]
\label{def:permutation-bucket-finding}
Let $k\ge2$, $n=kN$, and fix a publicly known partition
$[n]=B_1\sqcup\cdots\sqcup B_N$ with $|B_y|=k$.
The sets $B_1,\ldots,B_N$ are called \emph{public buckets}.
Given forward oracle access to an unknown $\pi\in S_n$, as in
Definition~\ref{def:permutation-query}, output a set $\rho\subseteq[n]$
satisfying
\[
 |\rho|=k\qquad\text{and}\qquad
 \pi(\rho)=B_y\quad\text{for some }y\in[N].
\]
\end{definition}

Throughout this subsection, use these public buckets and their label
map $b:[n]\to[N]$, defined by $b(u)=y$ for $u\in B_y$.
Use the target family $\mathscr B=\{B_1,\ldots,B_N\}$ in
Definition~\ref{def:target-family}. The following occupancies describe
partial and complete buckets in the database.

For $d\in\mathcal I_n$, $y\in[N]$, and $1\le j\le k$, define the
bucket occupancies and collision counts by
\[
 m_y(d):=|\operatorname{Im}d\cap B_y|,
 \qquad L_j(d):=|\{y\in[N]:m_y(d)\ge j\}|.
\]
The associated threshold projectors are operators on $\mathcal D$:
\[
 P_{j,\ge\ell}:=\sum_{\substack{d\in\mathcal I_n\\L_j(d)\ge\ell}}|d\rangle\langle d|,\qquad
 P_{j,=\ell}:=\sum_{\substack{d\in\mathcal I_n\\L_j(d)=\ell}}|d\rangle\langle d|,\qquad
 P_{j,\le\ell}:=\sum_{\substack{d\in\mathcal I_n\\L_j(d)\le\ell}}|d\rangle\langle d|.
\]

For this target family, $P_{\rm full}=P_{k,\ge1}$.
The count $L_j$ assigns one to each bucket with at least $j$
recorded slots, regardless of how many $j$-subsets it contains.
Because the database is injective, such a bucket gives $j$
distinct positions with the same bucket label. In particular,
$L_k(d)\ge1$ exactly when the database records a complete
$k$-collision, or equivalently when
$B_y\subseteq\operatorname{Im}d$ for some $y$.
The projector $P_{k,\ge1}$ tests this event.

To create another recorded $j$-collision, a query must add a
slot to a bucket that already contains $j-1$ recorded slots.
The next lemma bounds this increase by the square root of the
fraction of available slots that can complete such a step.
The bound allows erasures and replacements.

\begin{lemma}[One-query bound for creating partial collisions]
\label{adp:growth}
Fix integers $T,j,\ell$ with $0\le T<n$, $2\le j\le k$, and $\ell\ge1$,
and let $R\ge0$. Then
\begin{equation}
\label{adp:growth-bound}
 \bigl\|P_{j,\ge\ell}\,\mathcal O\,
       P_{j,=\ell-1}P_{j-1,\le R}\bigr\|_{\mathcal D_{\le T}\otimes\mathcal A}
 \le 2\sqrt{\frac{2kR}{n-T}}.
\end{equation}
Moreover, for every integer $h\ge0$,
\begin{equation}
\label{adp:one-step-increment}
 P_{j,\ge h+2}\,\mathcal O\,P_{j,\le h}=0.
\end{equation}
Thus one auxiliary query cannot increase $L_j$ by more than one.

\end{lemma}
\begin{proof}
Fix a query index $i\in[n]$, a Fourier answer label $c\in\mathbb Z_{n+1}$,
and a residual database $r\in\mathcal I_n$ with
$i\notin\operatorname{Dom}(r)$. The unitary $\mathcal O_{i,c}$ and all
database threshold projectors preserve $\mathcal D_{i,r}$. Denote the
restricted unitaries on $\mathcal D_{i,r}$ by
\[
 D_{i,c,r}:=D_{i,c}|_{\mathcal D_{i,r}},
 \qquad
 \mathcal O_{i,c,r}:=\mathcal O_{i,c}|_{\mathcal D_{i,r}}
                   =C_{i,r}D_{i,c,r}C_{i,r},
\]
and define the restricted crossing operator
$H_{i,c,r}:\mathcal D_{i,r}\to\mathcal D_{i,r}$ by
\[
 H_{i,c,r}:=\left.
   \bigl(P_{j,\ge\ell}\mathcal O_{i,c}
         P_{j,=\ell-1}P_{j-1,\le R}\bigr)
   \right|_{\mathcal D_{i,r}}.
\]
We first prove the estimates on this subspace, and then take the direct sums.

Every basis database in $\mathcal D_{i,r}$ is either $r$ or
$r\cup\{i\mapsto u\}$ for some $u\notin\operatorname{Im}(r)$, and
\[
 L_j(r\cup\{i\mapsto u\})
 =L_j(r)+\1_{\{m_{b(u)}(r)=j-1\}}.
\]
The added slot changes only the occupancy of bucket $b(u)$, which
crosses the threshold $j$ precisely when its previous occupancy was $j-1$.
Thus the only possible values of $L_j$ within $\mathcal D_{i,r}$ are
$L_j(r)$ and $L_j(r)+1$. In particular,
\[
 P_{j,\ge h+2}\mathcal O_{i,c,r}P_{j,\le h}=0
 \quad\text{on }\mathcal D_{i,r}\quad(h\ge0).
\]
An upward crossing from $L_j=\ell-1$ to $L_j\ge\ell$ is possible only if
$L_j(r)=\ell-1$; otherwise $H_{i,c,r}=0$.

Now consider the source restriction to databases of size at most $T$.
The input projector $P_{j-1,\le R}$ ensures that every basis database $d$
surviving the source projectors satisfies $L_{j-1}(d)\le R$.
Since $r\subseteq d$ for every such $d$ in $\mathcal D_{i,r}$,
\[
 |r|\le |d|\le T,
 \qquad L_{j-1}(r)\le L_{j-1}(d)\le R.
\]
Consequently, the restricted operator is zero unless
$|r|\le T$, $L_j(r)=\ell-1$, and $L_{j-1}(r)\le R$.
Assume these conditions hold. Define a set of available slots $E_{i,r}$
and its diagonal projector $P_{E_{i,r}}$ on $\mathcal D_{i,r}$ by
\[
\begin{aligned}
 E_{i,r}&:=\{y\in[n]:y\notin\operatorname{Im}(r),
                         m_{b(y)}(r)=j-1\},\\
 P_{E_{i,r}}&:=\sum_{y\in E_{i,r}}
   |r\cup\{i\mapsto y\}\rangle\langle r\cup\{i\mapsto y\}|.
\end{aligned}
\]
On $\mathcal D_{i,r}$, the output projector $P_{j,\ge\ell}$ is
$P_{E_{i,r}}$, and the input projector $P_{j,=\ell-1}$ is
$I_{\mathcal D_{i,r}}-P_{E_{i,r}}$. Each bucket with occupancy $j-1$ has exactly $k-j+1$
unused slots, so
\[
 |E_{i,r}|=(k-j+1)|\{y:m_y(r)=j-1\}|
          \le(k-j+1)L_{j-1}(r)\le kR,
 \qquad n-|r|\ge n-T.
\]

The compression formula~\eqref{adp:compressor} gives
\[
 P_{E_{i,r}}(C_{i,r}-I_{\mathcal D_{i,r}})
 =P_{E_{i,r}}\ket{u_{i,r}}
       \bigl(\langle r|-\langle u_{i,r}|\bigr),
 \qquad
 \|P_{E_{i,r}}(C_{i,r}-I_{\mathcal D_{i,r}})\|
 =\sqrt{\frac{2|E_{i,r}|}{n-|r|}}.
\]
The diagonal unitary $D_{i,c,r}$ commutes with $P_{E_{i,r}}$, and
\[
 \mathcal O_{i,c,r}-D_{i,c,r}
 =(C_{i,r}-I_{\mathcal D_{i,r}})D_{i,c,r}C_{i,r}
    +D_{i,c,r}(C_{i,r}-I_{\mathcal D_{i,r}}).
\]
Since $P_{E_{i,r}}D_{i,c,r}(I_{\mathcal D_{i,r}}-P_{E_{i,r}})=0$, the triangle inequality
and unitarity imply
\begin{align*}
 \|P_{E_{i,r}}\mathcal O_{i,c,r}(I_{\mathcal D_{i,r}}-P_{E_{i,r}})\|
 \le 2\|P_{E_{i,r}}(C_{i,r}-I_{\mathcal D_{i,r}})\|
 =2\sqrt{\frac{2|E_{i,r}|}{n-|r|}}
 \le2\sqrt{\frac{2kR}{n-T}}.
\end{align*}
The remaining input projector $P_{j-1,\le R}$ and the input size restriction
cannot increase this norm. Including the zero cases above, we have therefore
proved, for every $i,c,r$,
\[
 \|H_{i,c,r}\|_{\mathcal D_{i,r}\cap\mathcal D_{\le T}}
 \le2\sqrt{\frac{2kR}{n-T}}.
\]

Finally, include the workspace $\mathcal W_\pi$ in the algorithm
register $\mathcal A$. All database threshold projectors
preserve each $\mathcal D_{i,r}$, so the full crossing operator decomposes as
\[
 P_{j,\ge\ell}\mathcal O P_{j,=\ell-1}P_{j-1,\le R}
 =\bigoplus_{i\in[n],\,c\in\mathbb Z_{n+1}}
   \left[
     \left(\bigoplus_{\substack{r\in\mathcal I_n\\
                               i\notin\operatorname{Dom}(r)}}H_{i,c,r}\right)
     \otimes I_{\mathcal W_\pi}
   \right].
\]
The input size restriction also respects this orthogonal decomposition.
The norm of a direct sum is the maximum of the component norms, and tensoring
with $I_{\mathcal W_\pi}$ does not change them. Hence
\begin{align*}
 \bigl\|P_{j,\ge\ell}\mathcal O
        P_{j,=\ell-1}P_{j-1,\le R}\bigr\|_{\mathcal D_{\le T}\otimes\mathcal A}
 =\max_{\substack{i\in[n],\,c\in\mathbb Z_{n+1}\\
                   r\in\mathcal I_n,\ i\notin\operatorname{Dom}(r)}}
       \|H_{i,c,r}\|_{\mathcal D_{i,r}\cap\mathcal D_{\le T}}
 \le2\sqrt{\frac{2kR}{n-T}},
\end{align*}
which proves~\eqref{adp:growth-bound}. Taking the same direct sums of the
local zero identities proves~\eqref{adp:one-step-increment}.

\end{proof}

The growth bound controls the bucket amplitude added by verification.
We combine it with Lemma~\ref{bc:verification} to bound success
using the state before the checks.

For a joint state $\psi$ with a proposed ordered $k$-tuple stored
in $\mathcal A$, let $p_{\rm succ}$ denote the probability that
the tuple is a valid output for \permBucketFind.

\begin{lemma}[Success bound for \permBucketFind]
\label{lem:bucket-extraction}
If $\psi\in\mathcal P_{\le t}\otimes\mathcal A$ is a unit
state and $t+k+3\le n/8$, then
\begin{equation}
\label{eq:bucket-extraction}
 \sqrt{p_{\rm succ}}
 \le
 \|(P_{\rm full}\otimes I_{\mathcal A})
                   (V\otimes I_{\mathcal A})\psi\|
 +C_k\sqrt{\frac{t+k+1}{n}},
\end{equation}
where $C_k$ depends only on $k$.
\end{lemma}

\begin{proof}
Append $k$ genuine forward value queries into fresh answer
registers and check the proposed tuple, retaining invalid
tuples as rejecting branches. Let $\psi_{\rm ver}$ be the
resulting genuine state. The check preserves the candidate
and accepts with exactly its original success probability,
so
\[
 \|(I_{\mathcal P}\otimes P_{\rm ver})\psi_{\rm ver}\|
 =\sqrt{p_{\rm succ}}.
\]

Start an auxiliary checking process at
$\eta_0=(V\otimes I_{\mathcal A})\psi$ and use the same
algorithm-only operations and $k$ unprojected auxiliary
queries. Write $\eta_j$ for its state after $j$ checks.
These states have norm one and database size at most $t+j$.
The one-query bound in the proof of Lemma~\ref{bc:query-degree}
implies that the genuine state before check $j+1$ has permutation
degree at most $t+j$.
The raw comparison in Eq.~\eqref{bc:raw}, summed over these
$k$ queries, therefore gives
\[
 \|\eta_k-(V\otimes I_{\mathcal A})\psi_{\rm ver}\|
 \le Ck\sqrt{\frac{t+k+1}{n}}.
\]
Since $V$ is an isometry acting only on the oracle register,
comparison of acceptance norms and Lemma~\ref{bc:verification}
imply
\[
 \sqrt{p_{\rm succ}}
 \le
 \|(P_{\rm full}\otimes I_{\mathcal A})\eta_k\|
 +Ck\sqrt{\frac{t+k+1}{n}}
 +k\sqrt{\frac{2}{n-t-k}}.
\]

It remains to bound the bucket amplitude added by the checks.
On databases of size at most $s$, at most $s/(k-1)\le s$
buckets have occupancy at least $k-1$. Apply
Lemma~\ref{adp:growth} with $j=k$, threshold $\ell=1$,
and $R=s$. Since $P_{\rm full}=P_{k,\ge1}$, it gives
\[
 \|P_{\rm full}\mathcal O(I_{\mathcal D}-P_{\rm full})\|
                 _{\mathcal D_{\le s}\otimes\mathcal A}
 \le2\sqrt{\frac{2ks}{n-s}}.
\]
Identity factors are suppressed in this display and below.
The bound also holds for value queries, since the change
of answer basis acts only on the algorithm register.
Splitting the state into its $P_{\rm full}$ and
$I_{\mathcal D}-P_{\rm full}$ components, and using unitarity, yields
\[
 \|P_{\rm full}\eta_{j+1}\|
 \le \|P_{\rm full}\eta_j\|
       +2\sqrt{\frac{2k(t+j)}{n-t-j}}
 \qquad(0\le j<k).
\]
Algorithm-only operations commute with the target projector.
Summing these inequalities bounds the final bucket
amplitude by its initial value plus the displayed crossing
terms. In the stated parameter range, these terms and
the preceding comparison and consistency errors sum to
$C_k\sqrt{(t+k+1)/n}$. Substituting
$\eta_0=(V\otimes I_{\mathcal A})\psi$ proves the claim.
\end{proof}

The checks may depend on the proposed tuple, so the extended
circuit need not be one-round.

The remaining task in each finding proof is to bound the amplitude
on databases containing a full bucket before verification. For
adaptive algorithms, we compare the exact encoding with the projected
recorder and bound the growth of partial collisions. For one-round
algorithms, we bound the probability of recording the required values
directly in the exact encoding. The one-round assumption is used only
for the state before the checks.

\section{Adaptive \kFind}
\label{sec:adaptive}

We now prove the adaptive lower bound in
Theorem~\ref{thm:adaptive_find}. We first represent an exactly
$k$-to-one input by a random permutation. We then bound how quickly
the auxiliary oracle can record partial collisions of increasing
sizes. The oracle comparison and the verification argument turn
this bound into a lower bound for the original algorithm.

Recall that, for fixed $k\ge2$,
\begin{equation}
 \alpha_k:=\frac{2^{k-1}-1}{2^k-1}
     =\frac12-\frac{1}{2(2^k-1)}.
\end{equation}
Our goal is an $\Omega_k(N^{\alpha_k})$ lower bound for \kFind
with arbitrary adaptive quantum queries.
For example, the lower bound is $\Omega(N^{3/7})$ for $k=3$
and $\Omega(N^{7/15})$ for $k=4$ under the exact promise.

\subsection{Reduction to \permBucketFind}
\label{adp:permutation-queries}

We reduce \kFind to \permBucketFind
(Definition~\ref{def:permutation-bucket-finding}) using a uniformly
random $\pi\in S_n$. Success is averaged over $\pi$ and the algorithm's
measurement outcomes.
Use the public buckets and their label map $b:[n]\to[N]$ from
Section~\ref{sec:bucket-progress}. For an alphabet size $q\ge N$, fix a
publicly known~injection~$\iota:[N]\hookrightarrow\mathbb Z_q$.

\begin{definition}[The balanced input distribution]
\label{def:balanced-buckets}
The \emph{balanced input distribution}, for a fixed injection
$\iota:[N]\hookrightarrow\mathbb Z_q$, is uniform on the strings
\[
 X_{n,k}^{\iota}
 :=\left\{x\in\mathbb Z_q^n:
       |\{i:x_i=\iota(y)\}|=k\text{ for every }y\in[N]\right\}.
\]
Since $n=kN$, these conditions account for every coordinate of $x$;
no other symbol occurs.
\end{definition}

Equation~\eqref{eq:input-purification} purifies the balanced distribution
into the initial input-register vector
\[
 \frac1{\sqrt{|X_{n,k}^{\iota}|}}\sum_{x\in X_{n,k}^{\iota}}
                         \ket{x}_{\mathcal X}.
\]
Each input query is the operator $O$ from~\eqref{eq:oracleAvg}.
We allow arbitrary input-independent operations between successive
queries, so the algorithm may be adaptive.

A uniform permutation fills the public buckets with uniformly chosen
input positions, producing exactly the balanced input distribution.
Computing a queried position's slot, applying its bucket-label phase,
and uncomputing the slot then simulates an input query with two
permutation queries while preserving the success probability.

\begin{lemma}[Reduction to \permBucketFind]
\label{adp:reduction}
Let $n=kN$, $q\ge N$, and use the public buckets of
Definition~\ref{def:permutation-bucket-finding}. Suppose a quantum
algorithm makes at most $T$ calls to the combined input oracle $O$
of~\eqref{eq:oracleAvg} and solves \kFind with probability at least
a constant $p_0>0$ on the balanced input distribution of
Definition~\ref{def:balanced-buckets}.
Then there is a quantum algorithm for \permBucketFind in the model
of Figure~\ref{fig:QAPerm} which, with its permutation register initialized
in $\ket{\widehat{\emptyset}}_{\mathcal P}$, makes at most $2T$ calls
to $\mathsf O$ and succeeds with probability at least $p_0$, averaged
over uniform $\pi\in S_n$ and the algorithm's measurement outcomes.
\end{lemma}

The proof of Lemma~\ref{adp:reduction} uses three claims: a uniform
permutation induces the balanced input distribution, one input query
can be simulated by two permutation queries, and the success events
for \kFind and \permBucketFind coincide. For each $\pi\in S_n$, put
\[
 x^\pi=(x_1^\pi,\ldots,x_n^\pi)\in\mathbb Z_q^n,
 \qquad x_i^\pi:=\iota(b(\pi(i))).
\]

\begin{claim}[The balanced distribution induced by a permutation]
\label{adp:balanced-distribution}
For every fixed public bucket map $b$ and injection $\iota$,
$x^\pi$ belongs to $X_{n,k}^{\iota}$ for every $\pi\in S_n$.
If $\pi$ is uniform on $S_n$, then $x^\pi$ is uniform on
$X_{n,k}^{\iota}$.
\end{claim}
\begin{proof}
For every $y\in[N]$,
\[
 \{i:x_i^\pi=\iota(y)\}=\pi^{-1}(B_y).
\]
Since $\pi$ is a bijection and $|B_y|=k$, this set contains exactly $k$ positions.
Thus $x^\pi\in X_{n,k}^{\iota}$.

Fix $x\in X_{n,k}^{\iota}$ and put $A_y=\{i:x_i=\iota(y)\}$.
A permutation $\pi$ satisfies $x^\pi=x$ exactly when each restriction
$\pi|_{A_y}$ is a bijection $A_y\to B_y$.
Indeed, $b(\pi(i))=y$ for $i\in A_y$ forces $\pi(i)\in B_y$;
injectivity and the equal sizes then give a bijection. Conversely,
choosing these bijections gives a permutation of $[n]$, because the
$A_y$ partition the domain and the $B_y$ partition the range.
There are $k!$ choices for each restriction, independently for all $N$
sets $A_y$. Consequently,
\[
 \bigl|\{\pi\in S_n:x^\pi=x\}\bigr|=(k!)^N,
 \qquad
 \Pr_{\pi\sim\operatorname{Unif}(S_n)}[x^\pi=x]=\frac{(k!)^N}{n!}.
\]
This probability is independent of $x$, proving uniformity.
\end{proof}

\begin{claim}[Two permutation queries simulate one input query]
\label{adp:query-simulation}
For every $\pi\in S_n$, one call to $O_{x^\pi}$ can be implemented
with two calls to $O_\pi$, returning the additional workspace to its
initial state.
\end{claim}
\begin{proof}
Keep the original query registers $\mathcal I$ and $\mathcal C$, and
introduce a fresh register $W$ isomorphic to $\mathcal C_\pi$,
initialized to $\ket{0}_W$. In the simulation, $W$ serves as the
permutation answer register, while the original answer register
$\mathcal C$ is part of $\mathcal W_\pi$.
For $i\in[n]$ and $c\in\mathbb Z_q$, apply
\begin{align*}
 &\ket{i,c}\ket{0}_W
 \xrightarrow[O_\pi]{\text{compute}}
 \ket{i,c}\ket{\pi(i)}_W
 \xrightarrow[\text{no query}]{\text{phase}}
 \omega_q^{c\iota(b(\pi(i)))}\ket{i,c}\ket{\pi(i)}_W
 \xrightarrow[O_\pi]{\text{uncompute}}
 \omega_q^{c\iota(b(\pi(i)))}\ket{i,c}\ket{0}_W.
\end{align*}
The first step applies the additive form of $O_\pi$ to
$\mathcal I\otimes W$, leaving $\mathcal C$ unchanged.
The second step applies the known phase $\omega_q^{c\iota(b(u))}$
when $W$ contains $u\in[n]$, and may act trivially when $u=0$.
It costs no queries because $b$ and $\iota$ are public.
The last step reverses the additive query by conjugating it with
answer-register negation $\ket{z}\mapsto\ket{-z}$, so it also costs
one call to $O_\pi$.

Since $x_i^\pi=\iota(b(\pi(i)))$, the resulting phase is precisely
$\omega_q^{c x_i^\pi}$, as required by~\eqref{eq:oracle}.
The calculation holds on superpositions of $i,c$ and with arbitrary
additional workspace. Both calls are forward queries; no evaluation
of $\pi^{-1}$ is needed.
\end{proof}

\begin{claim}[Equivalence of the success events]
\label{adp:collision-bucket}
For every $\pi\in S_n$ and every $\rho\subseteq[n]$ with $|\rho|=k$,
$\rho$ is a valid output for \kFind on $x^\pi$ if and only if
it is a valid output for \permBucketFind on $\pi$.
\end{claim}
\begin{proof}
Because $\iota$ is injective, the values $x_i^\pi$ for $i\in\rho$ are
all equal exactly when $b(\pi(i))$ is constant on $\rho$.
This is equivalent to $\pi(\rho)\subseteq B_y$ for some $y$.
Since $\pi$ is injective, $|\pi(\rho)|=|\rho|=k=|B_y|$,
so this inclusion is equivalent to equality.
\end{proof}

\begin{proof}[Proof of Lemma~\ref{adp:reduction}]
We combine the three claims above.
The original combined oracle in~\eqref{eq:oracleAvg} acts on each
input-register branch $\ket{x}$ as the fixed-input oracle $O_x$.
For each oracle input $\pi$, simulate the original algorithm on
$x^\pi$. Claim~\ref{adp:balanced-distribution} ensures that $x^\pi$
is an exactly $k$-to-one input. Replace each call to $O_{x^\pi}$ by
the two-query implementation in Claim~\ref{adp:query-simulation},
keeping the original algorithm's intervening operations and final
measurement. This reproduces its output distribution on $x^\pi$
exactly. The implementation is independent of $\pi$ except for its
oracle calls and preserves the permutation-register basis label $\pi$.
By linearity, replacing these calls by $\mathsf O$ performs the
simulation simultaneously on all branches $\ket{\pi}$, using at most
$2T$ calls in total.
By Claim~\ref{adp:collision-bucket}, for each fixed $\pi$, the simulated
success probability for \permBucketFind equals the original success
probability for \kFind on $x^\pi$.

Starting the permutation register in $\ket{\widehat{\emptyset}}_{\mathcal P}$ averages these
success probabilities over uniform $\pi$, because the branches are
orthogonal. By Claim~\ref{adp:balanced-distribution}, this induces
precisely the balanced input distribution. The resulting
success probability therefore equals the original algorithm's success
probability on that distribution, which is at least $p_0$.
\end{proof}

An algorithm that succeeds with probability at least $p_0$ on every
promised input also satisfies this average-case hypothesis. Hence a
lower bound for \permBucketFind implies the required
worst-case lower bound for \kFind. In the remainder of the analysis,
$T$ counts calls to $\mathsf O$; $O_\pi$ denotes only the block
corresponding to a fixed permutation.

\subsection{Bounding partial collisions of successive sizes}
\label{adp:hierarchy-section}

We adapt the partial-collision hierarchy of
Liu and Zhandry~\cite[Section~4]{liu2019finding} to databases of
partial injections. At each collision size, we set a cutoff on the
number of recorded collisions and temporarily remove components
that exceed it. Below these cutoffs, few $(k-1)$-collisions are
available for extension, so each query adds only a small amplitude
to full $k$-collisions. Bounding the amplitude removed at every
cutoff then controls the original process. The argument allows
queries to erase or replace assignments.

\begin{lemma}[Cutoff bound for partial collisions]
\label{adp:hierarchy}
Let $0\le T<n$ be an integer, and let $R_1,\ldots,R_{k-1}$ be any
integer cutoffs satisfying $R_1\ge T$ and $R_j\ge0$ for
$2\le j\le k-1$. Define the one-step \mbox{growth parameters by}
\[
 \kappa_j:=2\sqrt{\frac{2kR_{j-1}}{n-T}}\qquad(2\le j\le k).
\]
Then any $T$-query auxiliary algorithm with final joint state $\phi_T$
satisfies
\begin{equation}
\label{adp:hierarchy-bound}
 \|P_{k,\ge1}\phi_T\|
 \le T \kappa_k+
       \sum_{j=2}^{k-1}\binom{T}{R_j+1}\kappa_j^{R_j+1}.
\end{equation}

\end{lemma}
\begin{proof}
Let $U_t$ be the algorithm-only unitary preceding query $t+1$, with
$U_0$ the initial unitary, as in~\eqref{eq:auxiliary-process}. Every $U_t$ commutes with the database projectors.
The final algorithm-only unitary may be omitted when bounding a database
measurement. Define the cutoff projector $P:\mathcal D\to\mathcal D$
and the corresponding truncated states by
\[
 P:=\prod_{j=2}^{k-1}P_{j,\le R_j},
 \qquad \widetilde\phi_0=\phi_0,
 \qquad \widetilde\phi_{t+1}:=P\mathcal O U_t\widetilde\phi_t.
\]
These are subnormalized states. They describe a mathematical truncation of
the computation, not extra measurements available to the algorithm. All the
factors in $P$ commute, and $\|\widetilde\phi_t\|\le1$.

For $2\le j\le k-1$ and $s\ge1$, the one-step property and
Lemma~\ref{adp:growth} give
\begin{equation}
 \|P_{j,\ge s}\widetilde\phi_{t+1}\|
 \le\|P_{j,\ge s}\widetilde\phi_t\|
    +\kappa_j\|P_{j,\ge s-1}\widetilde\phi_t\|.
\end{equation}
To see this, split $U_t\widetilde\phi_t$ into the three
orthogonal components with $L_j\ge s$, $L_j=s-1$, and
$L_j\le s-2$. Unitarity bounds the first contribution by
$\|P_{j,\ge s}\widetilde\phi_t\|$, since $U_t$ commutes with
the database projectors. The one-query increment bound makes the
last contribution zero. On the middle component,
$L_{j-1}\le R_{j-1}$ by the cutoff; when $j=2$, use
$L_1\le |d|\le t\le R_1$ instead. Lemma~\ref{adp:growth}
bounds this contribution by
$\kappa_j\|P_{j,=s-1}\widetilde\phi_t\|$, which is at most
the second term displayed above. The final projection $P$
commutes with the target projector and cannot increase the norm.

The initial database is empty. Induction using Pascal's identity and
$\|P_{j,\ge0}\widetilde\phi_t\|=\|\widetilde\phi_t\|\le1$ yields
\begin{equation}
 \|P_{j,\ge s}\widetilde\phi_t\|
 \le\binom{t}{s}\kappa_j^s\qquad(s\ge0).
\end{equation}
The binomial bound also covers thresholds larger than the
number of queries, where both sides vanish. For the cutoff
estimate below, we use it only at $s=R_j$.

At step $t+1$, the amplitude discarded for exceeding threshold $R_j$ is~at~most
\[
 \|P_{j,\ge R_j+1}\mathcal O U_t\widetilde\phi_t\|
 \le \kappa_j\|P_{j,\ge R_j}\widetilde\phi_t\|
 \le\binom{t}{R_j}\kappa_j^{R_j+1}.
\]
Only the source component with $L_j=R_j$ can exceed this
cutoff in one query; its lower collision count satisfies the
restriction used in Lemma~\ref{adp:growth}. This proves the first
inequality. For commuting projectors $P_2,\ldots,P_{k-1}$,
the identity
$I-\prod_jP_j=\sum_j(\prod_{a<j}P_a)(I-P_j)$ shows that
the norm removed by their product is at most the sum of the
norms removed by the individual projectors. Comparing the
original and truncated computations after each query therefore
gives
\begin{equation}
\label{adp:discarded}
 \|\phi_T-\widetilde\phi_T\|
 \le\sum_{t=0}^{T-1}\|(I_{\mathcal D}-P)\mathcal O U_t\widetilde\phi_t\|
 \le\sum_{j=2}^{k-1}\sum_{t=0}^{T-1}
         \binom{t}{R_j}\kappa_j^{R_j+1}
  =\sum_{j=2}^{k-1}\binom{T}{R_j+1}\kappa_j^{R_j+1}.
\end{equation}
The first inequality follows by induction from
$\|\phi_{t+1}-\widetilde\phi_{t+1}\|\le
\|\phi_t-\widetilde\phi_t\|+\|(I_{\mathcal D}-P)\mathcal O U_t\widetilde\phi_t\|$.

Finally, in the truncated evolution the amplitude in $L_k\ge1$ increases
by at most $\kappa_k$ per query, since $L_{k-1}\le R_{k-1}$ and
$\|\widetilde\phi_t\|\le1$. For $k=2$, use $L_1\le T\le R_1$ and $P=I_{\mathcal D}$ instead.
The initial amplitude is zero, so
$\|P_{k,\ge1}\widetilde\phi_T\|\le T\kappa_k$. Adding~(\ref{adp:discarded}) proves the
lemma.
\end{proof}

We choose smaller cutoffs as the collision size increases.
These choices make the discarded amplitude exponentially small
while keeping the amplitude of a full $k$-collision small whenever
the number of queries is a sufficiently small multiple of
$N^{\alpha_k}$.
In other words, constant probability of recording a full bucket requires
$\Omega_k(N^{\alpha_k})$ auxiliary queries.

\begin{lemma}[The auxiliary query lower bound]
\label{adp:aux-theorem}
Fix $k\ge2$. There is an absolute constant $c_0>0$
such that, for every $0<c\le c_0$ and all sufficiently large $N$, any
auxiliary algorithm making $T\le cN^{\alpha_k}$ queries satisfies
\begin{equation}
\label{adp:aux-main}
 \|P_{k,\ge1}\phi_T\|
 \le4\sqrt2\,c+(k-2)2^{-N^{1/(2^k-1)}}.
\end{equation}
\end{lemma}
\begin{proof}
Set
\[
 b_j:=\frac{2^{k-j}-1}{2^k-1}\quad(1\le j\le k),
 \qquad
 R_j:=\lceil N^{b_j}\rceil\quad(1\le j\le k-1).
\]
Then $b_1=\alpha_k$, $b_k=0$, and $R_1\ge T$ when $c\le1$.
Also $T\le n/2$ for sufficiently large $N$.
The constants in Lemma~\ref{adp:hierarchy} satisfy
\[
 \kappa_j\le4\sqrt{R_{j-1}/N}
       \le4\sqrt2\,N^{(b_{j-1}-1)/2}.
\]
Since $\alpha_k+(b_{j-1}-1)/2=b_j$, we obtain
\[
 T\kappa_j\le4\sqrt2\,cN^{b_j}\le4\sqrt2\,cR_j
       \quad(2\le j\le k-1),
 \qquad T\kappa_k\le4\sqrt2\,c.
\]
Choose $c_0\le1/(8\sqrt2 e)$. For each cutoff level,
\[
 \binom{T}{R_j+1}\kappa_j^{R_j+1}
 \le\left(\frac{eT\kappa_j}{R_j+1}\right)^{R_j+1}
 \le2^{-(R_j+1)}
 \le2^{-N^{1/(2^k-1)}}.
\]
When $R_j+1>T$ the binomial term is zero, so the same bound holds.
Substitution into~\eqref{adp:hierarchy-bound} proves~\eqref{adp:aux-main}.
For any fixed target success probability, taking $c$ sufficiently small and
then $N$ sufficiently large makes the square of its right-hand side smaller
than that probability. This proves the auxiliary-model query lower bound.
\end{proof}

\subsection{Completion of the adaptive lower bound}
\label{bc:final-section}

We complete the proof of Theorem~\ref{thm:adaptive_find} by transferring
the auxiliary lower bound to the genuine oracle.
We first show that inserting $K_D$ after queries preserves
the partial-collision estimates.

\begin{claim}[The projections preserve the partial-collision bounds]
\label{bc:hierarchy-projected}
The conclusions of Lemmas~\ref{adp:hierarchy} and
\ref{adp:aux-theorem} hold when any subset of the auxiliary queries
is followed by $K_D$.
\end{claim}
\begin{proof}
By Lemma~\ref{bc:domain-images}, a nonzero matrix element
$\langle d'|K_D|d\rangle$ requires
$\operatorname{Im}d'=\operatorname{Im}d$. Hence $|d'|=|d|$ and
$L_j(d')=L_j(d)$ for every $j$. In particular, $K_D$ commutes with every
size and occupancy projector used in the hierarchy.
For any source and target occupancy projectors $P_0,P_1$,
commutation gives
$\|P_1K_D\mathcal O P_0\|
 =\|K_DP_1\mathcal O P_0\|
 \le\|P_1\mathcal O P_0\|$.
The same argument applies after restricting the source database
size. Thus the one-query growth bound and the bound on the
maximum increase of $L_j$ remain valid. The proof of
Lemma~\ref{adp:hierarchy} uses only these bounds, commutation with
the cutoff projectors, and the fact that the evolution cannot
increase norms. These properties also hold for $K_D\mathcal O$.
Consequently, the recurrence, the discarded-amplitude bound, and
the exponentially small error in Lemma~\ref{adp:aux-theorem}
are unchanged.
\end{proof}

The partial-collision bounds therefore apply to the projected
recorder. We combine them with the simulation estimate and the
common success bound, which already accounts for witness verification.

\adaptivefindingtheorem*
\begin{proof}[Proof of Theorem~\ref{thm:adaptive_find}]
Let an algorithm use at most $T_0$ queries and succeed with
probability at least a fixed $p_0>0$ on every promised input.
It also succeeds with probability at least $p_0$ on the balanced
input distribution. Lemma~\ref{adp:reduction} gives an algorithm
for \permBucketFind using at most $2T_0$ forward permutation
queries and having the same average success probability.
Set $T=2T_0$, padding the simulated algorithm with queries of zero
Fourier charge if necessary. Let $\psi_T$ be its genuine final state
before witness verification, with the proposed tuple coherently
stored in the algorithm register, and let
$\phi_T$ be the corresponding auxiliary state using
$K_D\mathcal O$ for all $T$ queries. If $T+k+3>n/8$,
then $T=\Omega_k(n)$ and the claimed lower bound already
follows. Otherwise, Lemma~\ref{bc:query-degree},
Lemma~\ref{lem:bucket-extraction}, and
Lemma~\ref{bc:simulation} give
\begin{align}
 \sqrt{p_0}
 &\le
 \|(P_{\rm full}\otimes I_{\mathcal A})(V\otimes I_{\mathcal A})\psi_T\|
       +C_k\sqrt{\frac{T+k+1}{n}} \nonumber\\
 &\le
 \|(P_{\rm full}\otimes I_{\mathcal A})\phi_T\|
       +C\frac{T(T+1)}{n}
       +C_k\sqrt{\frac{T+k+1}{n}}.
 \label{bc:final-amplitude}
\end{align}

Suppose $T\le cN^{\alpha_k}$ for a sufficiently small
fixed $c>0$. Claim~\ref{bc:hierarchy-projected} and
Lemma~\ref{adp:aux-theorem} imply
\[
 \|(P_{\rm full}\otimes I_{\mathcal A})\phi_T\|
 \le4\sqrt2\,c+(k-2)2^{-N^{1/(2^k-1)}}.
\]
Since $n=kN$ and $\alpha_k<1/2$,
\[
 \frac{T(T+1)}n
 =O_k\!\left(N^{-1/(2^k-1)}\right)=o(1),
 \qquad
 \sqrt{\frac{T+k+1}{n}}
 =O_k\!\left(N^{-(1-\alpha_k)/2}\right)=o(1).
\]
Choose $c$ small enough that $4\sqrt2\,c<\sqrt{p_0}/2$,
and then take $N$ sufficiently large. This contradicts
Eq.~\eqref{bc:final-amplitude}.
Thus $T=2T_0=\Omega_k(N^{\alpha_k})$, proving the theorem.
\end{proof}

\section{Non-adaptive \kFind}
\label{sec:nonadaptive-promise}

We prove the non-adaptive lower bound by analyzing the original
parallel queries. The reduction in Lemma~\ref{adp:reduction} does
not preserve the number of rounds: it replaces each input query
by two sequential permutation queries, one to compute $\pi(i)$
and one to uncompute it. Thus it turns one parallel layer into
two successive layers and cannot directly prove a one-round lower
bound.

Use the public buckets of Definition~\ref{def:permutation-bucket-finding}
and the balanced input distribution of Definition~\ref{def:balanced-buckets}.
By Claim~\ref{adp:balanced-distribution}, this input can be generated
by choosing a uniform permutation $\pi\in S_n$ and setting
$x_i^\pi=\iota(b(\pi(i)))$. We use $\mathcal P$ as the hidden
input register, while the algorithm queries $x^\pi$ with the
original oracle from~\eqref{eq:oracle}. Its action is
\[
 \ket{\pi}_{\mathcal P}\ket{i,c}_{\mathcal I\otimes\mathcal C}
 \longmapsto
 \omega_q^{c\iota(b(\pi(i)))}
 \ket{\pi}_{\mathcal P}\ket{i,c}_{\mathcal I\otimes\mathcal C}.
\]
Applying the isometry $V$ from Definition~\ref{def:encoding-map}
to the hidden register represents this same state in $\mathcal D$.
This changes only the representation of the input, so it preserves
the algorithm's single query round.

The main lemma bounds the probability that the encoded database
contains a prescribed set of values. It applies to a bounded function
of the permutation at a fixed set of positions. The requirement that
the set be fixed is essential: a bound on permutation degree alone
does not give the same conclusion. The estimate is independent of
the output alphabet size $q$.

\subsection{Recorded values after one query round}
For $\tau\in S_n$, use the left action $L_\tau$ from
Section~\ref{sec:representation} for range relabelling on
both $\mathcal P$ and $\mathcal D$:
$L_\tau|\pi\rangle=|\tau\circ\pi\rangle$ and
$L_\tau|d\rangle=|\tau\circ d\rangle$.
The encoding $V$ intertwines these actions.
For $B\subseteq[n]$, let $P_B:\mathcal D\to\mathcal D$ be the diagonal
projector asserting $B\subseteq\operatorname{Im}d$.

For a bounded function of at most $t$ fixed permutation positions,
the following lemma bounds the probability that the encoded database
contains $k$ prescribed values. When $t<n/2$, the bound is
$O_k((t/n)^k)$, the same order as the probability that $t$ uniformly
sampled values contain all $k$ targets. The set $S$ must be fixed
independently of the permutation: the proof uses the fact that
$\pi(S)$ is a uniformly random set of size $|S|$.

\begin{lemma}[Probability of recording prescribed values]\label{lem:static}
Fix a constant $k\ge1$. There is a constant $C_k$ with the following
property. Let $n\ge2k$ and let $t$ be an integer with $0\le t<n/2$.
Fix $S\subseteq[n]$ with $|S|\le t$, and let
$h:S_n\to\mathbb C$ depend only on $\pi|_S$, with
$\|h\|_\infty\le1$.
For every fixed $k$-set $B\subseteq[n]$,
\begin{equation}
 \norm{P_BVh}^2\le C_k\left(\frac{t}{n-t+1}\right)^k.
 \label{eq:static-occupancy}
\end{equation}
For $t=0$, the left side is zero.
\end{lemma}

We prove Lemma~\ref{lem:static} in two steps.
First, Claim~\ref{lem:diff} bounds the desired weight using repeated
differences of the form $I-L_{(uv)}$, which compare a function
with the function obtained by swapping two values.
Then Claim~\ref{clm:forest-differences} counts the forests formed by
these swaps to bound the sum of differences.

\begin{claim}[Iterated harmonic difference bound]\label{lem:diff}
Let $k,s$ be integers with $1\le k\le s<n/2$. Fix distinct values
$u_1,\ldots,u_k$, and put $B=\{u_1,\ldots,u_k\}$ and
$T_j=\{u_j,\ldots,u_k\}$. For $w\in\mathcal D_R^{(s)}$,
\[
 \norm{P_Bw}^2\le \frac{1}{(n-s+1)^k}
 \sum_{\substack{v_j\in[n]\setminus T_j\\1\le j\le k}}
 \norm{(I_{\mathcal D}-L_{(u_kv_k)})\cdots(I_{\mathcal D}-L_{(u_1v_1)})w}^2.
\]
\end{claim}
\begin{proof}
First fix $u\in B$ and a size-$s$ database $d$ with
$B\subseteq\operatorname{Im} d$. The range-harmonic zero-sum condition,
applied at the unique position assigned the value $u$, gives
\[
 w(d)+\sum_{v\notin\operatorname{Im} d}w((u\ v)\circ d)=0.
\]
The zero-sum condition therefore gives
\[
 (n-s+1)w(d)=\sum_{v\notin\operatorname{Im} d}\bigl(w(d)-w((u\ v)\circ d)\bigr).
\]
There are $n-s$ terms. Cauchy--Schwarz and
$(n-s)/(n-s+1)^2\le1/(n-s+1)$ imply
\[
 |w(d)|^2\le \frac1{n-s+1}\sum_{v\notin\operatorname{Im} d}
 \left|\bigl((I_{\mathcal D}-L_{(uv)})w\bigr)(d)\right|^2.
\]
Every such $v$ lies outside $B$. Sum over databases containing $B$
in their image, then enlarge the sum to all databases containing
$B\setminus\{u\}$. This gives
\begin{equation}
 \norm{P_Bw}^2\le \frac1{n-s+1}\sum_{v\notin B}
 \norm{P_{B\setminus\{u\}}(I_{\mathcal D}-L_{(uv)})w}^2. \label{eq:one-difference}
\end{equation}
Every range permutation preserves $\mathcal D_R^{(s)}$, so the
difference vectors remain range-harmonic. We may therefore apply
Eq.~\eqref{eq:one-difference} successively with $u=u_1,\ldots,u_k$.
We keep the differences in this order and do not need them to commute.
\end{proof}

To bound the repeated differences, draw an edge between the two
values exchanged by each transposition. The allowed choices form
a forest. If the image of the queried set misses a component, swaps
within that component do not change the function, so its difference
vanishes. The next claim combines this cancellation with a count
of the possible forests.

\begin{claim}[A forest bound for mixed differences]
\label{clm:forest-differences}
Let $k\ge1$, $n\ge2k$, and $1\le t<n/2$ be integers.
Let $S\subseteq[n]$ have size at most $t$, and let
$h:S_n\to\mathbb C$ depend only on $\pi|_S$ and satisfy
$\|h\|_\infty\le1$. For distinct $u_1,\ldots,u_k\in[n]$, put
$T_j=\{u_j,\ldots,u_k\}$. Then
\[
 \sum_{\substack{v_j\in[n]\setminus T_j\\1\le j\le k}}
 \norm{(I_{\mathcal P}-L_{(u_kv_k)})\cdots(I_{\mathcal P}-L_{(u_1v_1)})h}^2
 \le C_kt^k,
\]
where $C_k$ depends only on $k$ and the norm is taken with respect
to the uniform measure on $S_n$.
\end{claim}
\begin{proof}
Write $B=\{u_1,\ldots,u_k\}$. For each tuple in the sum, use the
difference operators $\nabla_j:\mathcal P\to\mathcal P$ defined by
$\nabla_j:=I_{\mathcal P}-L_{(u_jv_j)}$, and draw the directed edges
$u_j\to v_j$. Each $v_j$ is either an earlier target $u_\ell$ with
$\ell<j$, or an auxiliary label outside $B$.
Let $r$ be the number of distinct auxiliary labels in the tuple.
Start with these $r$ vertices and insert the targets in increasing
index order. Each new target attaches to an existing vertex, so the
resulting graph is a forest with $r$ components $C_1,\ldots,C_r$,
each rooted at a single auxiliary label. Every component has at most
$k+1\le2k$ vertices. Figure~\ref{fig:difference-forest} illustrates
the construction with two components.

\begin{center}
\begin{minipage}{\linewidth}
\captionsetup{type=figure}
\centering
\begin{tikzpicture}[
 x=1cm,y=1cm,font=\small,
 target/.style={circle,draw,minimum size=6mm,inner sep=1pt},
 auxiliary/.style={rectangle,draw,minimum width=7mm,minimum height=6mm,inner sep=1pt},
 forest edge/.style={-{Latex[length=2.2mm]},line width=.55pt}
]
 \draw[rounded corners=4pt,draw=black!35]
   (-.5,-1.05) rectangle (7.1,1.05);
 \draw[rounded corners=4pt,draw=black!35]
   (7.9,-1.05) rectangle (14.5,1.05);
 \node at (3.3,1.4) {$C_1$};
 \node at (11.2,1.4) {$C_2$};
 \node[target] (u4) at (.4,.5) {$u_4$};
 \node[target,fill=black!18] (u2) at (2.3,.5) {$u_2$};
 \node[target] (u5) at (2.3,-.5) {$u_5$};
 \node[target] (u1) at (4.2,0) {$u_1$};
 \node[auxiliary] (z1) at (6.2,0) {$z_1$};
 \node[target,fill=black!18] (u6) at (8.8,0) {$u_6$};
 \node[target] (u3) at (11.2,0) {$u_3$};
 \node[auxiliary] (z2) at (13.6,0) {$z_2$};
 \draw[forest edge] (u4)--(u2);
 \draw[forest edge] (u2)--(u1);
 \draw[forest edge] (u5)--(u1);
 \draw[forest edge] (u1)--(z1);
 \draw[forest edge] (u6)--(u3);
 \draw[forest edge] (u3)--(z2);
\end{tikzpicture}
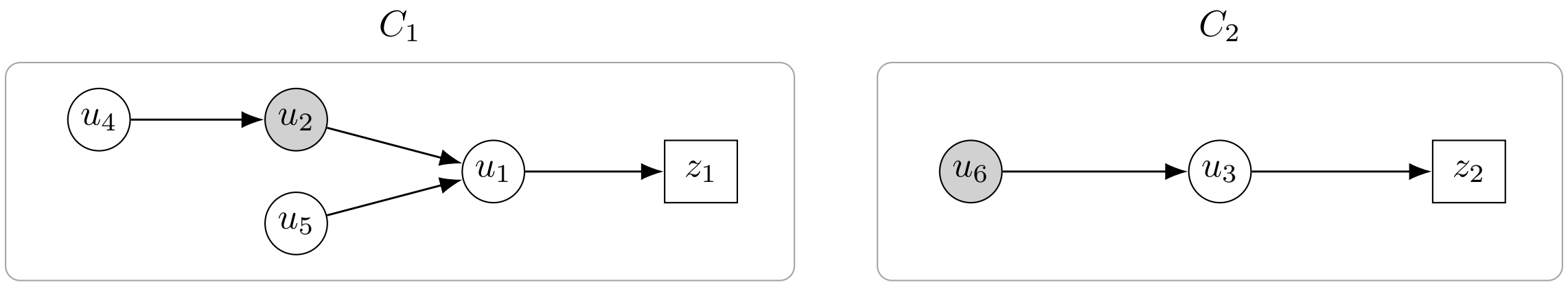
\caption{A transposition forest with $k=6$ and $r=2$,
given by $(v_1,\ldots,v_6)=(z_1,u_1,z_2,u_2,u_1,u_3)$,
where $z_1,z_2\notin B$ are distinct. Circles denote targets,
squares auxiliary roots, and arrows the choices $u_j\to v_j$.
Shading illustrates $\pi(S)\cap C_1=\{u_2\}$ and
$\pi(S)\cap C_2=\{u_6\}$. Every component must be hit for the mixed
difference to be nonzero.}
\label{fig:difference-forest}
\end{minipage}
\end{center}

For fixed $k$ there are $O_k(1)$ patterns:
each $v_j$ selects an earlier target or one of at most $k$ auxiliary
placeholders. A pattern with $r$ auxiliary vertices has at most
$n^r$ label assignments.

The mixed difference $\nabla_k\cdots\nabla_1h(\pi)$ vanishes unless
$\pi(S)$ meets every $C_a$. Indeed, every transposition $(u_jv_j)$
swaps two labels within a single component and fixes all labels
outside that component. Consequently, none of these transpositions
can move a queried image into a component that was initially missed.
To make the cancellation explicit, suppose $\pi(S)$ misses $C_a$,
and let $\pi'$ be any permutation obtained from $\pi$ by a sequence
of these transpositions. Then $\pi'(S)$ also misses $C_a$.
If $u_j,v_j\in C_a$, the permutations $\pi'$ and
$(u_jv_j)\circ\pi'$ agree on $S$. Since $h$ depends only on this
restriction,
\[
 (\nabla_jh)(\pi')
 =h(\pi')-h((u_jv_j)\circ\pi')=0.
\]
Difference operators from different components commute, so we may
move the factors from $C_a$ next to $h$ while preserving their order.
The first such factor applied to $h$ gives zero at every permutation
obtained this way. Every remaining factor takes a difference of
two such zero values, so the entire mixed difference vanishes at $\pi$.

For an arbitrary $\pi$, expand each factor
$\nabla_j=I_{\mathcal P}-L_{(u_jv_j)}$ by choosing either $I_{\mathcal P}$ or $-L_{(u_jv_j)}$.
The product has $2^k$ resulting terms, each equal up to sign to
$h$ evaluated at a permutation obtained from $\pi$.
Each term has absolute value at most $\|h\|_\infty\le1$.
The triangle inequality therefore gives the pointwise bound
\[
 |\nabla_k\cdots\nabla_1h(\pi)|\le2^k\|h\|_\infty\le2^k.
\]
To estimate the squared norm, take $\pi$ uniformly at random from $S_n$.
Since $S$ is fixed, $\pi(S)$ is a uniformly random $|S|$-subset of $[n]$.
A union bound over distinct positions in $S$ and their images in the
disjoint components gives
\[
 \Pr[\pi(S)\cap C_a\ne\varnothing\ \text{for all }a]
 \le \frac{(|S|)_r\prod_{a=1}^r|C_a|}{(n)_r}
 \le \left(\frac{4kt}{n}\right)^r.
\]
Here $(m)_r=m(m-1)\cdots(m-r+1)$, and $n\ge2k$ ensures
$(n)_r\ge(n/2)^r$. If $|S|<r$, the probability is zero.
Consequently, each tuple with $r$ auxiliary labels contributes at most
$4^k(4kt/n)^r$ to the sum in the claim. Summing over the at most
$n^r$ realizations of each pattern, all $O_k(1)$ patterns, and
$1\le r\le k$ gives
\[
 \sum_{\substack{v_j\in[n]\setminus T_j\\1\le j\le k}}
 \norm{\nabla_k\cdots\nabla_1h}^2
 \le C_k\sum_{r=1}^kn^r(t/n)^r
 =C_k\sum_{r=1}^kt^r\le C'_kt^k.
\]
The last step uses $t\ge1$ and fixed $k$.
\end{proof}

\begin{proof}[Proof of Lemma~\ref{lem:static}]
Recall that a vector of $\mathcal P$ has permutation degree at most
$t$ when it belongs to $\mathcal P_{\le t}$, the space from
Definition~\ref{def:permutation-degree}. We first show that $h$
belongs to this space. Identify $h$ with
$n!^{-1/2}\sum_\pi h(\pi)|\pi\rangle\in\mathcal P$.
For each partial injection $d$ with $\operatorname{Dom}d=S$,
the value $h(\pi)$ is constant over all completions $\pi\supseteq d$.
Grouping these completions shows that $h$ is a linear combination
of the vectors $|v_d\rangle$ from Eq.~\eqref{bc:completion} with
domain $S$.

Let $H_S\le S_n$ be the subgroup fixing every point of $S$;
it permutes the other $n-|S|$ positions. For $\sigma\in H_S$,
the domain action $R_\sigma$ permutes the completions of $d$ among
themselves, so $R_\sigma|v_d\rangle=|v_d\rangle$.
The domain and range isotypic spaces $\mathcal P_\lambda$ coincide
by Lemma~\ref{lem:pre-regular}. Each isotypic projection of
$|v_d\rangle$ is therefore also fixed by $H_S$.
By Lemma~\ref{lem:pre-branching}, a nonzero $H_S$-fixed vector of
type $\lambda$ can exist only if $\lambda_1\ge n-|S|$.
Consequently,
\[
 |v_d\rangle\in
 \bigoplus_{\substack{\lambda\vdash n\\\lambda_1\ge n-|S|}}
       \mathcal P_\lambda=\mathcal P_{\le |S|},
 \qquad h\in\mathcal P_{\le |S|}\subseteq\mathcal P_{\le t}.
\]

We now decompose $h$ by permutation degree. Eq.~\eqref{bc:isometry}
and Lemma~\ref{bc:encoding} identify the spaces containing its
encoded components:
\[
 h=\sum_{s=0}^t h_s,\qquad h_s:=P_{\mathcal P_s}h\in\mathcal P_s,
 \qquad Vh_s=V_sh_s\in\mathcal D_R^{(s)}\cap\mathcal D_D^{(s)}
 \subseteq\mathcal D_s.
\]
The projector $P_B$ preserves database size, so the vectors
$P_BVh_s$ for different $s$ remain orthogonal.
Moreover, $P_BVh_s=0$ for $s<k$, since a size-$s$ database
cannot contain all $k$ labels of $B$. Thus
\[
 \norm{P_BVh}^2=\sum_{s=0}^t\norm{P_BVh_s}^2
                 =\sum_{s=k}^t\norm{P_BVh_s}^2.
\]
If $t<k$, this sum is empty and the desired bound follows
immediately; this also proves the assertion for $t=0$.
We may therefore assume $t\ge k$.

Enumerate $B=\{u_1,\ldots,u_k\}$ and put
$T_j=\{u_j,\ldots,u_k\}$.
For each $k\le s\le t$, we have $s<n/2$ and
$Vh_s\in\mathcal D_R^{(s)}$, so Claim~\ref{lem:diff} applies.
The encoding $V$ intertwines range relabellings and is an isometry
by Lemma~\ref{bc:encoding}. Thus we can move $V$ to the left of
all the difference operators and remove it from the norm, obtaining
\[
 \norm{P_BVh_s}^2\le\frac1{(n-s+1)^k}
 \sum_{\substack{v_j\in[n]\setminus T_j\\1\le j\le k}}
 \norm{(I_{\mathcal P}-L_{(u_kv_k)})\cdots(I_{\mathcal P}-L_{(u_1v_1)})h_s}^2.
\]

To sum these estimates, fix a tuple $(v_1,\ldots,v_k)$.
Every range permutation preserves each $\mathcal P_s$, so the
mixed differences of the $h_s$ still lie in mutually orthogonal
spaces. Pythagoras therefore gives
\[
\begin{aligned}
 \norm{(I_{\mathcal P}-L_{(u_kv_k)})\cdots(I_{\mathcal P}-L_{(u_1v_1)})h}^2
 &=\sum_{s=0}^t
   \norm{(I_{\mathcal P}-L_{(u_kv_k)})\cdots(I_{\mathcal P}-L_{(u_1v_1)})h_s}^2\\
 &\ge\sum_{s=k}^t
   \norm{(I_{\mathcal P}-L_{(u_kv_k)})\cdots(I_{\mathcal P}-L_{(u_1v_1)})h_s}^2.
\end{aligned}
\]
Using $n-s+1\ge n-t+1$ and summing over $s$ therefore gives
\begin{equation}
\label{eq:sum-differences}
 \norm{P_BVh}^2\le\frac1{(n-t+1)^k}
 \sum_{\substack{v_j\in[n]\setminus T_j\\1\le j\le k}}
 \norm{(I_{\mathcal P}-L_{(u_kv_k)})\cdots(I_{\mathcal P}-L_{(u_1v_1)})h}^2.
\end{equation}
We can now apply Claim~\ref{clm:forest-differences} to $h$ itself.
This order matters: $h$ is bounded by one and depends only on
$\pi|_S$, whereas its individual degree components need not be bounded by
one. The claim bounds the sum in
Eq.~\eqref{eq:sum-differences} by $C_kt^k$, proving
Eq.~\eqref{eq:static-occupancy}.
\end{proof}

\subsection{Success probability for \kFind}
\label{sec:na-finding-proof}

We apply the preceding lemma to \kFind under the balanced
distribution of Definition~\ref{def:balanced-buckets}.
For one query round, the algorithm vectors for different query
records are orthogonal. Their contributions to the encoded
probability therefore add. We sum the bound over records and buckets,
then use the extraction lemma to account for outputs containing
positions whose values have not been recorded.

\begin{lemma}[One-round success under the balanced distribution]
\label{thm:na-success}
Let $k\ge2$ be fixed, $n=kN$, and $q\ge N$. Use the public
buckets and the balanced hard input
$x^\pi\in\mathbb Z_q^n$ with $x_i^\pi=\iota(b(\pi(i)))$,
for uniform $\pi\in S_n$.
If $t+k+3\le n/8$, the average success probability $p_{\rm succ}$
of every $t$-query one-round algorithm for \kFind satisfies
\begin{equation}
\label{eq:na-success}
 p_{\rm succ}\le C_k\left(\frac{t^k}{n^{k-1}}+\frac{t+1}{n}\right).
\end{equation}
\end{lemma}
\begin{proof}
Apply Lemma~\ref{lem:one-round-phases} with
$\nu=\operatorname{Unif}(X_{n,k}^{\iota})$ and use its permutation
purification from Claim~\ref{adp:balanced-distribution}.
After any final input-independent
unitary, its joint state has the form
\[
 \psi=\sum_a h_a\otimes u_a,\qquad
 h_a(\pi)=\prod_{i\in S_a}
   \omega_q^{c_{a,i}\iota(b(\pi(i)))},
 \qquad |S_a|\le t.
\]
The algorithm vectors $u_a$ are pairwise orthogonal and
$\sum_a\|u_a\|^2=1$. Each $h_a$ has modulus one and
depends only on the values of $\pi$ on $S_a$.

By Lemma~\ref{lem:static}, each bucket satisfies
\[
 \|P_{B_y}Vh_a\|^2
 \le C_k\left(\frac{t}{n-t+1}\right)^k.
\]
The diagonal projectors obey
$P_{k,\ge1}\preceq\sum_{y=1}^N P_{B_y}$.
Orthogonality of the algorithm vectors, followed by this
inequality on each branch, gives
\begin{align}
 \|(P_{k,\ge1}\otimes I_{\mathcal A})(V\otimes I_{\mathcal A})\psi\|^2=\sum_a\|u_a\|^2\,\|P_{k,\ge1}Vh_a\|^2
 \le C_kN\left(\frac{t}{n-t+1}\right)^k
 \le C_k\frac{t^k}{n^{k-1}}. \label{eq:na-initial-occupancy}
\end{align}
This estimate applies before the final measurement, so it does not
depend on which candidate tuple the algorithm eventually outputs.

The state has permutation degree at most $t$, since every
branch depends on at most $t$ permutation values. Coherently
storing the output tuple acts only on the algorithm registers
and preserves this degree bound. By
Claim~\ref{adp:collision-bucket}, the valid outputs for \kFind
on $x^\pi$ are exactly those for \permBucketFind on $\pi$.
Lemma~\ref{lem:bucket-extraction}, applied to the original state
before any witness checks, therefore implies
\[
 \sqrt {p_{\rm succ}}\le C_k
       \sqrt{\frac{t^k}{n^{k-1}}}
       +C_k\sqrt{\frac{t+k+1}{n}}.
\]
Squaring and using fixed $k$ gives~\eqref{eq:na-success}.
In the case $t=0$, Lemma~\ref{lem:static} gives zero
initial occupancy and the same estimate applies.
\end{proof}

A constant worst-case success guarantee gives constant average
success on the balanced distribution. The preceding bound therefore
yields Theorem~\ref{thm:find}.

\nonadaptivefindingtheorem*
\begin{proof}[Proof of Theorem~\ref{thm:find}]
Let $p_0>0$ be the fixed worst-case success guarantee. It implies
average success at least $p_0$ on the balanced hard distribution.
If $t+k+3>n/8$, then $t=\Omega_k(n)$ as $n$ grows and
the claimed lower bound follows. Otherwise, apply
Lemma~\ref{thm:na-success}. When
$t\le c\,n^{1-1/k}$, its first term is~at~most $C_kc^k$
and its second term tends to zero. Choosing $c>0$
small enough in terms of $k,p_0$ contradicts
$p_{\rm succ}\ge p_0$. Thus
$t=\Omega_k(n^{1-1/k})$.
\end{proof}

\section{Non-adaptive \kColl}
\label{sec:byproduct}

We prove Theorem~\ref{thm:test} by comparing the algorithm's states
under four input distributions. We first allow labels to repeat:
$\mathsf{Fun}$ assigns independent labels to individual positions,
whereas $\mathsf{Block}_k$ assigns independent labels to hidden
$k$-element blocks. We then require the labels to be distinct,
obtaining the promised distributions $\mathsf{Inj}$ and
$\mathsf{Exact}_k$. \decisionrev{The shared harmonic encoding from
Section~\ref{sec:harmonic-framework} controls the effect of this
replacement, giving the lower bound for every $q\ge n$.}
For a distribution $\mathsf{S}$,
write $\varrho_{\mathsf{S}}$ for the algorithm's reduced state
after the queries.

\subsection{Comparing the relaxed distributions}
\label{sec:direct-relaxed-decision}

Let $\mathsf{Fun}$ be the uniform distribution on strings
$x\in\mathbb Z_q^n$. Let $\mathsf{Block}_k$ be the distribution obtained
by choosing a uniform perfect $k$-matching $\mu$ of $[n]$ and assigning
an independent uniform element of $\mathbb Z_q$ to each block.
Here a perfect $k$-matching means a partition of $[n]$ into $n/k$
disjoint $k$-element blocks.
Different blocks may receive the same label. We first compare these two
distributions directly at the level of the algorithm's reduced states.
Section~\ref{sec:injection-switching} will then impose distinct labels.

In the block distribution, two queried positions may belong to
the same hidden block and therefore have the same label. The next lemma controls both this
event and the interference between query records, showing that the
distinguishing advantage is small when $t\ll\sqrt{n/k}$.

\begin{lemma}[One-round comparison of independent and block labels]
\label{lem:direct-relaxed-decision}
Let $k\ge2$, $k\mid n$, and $q\ge1$.
For any $t$-query non-adaptive quantum algorithm with
$t\le\sqrt{n/(8k)}$,
\begin{equation}
\label{eq:direct-relaxed-trace}
 \norm{\varrho_{\mathsf{Block}_k}-\varrho_{\mathsf{Fun}}}_1
 \le 2\sqrt\beta+\beta+\frac{8kt^2}{n}.
\end{equation}
Here $\beta=\binom t2(k-1)/(n-1)$.
\end{lemma}

The optimal success probability for distinguishing the two distributions
with equal prior probabilities is
$1/2+\norm{\varrho_{\mathsf{Block}_k}-\varrho_{\mathsf{Fun}}}_1/4$.
Using $\beta\le kt^2/n\le1/8$, the lemma bounds this probability by
\begin{equation}
\label{eq:direct-relaxed-success}
 \frac12+2t\sqrt{\frac kn}.
\end{equation}

The following disjointness estimate controls the coherent off-diagonal
terms between distinct query labels in the proof of
Lemma~\ref{lem:direct-relaxed-decision}. It bounds their total
contribution to an arbitrary measurement using orthogonality of the
algorithm vectors, without assuming that the measurement is diagonal.
Sparse labels are viewed as partial functions taking nonzero values.
For such a label $\eta$ and a multiset $C$ of $m$ nonzero charges, let
$\Omega(\eta,C)$ be all labels obtained by placing the charges in $C$
at $m$ distinct positions outside $\operatorname{Supp}\eta$, while
retaining $\eta$.

\begin{claim}[An orthogonal-vector disjointness estimate]
\label{lem:direct-disjointness}
Let $(v_\sigma)_{\sigma\in\Omega(\eta,C)}$ be pairwise orthogonal
vectors in a Hilbert space $\mathcal H$, and let
$H:\mathcal H\to\mathcal H$ be an operator with $\norm H\le1$.
For $m\ge1$,
\[
 \left|\sum_{\substack{\sigma,\sigma'\in\Omega(\eta,C)\\
       \operatorname{Supp}(\sigma\setminus\eta)
       \cap\operatorname{Supp}(\sigma'\setminus\eta)=\varnothing}}
       \langle v_{\sigma'},Hv_\sigma\rangle\right|
 \le (2m)^m\sum_{\sigma\in\Omega(\eta,C)}\norm{v_\sigma}^2.
\]
\end{claim}
\begin{proof}
Fix an ordering $(c_1,\ldots,c_m)$ of $C$. For each placement $\sigma$,
let $(i_1(\sigma),\ldots,i_m(\sigma))$ be its corresponding ordered
positions; within each equal-charge class, order the positions increasingly.
This gives a unique ordered tuple of distinct positions for every placement.
Inclusion--exclusion over partial injective maps
$\theta:\operatorname{Dom}\theta\subseteq[m]\hookrightarrow[m]$ gives
\[
 \mathbf 1_{\operatorname{Supp}(\sigma\setminus\eta)
                   \cap\operatorname{Supp}(\sigma'\setminus\eta)=\varnothing}
 =\sum_\theta(-1)^{|\operatorname{Dom}\theta|}
       \mathbf 1_{i_r(\sigma)=i_{\theta(r)}(\sigma')
                   \text{ for all }r\in\operatorname{Dom}\theta}.
\]
Indeed, if the two supports share $a$ positions, their equality relations
form a matching of size $a$, and the right side is
$\sum_{j=0}^a(-1)^j\binom aj$.

For a fixed $\theta$, group the placements by the tuple
$h=(i_r(\sigma))_{r\in\operatorname{Dom}\theta}$. Set
\[
 X_h=\{\sigma:i_r(\sigma)=h_r\text{ for every }r\in\operatorname{Dom}\theta\},
 \qquad
 Y_h=\{\sigma':i_{\theta(r)}(\sigma')=h_r
                    \text{ for every }r\in\operatorname{Dom}\theta\}.
\]
The sets $X_h$ and $Y_h$ may be empty. Each placement belongs to exactly one $X_h$
and exactly one $Y_h$. Orthogonality and Cauchy--Schwarz therefore give
\begin{align*}
 \left|\sum_h
   \left\langle\sum_{\sigma'\in Y_h}v_{\sigma'},
                    H\sum_{\sigma\in X_h}v_\sigma\right\rangle\right|
 \le\sum_h
    \left(\sum_{\sigma'\in Y_h}\norm{v_{\sigma'}}^2\right)^{1/2}
    \left(\sum_{\sigma\in X_h}\norm{v_\sigma}^2\right)^{1/2}
\le\sum_{\sigma\in\Omega(\eta,C)}\norm{v_\sigma}^2.
\end{align*}
The number of partial injective maps is
\[
 \sum_{j=0}^m\binom mj^2j!
 \le m!\binom{2m}{m}
 =\frac{(2m)!}{m!}\le(2m)^m.
\]
Summing their bounds proves the claim.
\end{proof}

\begin{proof}[Proof of Lemma~\ref{lem:direct-relaxed-decision}]
The case $t=0$ is immediate, so assume $t\ge1$; in particular $n\ge8k$.
Use the grouped decomposition of Lemma~\ref{lem:one-round-phases},
with aggregate charge labels $\sigma\in\mathbb Z_q^n$ supported on
at most $t$ positions. Its algorithm vectors $u_\sigma$ are pairwise
orthogonal and satisfy $\sum_\sigma\norm{u_\sigma}^2=1$.
The same vectors occur under both input distributions, and we may absorb the common final
input-independent unitary into them. Under $\mathsf{Fun}$,
orthogonality of the Fourier characters gives
\begin{equation}
\label{eq:direct-iid-state}
 \varrho_{\mathsf{Fun}}=\sum_\sigma u_\sigma u_\sigma^\dagger.
\end{equation}

For any partition $\mu$ and labels $y_B$ with $x_i=y_B$ for $i\in B$,
define the block charges by
\begin{equation}
\label{eq:hidden-block-charge}
 \tau_{\mu,\sigma}(B):=\sum_{i\in B}\sigma(i)\pmod q
 \quad(B\in\mu),\qquad
 \omega_q^{\sum_i\sigma(i)x_i}
 =\omega_q^{\sum_{B\in\mu}\tau_{\mu,\sigma}(B)y_B}.
\end{equation}
Use $\lvert\widehat\tau\rangle$ for the Fourier basis of the
independent block-label register $\mathbb C[\mathbb Z_q^\mu]$,
as in Eq.~\eqref{eq:product-fourier-basis} with coordinate set $A=\mu$.
A purification of the other state is
\[
 \Psi=\frac1{\sqrt{|\mathcal M|}}
       \sum_{\mu\in\mathcal M}|\mu\rangle
       \sum_\sigma|\widehat{\tau_{\mu,\sigma}}\rangle u_\sigma,
\]
where $\mathcal M$ is the set of perfect $k$-matchings. This follows
by multiplying the oracle phases and collecting the charges within
each block.

Let $G_\mu(\sigma)$ mean that $\operatorname{Supp}\sigma$ meets every
block of $\mu$ in at most one position. Define $\Psi_{\mathsf{G}}$ by retaining
only summands for which $G_\mu(\sigma)$ holds, and let
$\varrho_{\mathsf{G}}:\mathcal A\to\mathcal A$ be the subnormalized reduced operator
$\varrho_{\mathsf{G}}=\operatorname{Tr}_{\mathrm{env}}\Psi_{\mathsf{G}}\Psi_{\mathsf{G}}^\dagger$, where
the partial trace removes the matching and block-label registers.
For any fixed support of size at most $t$, a union bound over its pairs gives
\[
 \Pr_\mu[\neg G_\mu(\sigma)]
 \le\binom t2\frac{k-1}{n-1}=\beta.
\]
The orthogonality of the $u_\sigma$ gives
\[
 \norm{\Psi-\Psi_{\mathsf{G}}}^2
 =\sum_\sigma\Pr_\mu[\neg G_\mu(\sigma)]\norm{u_\sigma}^2
 \le\beta,
 \qquad \norm{\Psi_{\mathsf{G}}}\le1.
\]
The identity
$aa^\dagger-bb^\dagger=(a-b)a^\dagger+b(a-b)^\dagger$,
followed by contraction under partial trace, yields
\begin{equation}
\label{eq:direct-truncation}
 \norm{\varrho_{\mathsf{Block}_k}-\varrho_{\mathsf{G}}}_1\le2\sqrt\beta.
\end{equation}

It remains to compare $\varrho_{\mathsf{G}}$ with~\eqref{eq:direct-iid-state}.
For an arbitrary Hermitian $H:\mathcal A\to\mathcal A$ with $\norm H\le1$, tracing the block
Fourier register gives
\begin{align}
 \operatorname{Tr}\bigl(H(\varrho_{\mathsf{G}}-\varrho_{\mathsf{Fun}})\bigr)
 &=-\sum_\sigma\Pr_\mu[\neg G_\mu(\sigma)]
             \langle u_\sigma,Hu_\sigma\rangle+E_H,\label{eq:direct-diag-off}\\
 E_H&=\sum_{\sigma\ne\sigma'}
       \Pr_\mu[G_\mu(\sigma),G_\mu(\sigma'),
                     \tau_{\mu,\sigma}=\tau_{\mu,\sigma'}]
                   \langle u_{\sigma'},Hu_\sigma\rangle.\nonumber
\end{align}
The first term has absolute value at most $\beta$.

To estimate $E_H$, let $\eta=\sigma\cap\sigma'$ consist of the
positions where both labels have the same nonzero charge.
On the event in~\eqref{eq:direct-diag-off}, each block contains
at most one nonzero entry of each label. Equality of the block
charges then implies that a block containing a nonzero entry of
one label contains exactly one entry of the other, with the same
charge. In particular, the two labels cannot have different
nonzero charges at the same position. After removing $\eta$, their
supports are disjoint, have a common size $m\ge1$, and carry the
same multiset $C$ of charges. Every contributing pair therefore
belongs to exactly one class $(m,\eta,C)$.

Fix $(m,\eta,C)$. The probability coefficient in
Eq.~\eqref{eq:direct-diag-off} has the same value
$p_{\eta,C,m}$ for every pair of disjoint placements in
$\Omega(\eta,C)$. Indeed, a permutation of the positions that fixes
$\operatorname{Supp}\eta$ can carry any such ordered pair to any
other while preserving all charges, and the random matching has an
invariant distribution. We claim that
\begin{equation}
\label{eq:direct-switch-probability}
 p_{\eta,C,m}\le m!\left(\frac{2k}{n}\right)^m.
\end{equation}
For this event to occur, the $m$ entries outside $\eta$ in one
label must be paired bijectively with entries of the other label
having the same charges. Each pair must lie in its own block.
There are at most $m!$ possible bijections. Fix one and reveal the
pairs in order, exposing the entire block of each successful pair.
After $\ell$ successes, $k\ell$ positions have been exposed.
If the next pair meets an exposed block, it cannot occupy a new
block. Otherwise the remaining matching is uniform, and the
conditional probability that its endpoints share a block is at most
\[
 \frac{k-1}{n-k\ell-1}\le\frac{2k}{n}.
\]
Here $\ell<m\le t$ and
$kt\le n/8$, which follows from $n\ge8k$ and
$t\le\sqrt{n/(8k)}$. Multiplying these conditional upper bounds and
then taking the union bound over bijections proves
\eqref{eq:direct-switch-probability}.

Apply Claim~\ref{lem:direct-disjointness} to each class. This gives
\[
 |E_H|\le\sum_{m=1}^t m!\left(\frac{2k}{n}\right)^m(2m)^m
        \sum_{\eta,C:\,|C|=m}
               \sum_{\sigma\in\Omega(\eta,C)}\norm{u_\sigma}^2.
\]
For a fixed $\sigma$, choosing the $m$ entries outside $\eta$
uniquely determines both $\eta$ and their multiset $C$ of charges.
Thus $\|u_\sigma\|^2$ occurs at most $\binom tm$ times in the
inner sums. Since $\sum_\sigma\|u_\sigma\|^2=1$,
$m!\binom tm\le t^m$, $m\le t$, and $4kt^2/n\le1/2$,
we obtain
\begin{equation}
\label{eq:direct-off-bound}
 |E_H|\le\sum_{m=1}^t\left(\frac{4kt^2}{n}\right)^m
       \le\frac{8kt^2}{n}.
\end{equation}
The dual characterization of the trace norm for Hermitian operators,
together with~\eqref{eq:direct-diag-off}, yields
\[
 \norm{\varrho_{\mathsf{G}}-\varrho_{\mathsf{Fun}}}_1\le\beta+8kt^2/n.
\]
Combining this with~\eqref{eq:direct-truncation} proves
\eqref{eq:direct-relaxed-trace}.
\end{proof}

\subsection{Replacing independent labels with distinct labels}
\label{sec:injection-switching}

The relaxed comparison allows different positions or blocks to
receive the same label. We now impose distinct labels.
Let $\mathsf{Inj}$ be the uniform distribution on injective
strings in $\mathbb Z_q^n$, and let $\mathsf{Exact}_k$ choose
a uniform perfect $k$-matching and assign its blocks uniformly
chosen pairwise distinct labels. These are the two promised
input distributions when $q\ge n$.

The next lemma bounds how much a one-round algorithm's state
changes when independent labels are replaced by distinct labels.
The bound depends on the number of queries and the alphabet size,
and is uniform over the number $r\le q$ of label variables. It
therefore applies even when $q$ is only linear in $n$. We then
extend the comparison to hidden blocks, replacing $\mathsf{Block}_k$
by $\mathsf{Exact}_k$.

We also use $\mathsf{Fun}$ and $\mathsf{Inj}$ for the analogous
distributions on strings in $\mathbb Z_q^r$.

\begin{lemma}[One-round injection switching]\label{lem:switch}
Let $1\le r\le q$. For any $t$-query non-adaptive quantum algorithm
on $\mathbb Z_q^r$ with $t\le q/8$,
\[
 \frac12\norm{\varrho_{\mathsf{Inj}}-\varrho_{\mathsf{Fun}}}_1
 \le\frac{4t}{\sqrt q}.
\]
\end{lemma}

By contractivity of trace distance, this bound also holds after
input-independent processing and bounds the change in probability
of any output event.

To prove Lemma~\ref{lem:switch}, we first compare the two encodings
of a single Fourier vector. The following two claims bound the harmonic
projection loss and then the distance between the~encoded~vectors.

In this subsection, we use the database spaces, harmonic projectors,
and encoding from Definitions~\ref{def:database-space},
\ref{def:harmonic-spaces}, and~\ref{def:encoding-map}, respectively,
on $m$ permutation labels instead of $n$. Thus $\mathcal P=\mathbb C[S_m]$,
$\mathcal D_s$ contains size-$s$ partial injections on $[m]$,
and $\Gamma_m$ is the frame operator.
The domain version of the spectral-gap inequality
\eqref{bc:gap} follows by replacing every partial injection
by its inverse. We use the encoding identities proved in
Section~\ref{sec:harmonic-framework}.

Let $m\ge8$ and $1\le s\le m/8$. Fix a set $S\subseteq[m]$ of
size $s$ and functions $g_i:[m]\to\mathbb C$, for $i\in S$, satisfying
$|g_i(y)|=1$ and $\sum_{y=1}^m g_i(y)=0$.
Define the unit vector
\[
 a=\frac1{\sqrt{(m)_s}}
 \sum_{\substack{d:S\hookrightarrow[m]}}
 \left(\prod_{i\in S}g_i(d(i))\right)|d\rangle.
\]

\begin{claim}[Projection error for a fixed set of positions]\label{lem:fixed}
\begin{equation}
\label{eq:joint-loss}
 \norm{a-K_R^{(s)}K_D^{(s)}a}^2\le\frac{2s}{m}.
\end{equation}
\end{claim}
\begin{proof}
We bound the range and domain projection errors separately.
Only positions in $S$ contribute to the range extension sums.
Fix $i\in S$ and an injection
$r:S\setminus\{i\}\hookrightarrow[m]$. Since $g_i$ has zero mean,
\[
 \langle r|\mathcal L_i a\rangle
 =-\frac{\prod_{j\ne i}g_j(r(j))}
 {\sqrt{(m)_s(m-s+1)}}
 \sum_{y\in\operatorname{Im} r}g_i(y).
\]
For a uniform $(s-1)$-element subset $Y\subseteq[m]$,
\[
\begin{aligned}
 \E\left|\sum_{y\in Y}g_i(y)\right|^2
 &=s-1+\frac{(s-1)(s-2)}{m(m-1)}
       \sum_{y\ne z}g_i(y)\overline{g_i(z)}\\
 &=s-1-\frac{(s-1)(s-2)}{m-1}
 =\frac{(s-1)(m-s+1)}{m-1}.
\end{aligned}
\]
Consequently, the contraction norms satisfy
\[
 \norm{\mathcal L_i a}^2
 =\frac{s-1}{(m-1)(m-s+1)},\qquad
 \sum_i\norm{\mathcal L_i a}^2
 =\frac{s(s-1)}{(m-1)(m-s+1)}.
\]
Applying \eqref{bc:gap} gives
\begin{equation}
\label{eq:range-loss}
 \norm{(I_{\mathcal D_s}-K_R^{(s)})a}^2
 \le\frac{s(s-1)}{(m-1)(m-2s+2)}.
\end{equation}

For the domain estimate, interchange the roles of positions and
values. Let $M_y:\mathcal D_s\to\mathcal D_{s-1}$ remove output
slot $y$ and sum over its available input positions, with
normalization $1/\sqrt{m-s+1}$. This is the domain analogue of
$\mathcal L_i$ restricted to size $s$.

Every database with a nonzero coefficient in $a$ has domain $S$.
After one assignment is removed, the residual domain identifies the
missing position uniquely. Thus each extension sum has at most one
nonzero term. Each original coefficient is counted once for each of
its $s$ assigned output slots, giving
\[
 \sum_{y=1}^m\norm{M_y a}^2
 =\frac{s}{m-s+1}\norm a^2
 =\frac{s}{m-s+1}.
\]
The inverse version of \eqref{bc:gap} similarly gives
\begin{equation}
\label{eq:domain-loss}
 \norm{(I_{\mathcal D_s}-K_D^{(s)})a}^2\le\frac{s}{m-2s+2}.
\end{equation}
Since $K_R^{(s)}$ and $K_D^{(s)}$ are commuting orthogonal projectors,
\[
\begin{aligned}
 \norm{a-K_R^{(s)}K_D^{(s)}a}^2
 &\le\norm{(I_{\mathcal D_s}-K_R^{(s)})a}^2
     +\norm{(I_{\mathcal D_s}-K_D^{(s)})a}^2\\
 &\le\frac{s(m+s-2)}{(m-1)(m-2s+2)}
 \le\frac{2s}{m},
\end{aligned}
\]
where the last inequality uses $m\ge8$ and $s\le m/8$.
\end{proof}

For the same set $S$ and functions $g_i$, define
\[
 h=\frac1{\sqrt{m!}}\sum_{\pi\in S_m}
 \left(\prod_{i\in S}g_i(\pi(i))\right)|\pi\rangle.
\]
For comparison with independent labels, allow all partial maps
$[m]\rightharpoonup[m]$ as orthonormal database basis vectors,
including those that are not injective. In this larger space, define
\[
 a_0=m^{-s/2}\sum_{d:S\to[m]}
 \left(\prod_{i\in S}g_i(d(i))\right)|d\rangle.
\]
These definitions also apply when $s=0$ and $S=\varnothing$.

\begin{claim}[Comparing one Fourier vector]\label{lem:vector}
For $0\le s\le m/8$,
\[
 \norm{Vh-a_0}\le\frac{4s}{\sqrt m}.
\]
\end{claim}
\begin{proof}
For $s=0$, both $Vh$ and $a_0$ are the empty database, so the bound
is immediate. For $s\ge1$, we first compare $Vh$ with the injective
vector $a$, then compare $a$ with the unrestricted vector $a_0$.
Each permutation extends exactly one injection with domain $S$.
Together with $(m)_s(m-s)!=m!$, this gives $A_sa=h$.
In particular, $h$ is a unit vector of degree at most $s$.
Set $w=K_R^{(s)}K_D^{(s)}a$ and $e=\norm{a-w}^2$.
By Eq.~\eqref{bc:physical}, $P_{\mathcal P_s}h=A_sw$.
On the harmonic subspace $\mathcal D_R^{(s)}\cap\mathcal D_D^{(s)}$,
define the operator $\widehat{\Gamma}_s:=V_s\Gamma_m V_s^\dagger$.
The exact \mbox{frame identities imply}
\[
 P_{\mathcal D_s}Vh=V_sA_sw=\widehat{\Gamma}_s^{1/2}w.
\]
Indeed, $V_s=A_s^\dagger\Gamma_m^{-1/2}$ is a unitary map from
$\mathcal P_s$ onto this harmonic subspace. By Eq.~\eqref{bc:gamma-bounds},
\[
 \norm{P_{\mathcal D_{<s}}Vh}^2
 =1-\norm{\widehat{\Gamma}_s^{1/2}w}^2
 \le 1-\norm w^2=e.
\]
The frame spectrum also gives
\[
 \delta:=\norm{\widehat{\Gamma}_s^{1/2}-I_{\mathcal D_R^{(s)}\cap\mathcal D_D^{(s)}}}
 \le \exp\!\left(\frac{s}{2(m-2s+1)}\right)-1
 \le\frac{s}{m}
\]
in the stated parameter range. The vector $a-w$ is supported on
size-$s$ databases, so it is orthogonal to $P_{\mathcal D_{<s}}Vh$.
The triangle inequality and Eq.~\eqref{eq:joint-loss} give
\[
 \norm{Vh-a}
 \le\sqrt{2e}+\delta
 \le2\sqrt{\frac{s}{m}}+\frac{s}{m}
 \le3\sqrt{\frac{s}{m}}.
\]

The injective part of $a_0$ is $\sqrt p\,a$, where $p=(m)_s/m^s$. Thus
\[
 \norm{a-a_0}^2=2(1-\sqrt p)
 \le2(1-p)\le\frac{s(s-1)}m,
\]
where the final inequality is a union bound for collisions among
$s$ independent uniform values. Combining the two comparisons and
using $\sqrt{s}\le s$ proves the assertion.
\end{proof}

\begin{proof}[Proof of Lemma~\ref{lem:switch}]
If $q<8$, the hypothesis forces $t=0$, so the assertion is immediate.
Assume $q\ge8$ and fix a bijection between $[q]$ and $\mathbb Z_q$.
We extend the domain to $[q]$ for the analysis. A uniform injection
on $[r]$ is the restriction of a uniform permutation on $[q]$;
independent labels also extend to $q$ coordinates. Queries remain
restricted to $[r]$.

By Lemma~\ref{lem:one-round-phases}, grouping basis query records by
their aggregate charge labels gives algorithm vectors $u_\sigma$,
$\sigma\in\mathbb Z_q^q$, such that the state after the parallel
query layer on a fixed input $x$ is
\[
 \sum_{|\sigma|\le t}\omega_q^{\sum_i\sigma(i)x_i}u_\sigma,
 \qquad \langle u_\sigma,u_{\sigma'}\rangle=0\quad(\sigma\ne\sigma'),
 \qquad \sum_\sigma\|u_\sigma\|^2=1.
\]
These vectors depend only on the algorithm, so the same $u_\sigma$
occur for both input distributions.

For independent input values, encode each coordinate separately.
Map its zero Fourier label to a blank and each nonzero label $c$ to
\[
 q^{-1/2}\sum_{y\in\mathbb Z_q}\omega_q^{cy}|y\rangle.
\]
The blank vector and these $q-1$ vectors are orthonormal, so this
map is an isometry. The encoded purification is
$\sum_\sigma a_{0,\sigma}\otimes u_\sigma$,
where $a_{0,\sigma}$ is the vector of Claim~\ref{lem:vector}, with
$S=\Supp{\sigma}$ and $g_i(y)=\omega_q^{\sigma(i)y}$.
Each character has zero mean and modulus one.

For $h_\sigma(\pi):=\prod_{i\in\Supp{\sigma}}g_i(\pi(i))$,
encode the permutation input register using $V$. This gives the
purification $\sum_\sigma Vh_\sigma\otimes u_\sigma$ in the same
database space. Pairwise orthogonality of the $u_\sigma$ and
Claim~\ref{lem:vector} give
\[
 \norm{\sum_\sigma(Vh_\sigma-a_{0,\sigma})\otimes u_\sigma}^2
 =\sum_\sigma\norm{Vh_\sigma-a_{0,\sigma}}^2\norm{u_\sigma}^2
 \le\frac{16t^2}{q}.
\]
Both purifications have norm one, and the encodings act only on
the inaccessible input register. The trace distance between the
reduced algorithm states is at most the Euclidean distance between
these purifications, hence at most $4t/\sqrt q$.
\end{proof}

The same bound applies when several input positions share a block label.
Fix a partition $\mu$ of $[n]$ into $r\le q$ blocks, hidden from
the algorithm. Assign a label $y_B\in\mathbb Z_q$ to each block
$B\in\mu$, and set $x_i=y_B$ for $i\in B$.
Write $\mathsf{Block}_\mu$ for the resulting distribution with
independent uniform block labels, and $\mathsf{Exact}_\mu$ for
that with uniformly chosen pairwise distinct block labels.

\begin{corollary}[Injection switching for hidden block labels]
\label{cor:hidden-partition-switching}
For any $t$-query non-adaptive quantum algorithm with $t\le q/8$,
\[
 \frac12\norm{\varrho_{\mathsf{Exact}_\mu}-\varrho_{\mathsf{Block}_\mu}}_1
 \le\frac{4t}{\sqrt q}.
\]
\end{corollary}

By convexity of the trace norm, the bound also holds after averaging
over any common distribution on $\mu$. In particular, averaging over
uniform perfect $k$-matchings gives the comparison between
$\mathsf{Block}_k$ and $\mathsf{Exact}_k$ needed below.

\begin{proof}
The case $t=0$ is immediate; otherwise $q\ge8$.
Fix $\mu$ and use the original charge labels
$\sigma\in\mathbb Z_q^n$ from Lemma~\ref{lem:one-round-phases}.
The corresponding algorithm vectors $u_\sigma$ are pairwise
orthogonal and satisfy $\sum_\sigma\|u_\sigma\|^2=1$.
Use the block charges $\tau_{\mu,\sigma}$ defined in
Eq.~\eqref{eq:hidden-block-charge}.
We have $|\operatorname{Supp}\tau_{\mu,\sigma}|
\le|\operatorname{Supp}\sigma|\le t$.
Enumerate the $r$ blocks and pad their label register to $q$
coordinates, as in the proof of Lemma~\ref{lem:switch}.
Apply Claim~\ref{lem:vector} with $m=q$ to each branch, using
the zero-mean, unit-modulus characters
$g_B(y)=\omega_q^{\tau_{\mu,\sigma}(B)y}$ on its nonzero block charges.
Although different $\sigma$ can give the same block charge,
their algorithm vectors $u_\sigma$ remain orthogonal.
Consequently, the distance between the two encoded purifications
is at most
\[
 \left(\sum_\sigma
       \frac{16|\operatorname{Supp}\tau_{\mu,\sigma}|^2}{q}
       \|u_\sigma\|^2\right)^{1/2}
 \le\frac{4t}{\sqrt q}.
\]
The block indexing is used only in the environment purification;
the algorithm still queries positions in $[n]$ and receives $x_i$.
Taking reduced states gives the claimed trace-distance bound.
\end{proof}

\subsection{Completion of the decision lower bound}
\label{sec:decision-completion}

We now combine the two comparisons.
Section~\ref{sec:direct-relaxed-decision} bounds the success
probability for distinguishing $\mathsf{Fun}$ from $\mathsf{Block}_k$.
Section~\ref{sec:injection-switching} bounds the change in this
probability when we replace them by $\mathsf{Inj}$ and
$\mathsf{Exact}_k$.

\nonadaptivecollisiontheorem*
\begin{proof}[Proof of Theorem~\ref{thm:test}]
Fix a $t$-query one-round algorithm satisfying the stated
worst-case success guarantee. Write $\rho\in\{0,1\}$ for its
output, with $0$ declaring that the input is injective and $1$
declaring that it is exactly $k$-to-one.
If $t=0$, its output distribution is independent of the input,
so it cannot succeed with probability greater than $1/2$ on
both promised cases. Thus $t\ge1$.
If $t>\sqrt{n/(8k)}$, then $\varepsilon<1/2$ already gives
$t\ge(\varepsilon/6)\sqrt{n/k}$.
We may therefore assume $1\le t\le\sqrt{n/(8k)}$.
This implies $n\ge8k$ and
$t\le\sqrt{n/(8k)}\le n/(8k)\le q/8$, so all comparison bounds apply.

First run the circuit on inputs drawn from $\mathsf{Fun}$
and $\mathsf{Block}_k$. These inputs may fall outside the promise,
but the circuit and its output probabilities are still well defined.
The equal-prior bound~\eqref{eq:direct-relaxed-success} gives
\[
 p_{\rm relaxed}:=
 \frac12\Pr_{x\sim\mathsf{Fun}}[\rho=0]
 +\frac12\Pr_{x\sim\mathsf{Block}_k}[\rho=1]
 \le\frac12+2t\sqrt{\frac kn}.
\]

Next replace the independent labels by distinct labels.
Lemma~\ref{lem:switch}, with $r=n$, and
Corollary~\ref{cor:hidden-partition-switching}, averaged over
uniform perfect $k$-matchings, give
\[
\begin{aligned}
 \frac12\norm{\varrho_{\mathsf{Inj}}-\varrho_{\mathsf{Fun}}}_1
 \le\frac{4t}{\sqrt q}, \qquad
 \frac12\norm{\varrho_{\mathsf{Exact}_k}
                    -\varrho_{\mathsf{Block}_k}}_1
 \le\frac{4t}{\sqrt q}.
\end{aligned}
\]
Each bound controls the change in the probability of every
output event. Define the success probability under the promised distributions by
\[
 p_{\rm strict}:=
 \frac12\Pr_{x\sim\mathsf{Inj}}[\rho=0]
 +\frac12\Pr_{x\sim\mathsf{Exact}_k}[\rho=1].
\]
The probabilities include the algorithm's measurement outcomes.
Since each distribution has prior probability $1/2$, the two switching
errors contribute
\[
 |p_{\rm strict}-p_{\rm relaxed}|
 \le\frac12\frac{4t}{\sqrt q}+\frac12\frac{4t}{\sqrt q}
 =\frac{4t}{\sqrt q}.
\]
Every input from $\mathsf{Inj}$ or $\mathsf{Exact}_k$ satisfies
the corresponding promise. Hence the worst-case guarantee
implies $p_{\rm strict}\ge1/2+\varepsilon$, and
\[
 \frac12+\varepsilon
 \le p_{\rm strict}
 \le\frac12+2t\sqrt{\frac kn}+\frac{4t}{\sqrt q}
 \le\frac12+6t\sqrt{\frac kn},
\]
where the final inequality uses $q\ge n$.
Thus $t\ge(\varepsilon/6)\sqrt{n/k}$ in all cases, with a constant
independent of $k$.
\end{proof}

\clearpage

\appendix

\begin{center}
\LARGE\bfseries Appendix
\end{center}

\section{Linear Algebra and Representation-Theoretic Background}
\label{app:background}

This appendix expands the notation introduced in
Sections~\ref{pre:linear-algebra} and~\ref{sec:representation}.
It also records supplementary induction formulas and proves
Lemmas~\ref{lem:pre-partial-bijections} and~\ref{lem:pre-point-multiplier}.

\subsection{Detailed linear-algebra conventions}
\label{app:linear-algebra}

\paragraph{Spaces and operators.}
For a finite set \(X\), let
\[
 \mathcal X=\mathbb C[X]
 :=\Span\{|x\rangle_{\mathcal X}:x\in X\}
\]
with the displayed basis orthonormal. A subscript outside a ket or bra
names its register; parameters specifying the vector are
written inside the ket or bra. We omit space subscripts when the
ambient space is clear. The bra \(\langle x|\) is the conjugate transpose
of \(|x\rangle\), and \(\mathcal L(\mathcal X)\) is the space of linear
operators \(\mathcal X\to\mathcal X\), spanned by the operators
\(|x\rangle\langle y|\). For a finite group \(X\), \(\mathbb C[X]\) is
also its group algebra with this inner product.
We write \(I_{\mathcal X}\) for the identity on
\(\mathcal X\).
For vectors, \(\langle u,v\rangle=u^\dagger v\) and \(\|u\|\) is the
Euclidean norm. For a linear map \(H:\mathcal X\to\mathcal Y\),
\(H^\dagger:\mathcal Y\to\mathcal X\) is its adjoint,
and its operator and trace norms are, respectively,
\[
 \|H\|:=\sup_{\|v\|=1}\|Hv\|,\qquad
 \|H\|_1:=\operatorname{Tr}\sqrt{H^\dagger H}.
\]
A density operator $\varrho:\mathcal X\to\mathcal X$ is positive
semidefinite and has trace one. We reserve $\varrho$ for density
operators and $\rho$ for measurement outcomes. The trace distance of
two density operators is $\frac12\|\varrho-\varrho'\|_1$.
It bounds the difference in probability of any measurement event
and cannot increase under partial trace.
An operator on \(\mathcal X\) is unitary if
\(H^\dagger H=H H^\dagger=I_{\mathcal X}\), and Hermitian if
\(H=H^\dagger\). A linear map \(H:\mathcal X\to\mathcal Y\) is an
isometry if \(H^\dagger H=I_{\mathcal X}\), and a
contraction if \(\|H\|\le1\).
For a linear map \(H:\mathcal X\to\mathcal Y\), its \emph{kernel}
(or \emph{null space}) is the subspace
\[
\ker H:=\{v\in\mathcal X:Hv=0_{\mathcal Y}\}.
\]
It consists of all inputs that \(H\) maps to the zero vector of
\(\mathcal Y\). We write \([H,J]=HJ-JH\) for a commutator. A Hermitian operator
\(H\) is positive semidefinite if \(\langle v,Hv\rangle\ge0\) for all
\(v\). For Hermitian \(H,J\), the relation \(H\preceq J\) means that
\(J-H\) is positive semidefinite.

\paragraph{Projections, decompositions, and events.}
For a subspace \(\mathcal S\subseteq\mathcal X\), let
\(P_{\mathcal S}:\mathcal X\to\mathcal X\) be the orthogonal projector
onto \(\mathcal S\), and write \(\mathcal S^\perp\) for its orthogonal
complement. For \(H:\mathcal X\to\mathcal Y\), its restriction
\(H|_{\mathcal S}:\mathcal S\to\mathcal Y\) and its restricted norm are
\[
 (H|_{\mathcal S})v:=Hv\quad(v\in\mathcal S),\qquad
 \|H\|_{\mathcal S}:=\|H|_{\mathcal S}\|
 =\sup_{\substack{v\in\mathcal S\\\|v\|\le1}}\|Hv\|.
\]
For an orthogonal decomposition \(\mathcal H=\bigoplus_s\mathcal H_s\), put
\[
 \mathcal H_{\le t}:=\bigoplus_{s\le t}\mathcal H_s,
 \qquad \mathcal H_{<t}:=\bigoplus_{s<t}\mathcal H_s,
 \qquad P_{\mathcal H_{\le t}}:=\sum_{s\le t}P_{\mathcal H_s}.
\]
For a family of operators \(H_a:\mathcal H\to\mathcal H\) indexed by
a finite set \(A\), the controlled direct sum \(\bigoplus_a H_a\)
acts on \(\mathbb C[A]\otimes\mathcal H\) as
\(\sum_a|a\rangle\langle a|\otimes H_a\). An operator specified on one
register is extended by the identity on all other registers when used in a
joint state.

For a predicate \(E\) on basis labels \(X\), its diagonal projector
\(P_E:\mathbb C[X]\to\mathbb C[X]\) is
\[
P_E:=\sum_{x\in X:\,E(x)}|x\rangle\langle x|.
\]
For a count function \(g:X\to\mathbb R\), the notation \(g\ge\ell\)
denotes the event \(\{x\in X:g(x)\ge\ell\}\), and similarly for
\(=\) and \(\le\). A vector of norm at most one is called
\emph{subnormalized}. We use the squared norm of a projected
component as its probability mass, without renormalization.

\paragraph{Traces and measurements.}
The trace maps \(\mathcal L(\mathcal H)\) to \(\mathbb C\).
The partial trace
\(\Tr_{\mathcal Y}:\mathcal L(\mathcal Y\otimes\mathcal A)
\to\mathcal L(\mathcal A)\) is defined, for
\(U\in\mathcal L(\mathcal Y\otimes\mathcal A)\), as follows. If
\(\{|y\rangle:y\in Y\}\) is an orthonormal basis of \(\mathcal Y\),
then
\[
 \Tr_{\mathcal Y}(U)
 :=\sum_{y\in Y}(\langle y|\otimes I_{\mathcal A})
                  U(|y\rangle\otimes I_{\mathcal A}).
\]
This operator acts on \(\mathcal A\) and satisfies
\(\Tr(\Tr_{\mathcal Y}(U))=\Tr(U)\). A positive operator-valued
measurement (POVM) on \(\mathcal H\) is a collection of positive
semidefinite operators \(E_\rho:\mathcal H\to\mathcal H\) with
\(\sum_\rho E_\rho=I_{\mathcal H}\). A projection-valued
measurement (PVM) consists of pairwise orthogonal projectors summing to
\(I_{\mathcal H}\). We may assume that final measurements are PVMs:
Naimark's dilation theorem implements any POVM as a PVM after adjoining
workspace initialized independently of the input. Our model allows
arbitrary such workspace.

\subsection{Representation definitions}
\label{app:representation-definitions}

\subsubsection{Representations and invariant subspaces}

An \emph{action} of $G$ on a set $X$ is a map
$(g,x)\mapsto g\cdot x$ from $G\times X$ to $X$ satisfying
$e\cdot x=x$ and $(gh)\cdot x=g\cdot(h\cdot x)$.
The action is \emph{transitive} if, for every $x,y\in X$, some $g\in G$
satisfies $g\cdot x=y$. The \emph{stabilizer} of $x$ is
$G_x:=\{g\in G:g\cdot x=x\}$.

In particular, $S_m$ is the group of permutations of $[m]$, with
composition $(\sigma\tau)(i)=\sigma(\tau(i))$.
For $G=S_m$, the natural action on $[m]$ is $\sigma\cdot i=\sigma(i)$.
It also acts on subsets by $\sigma(S)=\{\sigma(i):i\in S\}$.
The stabilizer of a point is isomorphic to $S_{m-1}$, whereas the
stabilizer of a set of size $s$ is isomorphic to $S_{m-s}\times S_s$.

\begin{definition}[Representations]
\label{def:representation}
A \emph{representation} of $G$ on a vector space $W$ is a map~$\chi:G\to\mathcal L(W)$~such~that
\[
 \chi(e)=I_W,\qquad \chi(gh)=\chi(g)\chi(h)\quad(g,h\in G).
\]
We denote it by $(W,\chi)$; $W$ is its \emph{representation space}
and $\dim W$ its dimension.
\end{definition}

Each $\chi(g):W\to W$ is an invertible linear operator, with inverse
$\chi(g^{-1})$. The associated action on vectors is
$(g,w)\mapsto\chi(g)w$: a group element $g$ specifies an operator,
which is then applied to $w$. When the meaning is clear, we use
either $W$ or $\chi$ to denote the representation. We also extend $\chi$ linearly
to the group algebra by
$\chi(\sum_g a_g g):=\sum_g a_g\chi(g)$.
In particular, for $G=S_m$, permutation composition is represented by
operator composition:
$\chi(\sigma\tau)w=\chi(\sigma)(\chi(\tau)w)$.

A representation $(W,\chi)$ on an inner-product space is \emph{unitary}
if $\chi(g)^\dagger=\chi(g^{-1})$ for every $g\in G$.
The \emph{trivial representation} $\mathbf1$ of $G$ is the
one-dimensional representation on $\mathbb C$ with
$\mathbf1(g)z=z$ for all $g\in G$ and $z\in\mathbb C$.

Throughout the paper we use unitary representations. For finite groups,
this entails no loss of generality: averaging an inner product over $G$
makes it invariant. In particular, the trivial representation of $S_m$
assigns the identity on $\mathbb C$ to every permutation.

\begin{definition}[Invariant subspaces and irreducibility]
\label{def:subrepresentation}
A subspace $U\subseteq W$ is \emph{invariant} under $(W,\chi)$ if
$\chi(g)U\subseteq U$ for all $g\in G$.
The restricted operators $\chi_U(g):U\to U$, given by
$\chi_U(g):=\chi(g)|_U$, define the
\emph{subrepresentation} $(U,\chi_U)$.
A nonzero representation is \emph{irreducible} if its only invariant
subspaces are $\{0\}$ and $W$, and \emph{reducible} otherwise.
An irreducible summand in a direct-sum decomposition is called a
\emph{constituent}.
\end{definition}

For a unitary representation, the orthogonal complement of an invariant
subspace is invariant. Repeatedly splitting $W=U\oplus U^\perp$ therefore
decomposes $W$ into an orthogonal direct sum of irreducible
representations; see~\cite[Section~1.5]{sagan2001symmetric}.

\begin{definition}[Isomorphism and irreducible type]
\label{def:representation-isomorphism}
An \emph{isomorphism} between representations $(W,\chi)$ and
$(\widetilde W,\widetilde\chi)$ of $G$ is an invertible linear map
$T:W\to\widetilde W$ satisfying
\[
 T\chi(g)=\widetilde\chi(g)T\qquad(g\in G).
\]
We then write $(W,\chi)\cong(\widetilde W,\widetilde\chi)$, or simply
$W\cong\widetilde W$. An \emph{irreducible type} is an isomorphism
class of~irreducible~representations.
\end{definition}

Thus an isomorphism must respect the group action: the same change of
coordinates gives $\widetilde\chi(g)=T\chi(g)T^{-1}$ for every $g$.
Equal dimensions alone do not imply isomorphism.

\begin{definition}[Multiplicity and isotypic subspaces]
\label{def:isotypic-subspace}
Let $W$ be a representation of $G$, and let $U$ be an irreducible
representation of $G$. The \emph{multiplicity} of $U$ in $W$ is the
number of constituents isomorphic to $U$ in an irreducible decomposition
of $W$. The \emph{$U$-isotypic subspace} of $W$ is the sum of all its
irreducible subrepresentations isomorphic to $U$.
\end{definition}

The multiplicities and isotypic subspaces are independent of the chosen
irreducible decomposition. Distinct isotypic subspaces are orthogonal,
and their direct sum is $W$. Individual irreducible summands within one
isotypic subspace need not be uniquely determined.

\subsubsection{Partitions and irreducible types of the symmetric group}

\begin{definition}[Partitions and Young diagrams]
\label{def:young-diagram}
An \emph{integer partition} $\lambda\vdash m$ is a weakly decreasing
sequence of positive integers summing to $m$, extended by zero parts.
Its \emph{Young diagram} has $\lambda_i$ boxes in row $i$; write
$(i,j)\in\lambda$ for a box in row $i$ and column $j$.
Its \emph{conjugate partition} has parts
$\lambda'_j=|\{i:\lambda_i\ge j\}|$.
For $\mu\subseteq\lambda$, the \emph{skew diagram} $\lambda/\mu$
consists of the boxes outside $\mu$; it is a \emph{horizontal strip}
if no two of its boxes lie in the same column.
We write $\nu\nearrow\lambda$ when $\lambda$ adds one~box~to~$\nu$.
\end{definition}

Here $\lambda_1$ is the first-row length, $\lambda'_j$ is a column
length, and $|\lambda|=m$ is the number of boxes.
The notation $\lambda=(m-s,\nu)$ places $\nu$ below a first row of
length $m-s$. Write $(a)$ for the one-row partition of size $a$, with
$(0)=\emptyset$, the partition of zero.

\begin{definition}[Tableaux, content, and hook length]
\label{def:tableaux-content}
A \emph{standard Young tableau} of shape $\lambda/\mu$ is a filling
by $1,\ldots,|\lambda|-|\mu|$ increasing along rows and columns.
Let $f^{\lambda/\mu}$ count these tableaux, and put
$f^\lambda:=f^{\lambda/\emptyset}$.
The \emph{content sum} of $\lambda$ and the \emph{hook length}
of its box $(i,j)$ are
\[
 c(\lambda):=\sum_{(i,j)\in\lambda}(j-i),\qquad
 h_\lambda(i,j):=\lambda_i-j+\lambda'_j-i+1.
\]
\end{definition}

The content sum adds the column index minus the row index over all boxes.
We use the conventions $f^\emptyset=1$ and $f^{\lambda/\mu}=0$
when $\mu\not\subseteq\lambda$.

For $G=S_m$, the irreducible types are indexed by partitions
$\lambda\vdash m$; see~\cite[Section~2.5]{sagan2001symmetric}
and~\cite[Theorem~5.12.2]{etingof2011representation}.
Choose a representative $(V_\lambda,\chi_\lambda)$ of each type.
Their isomorphism classes and dimensions are given in
Eq.~\eqref{eq:irreducible-types}.
We call the $V_\lambda$-isotypic subspace the \emph{$\lambda$-isotypic
subspace}. The trivial representation has type $V_{(m)}$.
A concrete realization of $(V_\lambda,\chi_\lambda)$ and a worked
example appear in Appendix~\ref{app:representation-theory}.

\subsubsection{Products, regular representations, restriction, and induction}

For representations $(W_1,\chi_1)$ of $G_1$ and $(W_2,\chi_2)$ of $G_2$,
their \emph{tensor product representation} of $G_1\times G_2$ is
$W_1\otimes W_2$ with action
\[
 (g_1,g_2)\cdot(w_1\otimes w_2)
 :=\chi_1(g_1)w_1\otimes\chi_2(g_2)w_2.
\]

The product group has componentwise multiplication.
In particular, for $G_1=G_2=S_m$, the joint type $(\lambda,\mu)$
means the representation $V_\lambda\otimes V_\mu$: the first
permutation acts on $V_\lambda$ and the second on $V_\mu$.
This convention will distinguish range and domain relabellings.

\begin{definition}[Left and right regular representations]
\label{def:regular-actions}
The \emph{left} and \emph{right regular representations} of $G$ on
$\mathbb C[G]$ are the operators
\[
 L_a|g\rangle:=|ag\rangle,\qquad
 R_b|g\rangle:=|gb^{-1}\rangle
 \qquad(a,b,g\in G).
\]
\end{definition}

The inverse in the right action ensures
$R_aR_b=R_{ab}$, and the left and right actions commute.
In particular, when $G=S_m$, their joint action is
\[
 L_\tau R_\sigma|\pi\rangle
 =|\tau\circ\pi\circ\sigma^{-1}\rangle.
\]
These are range and domain relabellings, respectively.
Write $\mathcal P_\lambda$ for the $\lambda$-isotypic subspace of the
left regular representation. Lemma~\ref{lem:pre-regular} shows that
the right regular $\lambda$-isotypic subspace is the same space.
For partial injections the corresponding relabelling action is
$|d\rangle\mapsto|\tau\circ d\circ\sigma^{-1}\rangle$.

\begin{definition}[Restriction and fixed subspaces]
\label{def:restriction}
For a representation $(W,\chi)$ of $G$ and a subgroup $H\le G$,
the \emph{restriction} $\operatorname{Res}_H^G W$ is $W$ with the
action $h\mapsto\chi(h)$ for $h\in H$. Its \emph{fixed subspace} is
\[
 W^H:=\{w\in W:\chi(h)w=w\text{ for every }h\in H\}.
\]
\end{definition}

Restriction may be reducible even when $W$ is irreducible for $G$.
For symmetric groups we will restrict both to point stabilizers
$S_{m-1}$ and to subgroups $S_a$ permuting a specified set of labels.

\begin{definition}[Induced representations]
\label{def:restriction-induction}
For a representation $(U,\chi_U)$ of $H\le G$, the \emph{induced
representation} $\operatorname{Ind}_H^G U$ is the function space
\[
 \{f:G\to U:f(gh)=\chi_U(h^{-1})f(g)\text{ for all }g\in G,\ h\in H\},
\]
with $G$-action $(a\cdot f)(g):=f(a^{-1}g)$.
\end{definition}

A function in this space is determined freely by one value in $U$ for
each left coset $gH=\{gh:h\in H\}$. Write $G/H$ for the set of these
cosets. Thus induction has one copy of $U$ for each coset, and
dimension $|G/H|\dim U$.
For $G=S_m$ and the set stabilizer $H\cong S_{m-s}\times S_s$,
the cosets correspond to the $s$-element subsets of $[m]$.

\subsection{Standard tools and induction formulas}
\label{app:representation-tools}

We record the induction and iterated restriction formulas complementing
the lemmas in Section~\ref{sec:representation}; standard references are
\cite[Section~6.1]{rosmanis2021tight} and
\cite[Chapter~1]{sagan2001symmetric}.
In particular, Lemma~\ref{lem:pre-schur} implies that an intertwiner
between nonisomorphic irreducible representations is zero.

\paragraph{Branching and fixed subspaces.}
Restricting an irreducible representation to the subgroup fixing
one label gives types obtained by removing one box from its Young
diagram. Deleting one box leaves the first-row length unchanged or
reduces it by one. Hence two $S_m$ types with a common $S_{m-1}$
constituent have first-row lengths differing by at most one.
For $\nu\vdash m-1$ and $0\le a\le m$, the induction form of the
branching rule and its iterated restriction form are
\[
 \operatorname{Ind}_{S_{m-1}}^{S_m}V_\nu
 \cong\bigoplus_{\nu\nearrow\lambda}V_\lambda,
 \qquad
 \operatorname{Res}_{S_a}^{S_m}V_\lambda
 \cong\bigoplus_{\mu\vdash a}V_\mu^{\oplus f^{\lambda/\mu}}.
\]
The fixed-space formula in Lemma~\ref{lem:pre-branching} follows by
taking the multiplicity of the trivial type $(a)$; it also shows that
$V_\lambda^{S_a}$ is nonzero exactly when $\lambda_1\ge a$.

\paragraph{Pieri rule and injections.}
For $0\le s\le m$ and $\mu\vdash s$, the induction form of the
Pieri rule is
\begin{equation}
\label{pre:pieri-rule}
 \operatorname{Ind}_{S_s\times S_{m-s}}^{S_m}
       (V_\mu\otimes\mathbf1)
 \cong
 \bigoplus_{\substack{\lambda\vdash m\\
                      \lambda/\mu\ \text{a horizontal strip}}}V_\lambda.
\end{equation}
This gives the injection decomposition in Lemma~\ref{lem:pre-pieri}.
The horizontal-strip condition is equivalent to
$\lambda_i\ge\mu_i\ge\lambda_{i+1}$ for every $i\ge1$.
For $\lambda=(m-r,\nu)$, it implies $r\le s$ and
$\nu\subseteq\mu$, with $\nu=\mu$ when $r=s$.

\subsection{Elementary induction facts}
\label{app:induction-details}

The following two facts justify the induction formulas in the proof
of Lemma~\ref{lem:pre-partial-bijections}.

\begin{claim}[Product representations and induction]
\label{lem:pre-induction-products}
If $U_i$ is an irreducible representation of $G_i$ for $i=1,2$, then
$U_1\otimes U_2$ is irreducible as a representation of $G_1\times G_2$.
For a finite family of representations $U_\alpha$ of $H\le G$,
\begin{equation}
\label{pre:induction-sum}
 \operatorname{Ind}_H^G\Bigl(\bigoplus_\alpha U_\alpha\Bigr)
 \cong\bigoplus_\alpha\operatorname{Ind}_H^G U_\alpha.
\end{equation}
For $H_i\le G_i$ and representations $U_i$ of $H_i$, $i=1,2$,
\begin{equation}
\label{pre:induction-product}
 \operatorname{Ind}_{H_1\times H_2}^{G_1\times G_2}(U_1\otimes U_2)
 \cong
 \operatorname{Ind}_{H_1}^{G_1}U_1\otimes
 \operatorname{Ind}_{H_2}^{G_2}U_2.
\end{equation}
\end{claim}
\begin{proof}
For irreducible $U_1,U_2$, write an operator commuting with the product
action as a matrix of operators on $U_1$, using a basis of $U_2$.
Lemma~\ref{lem:pre-schur} makes every matrix entry scalar, so the operator
is $I_{U_1}\otimes B$. Commutation with $G_2$ then makes $B$ scalar.
A proper invariant subspace would yield a nonscalar commuting orthogonal
projector, so the product representation is irreducible.

In Definition~\ref{def:restriction-induction}, the covariance condition
for a function with values in $\bigoplus_\alpha U_\alpha$ holds in each
component separately. This gives~\eqref{pre:induction-sum}.
For~\eqref{pre:induction-product}, use the map
\[
 f_1\otimes f_2\longmapsto
 \bigl[(g_1,g_2)\longmapsto f_1(g_1)\otimes f_2(g_2)\bigr].
\]
It satisfies the covariance condition for $H_1\times H_2$ and respects
the $G_1\times G_2$ action. Choose coset representatives in each $G_i/H_i$
and a basis of each $U_i$. Induced functions supported on one coset,
with one basis vector as their value at its representative, form a basis.
The displayed map sends tensor products of these basis functions
bijectively to the corresponding functions on pairs of cosets.
\end{proof}

\begin{claim}[Induction from transitively permuted subspaces]
\label{lem:pre-induced-subspaces}
Let $(W,\chi)$ be a representation of $G$, and suppose
$W=\bigoplus_{x\in X}W_x$, where $G$ acts transitively on $X$ and
$\chi(g)W_x=W_{g\cdot x}$. For~$x_0\in X$~and~$H=G_{x_0}$,
\[
 W\cong\operatorname{Ind}_H^G W_{x_0}.
\]
An isomorphism from the induced space to $W$ is
$f\mapsto\sum_{gH\in G/H}\chi(g)f(g)$, with one representative chosen
from each coset.
\end{claim}
\begin{proof}
The restricted action on $W_{x_0}$ is a representation of $H$.
Replacing $g$ by $gh$ leaves a \mbox{summand unchanged, since}
\[
 \chi(gh)f(gh)=\chi(g)\chi(h)\chi(h^{-1})f(g)=\chi(g)f(g).
\]
The cosets correspond bijectively to $g\cdot x_0\in X$.
Their freely chosen values $f(g)\in W_{x_0}$ map bijectively, under
$\chi(g)$, to the summands $W_{g\cdot x_0}$; hence the map is bijective.
For $a\in G$, reindexing cosets by $g'=a^{-1}g$ gives
\[
 \sum_{gH}\chi(g)(a\cdot f)(g)
 =\chi(a)\sum_{g'H}\chi(g')f(g'),
\]
which proves compatibility with the group action.
\end{proof}

\subsection{Partial bijections and simultaneous first-row types}
\label{app:partial-bijections-proof}
\label{pre:induction-tools}

A partial bijection is specified by its domain, image, and the
bijection between them. Range and domain relabellings act by
$(\tau,\sigma)|d\rangle=|\tau\circ d\circ\sigma^{-1}\rangle$.

\begin{proof}[Proof of Lemma~\ref{lem:pre-partial-bijections}]
We first express the full space of partial bijections by induction.
For $0\le s\le n$, let
$H=\{h\in S_n:h([s])=[s]\}\cong S_{n-s}\times S_s$.
Regard $\mathbb C[S_s]$ as an~$H\times H$~representation~via
\[
 (h_R,h_D)\cdot|\pi\rangle
 :=|(h_R|_{[s]})\circ\pi\circ(h_D|_{[s]})^{-1}\rangle.
\]
Thus both $S_{n-s}$ factors act trivially, while the two $S_s$ factors
act by the left and right regular representations.
Using the elementary induction facts in
Appendix~\ref{app:induction-details}, we obtain
\begin{equation}
\label{pre:partial-bijection-decomposition}
\begin{aligned}
 \mathbb C[\{d\in\mathcal I_n:|d|=s\}]\;
 \cong\;\operatorname{Ind}_{H\times H}^{S_n\times S_n}\mathbb C[S_s]\;
 \cong\;\bigoplus_{\eta\vdash s}
 \operatorname{Ind}_{S_{n-s}\times S_s}^{S_n}(\mathbf1\otimes V_\eta)
 \otimes
 \operatorname{Ind}_{S_{n-s}\times S_s}^{S_n}(\mathbf1\otimes V_\eta).
\end{aligned}
\end{equation}

For $s$-element sets $S,T\subseteq[n]$, let $W_{S,T}$ be the span of
bijections $S\to T$. These spaces form an orthogonal direct sum, and
$(\tau,\sigma)$ sends $W_{S,T}$ to $W_{\sigma(S),\tau(T)}$.
This action on pairs of support sets is transitive.
The reference subspace $W_{[s],[s]}$ is $\mathbb C[S_s]$, and the
stabilizer of its domain and image sets is $H\times H$.
Apply Claim~\ref{lem:pre-induced-subspaces} with group $S_n\times S_n$,
index set the pairs $(T,S)$, and reference index $([s],[s])$.
It gives the first isomorphism in~\eqref{pre:partial-bijection-decomposition}.

By Lemma~\ref{lem:pre-regular}, Eq.~\eqref{pre:regular-decomposition},
with $m=s$, the reference representation decomposes as
\[
 \mathbb C[S_s]\cong\bigoplus_{\eta\vdash s}
 (\mathbf1\otimes V_\eta)\otimes(\mathbf1\otimes V_\eta)
 \qquad\text{as an $H\times H$ representation}.
\]
Here $\mathbf1$ is the trivial representation of $S_{n-s}$, since
permuting unused labels fixes every reference bijection.
Apply Claim~\ref{lem:pre-induction-products},
Eqs.~\eqref{pre:induction-sum} and~\eqref{pre:induction-product},
with $G_1=G_2=S_n$, $H_1=H_2=H$, and
$U_1=U_2=\mathbf1\otimes V_\eta$ in each summand.
This gives the second isomorphism
in~\eqref{pre:partial-bijection-decomposition}.

By the induced form of the Pieri rule, Eq.~\eqref{pre:pieri-rule},
with $m=n$ and the subgroup factors interchanged, each induced factor contains
$V_\lambda$ once precisely when $\lambda/\eta$ is a horizontal
$(n-s)$-strip. If $\lambda_1=n-s$, the interlacing inequalities give
$\eta_i\ge\lambda_{i+1}$ for every $i$. Both $\eta$ and the lower
diagram $(\lambda_2,\lambda_3,\ldots)$ have $s$ boxes, so
$\eta_i=\lambda_{i+1}$ for every $i$.
Conversely, $\lambda=(n-s,\eta)$ is a partition and satisfies the
horizontal-strip condition whenever $\eta_1\le n-s$.
Requiring first row $n-s$ in both induced factors therefore leaves
exactly~\eqref{pre:simultaneous-first-row}, once for each valid $\eta$.
Each tensor product is irreducible for the joint action by
Claim~\ref{lem:pre-induction-products}.
\end{proof}

\subsection{Detailed point-evaluation support argument}
\label{app:point-multiplier}
\label{pre:point-multipliers}

\begin{proof}[Proof of Lemma~\ref{lem:pre-point-multiplier}]
Let $G_i=\{\sigma\in S_m:\sigma(i)=i\}\cong S_{m-1}$.
For $\sigma\in G_i$, its inverse fixes $i$, so
\[
 (\pi\circ\sigma^{-1})(i)=\pi(\sigma^{-1}(i))=\pi(i).
\]
Thus $M_{i,g}R_\sigma=R_\sigma M_{i,g}$ for every $\sigma\in G_i$.
The isotypic projectors $P_{\mathcal P_\lambda}$ and
$P_{\mathcal P_\mu}$ also commute with $R_\sigma$, since their
subspaces and orthogonal complements are invariant under the domain action.

Define $T:\mathcal P_\lambda\to\mathcal P_\mu$ and restricted
actions $\chi(\sigma):\mathcal P_\lambda\to\mathcal P_\lambda$,
$\widetilde\chi(\sigma):\mathcal P_\mu\to\mathcal P_\mu$ by the
formulas below, and apply Lemma~\ref{lem:pre-schur} with
\[
\begin{gathered}
 G=G_i,\quad W=\mathcal P_\lambda,\quad
 \widetilde W=\mathcal P_\mu,\quad
 T=\left.P_{\mathcal P_\mu}M_{i,g}P_{\mathcal P_\lambda}
                                        \right|_{\mathcal P_\lambda},\\
 \chi(\sigma)=R_\sigma|_{\mathcal P_\lambda},\qquad
 \widetilde\chi(\sigma)=R_\sigma|_{\mathcal P_\mu}
 \quad(\sigma\in G_i).
\end{gathered}
\]
These are the restrictions of the domain representations to $G_i$.
For $w\in\mathcal P_\lambda$, the commutation relations give
\[
\begin{aligned}
 T\chi(\sigma)w
 &=P_{\mathcal P_\mu}M_{i,g}R_\sigma w
  =P_{\mathcal P_\mu}R_\sigma M_{i,g}w\\
 &=R_\sigma P_{\mathcal P_\mu}M_{i,g}w
  =\widetilde\chi(\sigma)Tw.
\end{aligned}
\]
The representations $W,\widetilde W$ need not be irreducible.
Since an intertwiner preserves isotypic subspaces
(Lemma~\ref{lem:pre-schur}), a nonzero $T$ requires a common $G_i$ type:
otherwise every isotypic subspace of $W$ is mapped to zero.

By Lemma~\ref{lem:pre-regular}, the domain representation on
$\mathcal P_\lambda$ consists of $f^\lambda$ copies of $V_\lambda$,
and similarly for $\mathcal P_\mu$. Thus a common $G_i$ type must occur
in both $\operatorname{Res}_{G_i}^{S_m}V_\lambda$ and
$\operatorname{Res}_{G_i}^{S_m}V_\mu$.
Lemma~\ref{lem:pre-branching}, for the point stabilizer $G_i$,
identifies such a type with a diagram $\nu\vdash m-1$ obtained by
removing one box from each of $\lambda$ and $\mu$.
Consequently
\[
 \lambda_1-\nu_1,\ \mu_1-\nu_1\in\{0,1\},
 \qquad |\lambda_1-\mu_1|\le1.
\]
Every block violating this condition vanishes, proving
\eqref{pre:point-multiplier-support}.
\end{proof}

\section{Concrete Irreducible Models and a Worked Example}
\label{app:representation-theory}

This appendix gives a concrete model for the irreducible type
\(V_\lambda\) from Eq.~\eqref{eq:irreducible-types}. A worked example
illustrates its dimension, the action of permutations, and the
distinction between an irreducible constituent and an isotypic subspace.

\subsection{Young symmetrizer construction}

Fix \(\lambda\vdash m\). To construct a concrete model for
\(V_\lambda\), fill the Young diagram of
\(\lambda\) with \(1,\ldots,m\), each appearing once, to obtain a
tableau \(T\). Let \(R_T\le S_m\) be the subgroup permuting entries
within each row of \(T\), and let \(C_T\le S_m\) be the subgroup
permuting entries within each column. In the group algebra
\(\mathbb C[S_m]\), whose elements are complex linear combinations of
permutations and whose multiplication extends permutation composition,
define the \emph{Young symmetrizer} \(y_T\), the space \(V_\lambda\),
and the action operators \(\chi_\lambda(\tau):V_\lambda\to V_\lambda\) by
\[
\begin{gathered}
 y_T:=
 \left(\sum_{r\in R_T}r\right)
 \left(\sum_{c\in C_T}\operatorname{sgn}(c)c\right),
 \qquad V_\lambda:=\mathbb C[S_m]y_T,\\
 \chi_\lambda(\tau)(\sigma y_T):=(\tau\sigma)y_T
 \qquad(\sigma,\tau\in S_m).
\end{gathered}
\]
Thus \(V_\lambda=\operatorname{span}_{\mathbb C}
\{\sigma y_T:\sigma\in S_m\}\).
For each fixed \(\tau\), the map \(\chi_\lambda(\tau):V_\lambda\to V_\lambda\)
is the linear operator given by left multiplication by \(\tau\).
Here \(\operatorname{sgn}(c)\) is the sign of the permutation \(c\).
This space is irreducible, and different choices of \(T\) give
isomorphic models of \(V_\lambda\);
see~\cite[Theorem~5.12.2]{etingof2011representation}.
The inner product inherited from \(\mathbb C[S_m]\),
with its permutation basis orthonormal, makes this action unitary.

\subsection{A worked example}

\begin{samepage}
For a concrete example, take \(\lambda=(3,2)\vdash5\) and fill its
Young diagram with the tableau
\[
 T=
 \begin{array}{|c|c|c|}
 \hline
 1&2&3\\ \hline
 4&5&\multicolumn{1}{c}{}\\ \cline{1-2}
 \end{array}
 \qquad
 |\lambda|=5,\quad \lambda_1=3,\quad \lambda'=(2,2,1).
\]
\end{samepage}
Thus the row lengths are \(3,2\), and the column lengths are \(2,2,1\).
The contents \(j-i\) are \(0,1,2\) in the first row and \(-1,0\) in
the second, giving
\[
 c(\lambda)=\sum_{(i,j)\in\lambda}(j-i)
           =(0+1+2)+(-1+0)=2.
\]
The row subgroup preserves the sets \(\{1,2,3\}\) and \(\{4,5\}\);
the column subgroup permutes within \(\{1,4\}\), \(\{2,5\}\), and
\(\{3\}\). Explicitly,
\[
\begin{aligned}
 R_T&=\{r\in S_5:r(\{1,2,3\})=\{1,2,3\},\
                   r(\{4,5\})=\{4,5\}\}, &&|R_T|=3!\,2!=12,\\
 C_T&=\{r\in S_5:r(\{1,4\})=\{1,4\},\
                   r(\{2,5\})=\{2,5\}\}, &&|C_T|=2!\,2!\,1!=4.
\end{aligned}
\]
Here \(e\) denotes the identity permutation.
The two transpositions in \(C_T\) have sign \(-1\), and their product
has sign \(+1\). Hence the corresponding Young symmetrizer is
\[
\begin{aligned}
 y_T
 &=\left(\sum_{r\in R_T}r\right)
          (e-(1\,4))(e-(2\,5))\\
 &=\sum_{r\in R_T}
       \bigl(r-r(1\,4)-r(2\,5)+r(1\,4)(2\,5)\bigr).
\end{aligned}
\]
Here \(r(1\,4)\) denotes composition of permutations, with the
rightmost factor applied first, and the sum is a vector in
\(\mathbb C[S_5]\). In this example, the space and action are
\[
 V_{(3,2)}
 =\operatorname{span}_{\mathbb C}\{\sigma y_T:\sigma\in S_5\},
 \qquad
 \chi_{(3,2)}((1\,4))(\sigma y_T)=((1\,4)\sigma)y_T.
\]
For instance, \(\chi_{(3,2)}((1\,2))y_T=y_T\), because left
multiplication by \((1\,2)\in R_T\) permutes the terms in the row sum.

\paragraph{Counting \(f^\lambda\).}
For \(\lambda=(3,2)\), the five standard Young tableaux are
\[
\begin{gathered}
 T=
 \begin{array}{|c|c|c|}
 \hline
 1&2&3\\ \hline
 4&5&\multicolumn{1}{c}{}\\ \cline{1-2}
 \end{array}
 \qquad
 T'=
 \begin{array}{|c|c|c|}
 \hline
 1&2&4\\ \hline
 3&5&\multicolumn{1}{c}{}\\ \cline{1-2}
 \end{array}
 \qquad
 \begin{array}{|c|c|c|}
 \hline
 1&2&5\\ \hline
 3&4&\multicolumn{1}{c}{}\\ \cline{1-2}
 \end{array}
 \qquad
 \begin{array}{|c|c|c|}
 \hline
 1&3&4\\ \hline
 2&5&\multicolumn{1}{c}{}\\ \cline{1-2}
 \end{array}
 \qquad
 \begin{array}{|c|c|c|}
 \hline
 1&3&5\\ \hline
 2&4&\multicolumn{1}{c}{}\\ \cline{1-2}
 \end{array}.
\end{gathered}
\]
Every row and column is increasing, and these exhaust the standard fillings
of this shape. Thus \(f^\lambda=f^{(3,2)}=5\).
The first-row hook lengths are \(4,3,1\), and the second-row hook
lengths are \(2,1\). The hook-length formula of Lemma~\ref{lem:pre-hook}
gives the same count and identifies the~representation~dimension:
\[
 \dim V_{(3,2)}=f^{(3,2)}
 =\frac{5!}{4\cdot3\cdot1\cdot2\cdot1}=5.
\]
The family \((\sigma y_T)_{\sigma\in S_5}\), indexed by the
\(5!=120\) permutations, spans a five-dimensional space.

\paragraph{An explicit isomorphism between two models.}
The second tableau \(T'\) above is obtained from \(T\) by exchanging the
entries \(3\) and \(4\). Put \(\alpha=(3\,4)\).
Relabelling the entries conjugates the row and column groups, and hence
\[
 y_{T'}=\alpha y_T\alpha^{-1}.
\]
The two representation spaces are
\[
 V_T:=\mathbb C[S_5]y_T,\qquad
 V_{T'}:=\mathbb C[S_5]y_{T'},
\]
with left multiplication actions on \(V_T\) and \(V_{T'}\), respectively,
\(\chi_T(\tau)v:=\tau v\) and \(\chi_{T'}(\tau)w:=\tau w\).
Right multiplication by \(\alpha^{-1}\) defines a linear map
\[
 \Phi_\alpha:V_T\longrightarrow V_{T'},\qquad
 \Phi_\alpha(v):=v\alpha^{-1}.
\]
Its values belong to \(V_{T'}\), since
\[
 xy_T\alpha^{-1}=(x\alpha^{-1})y_{T'}\in V_{T'}
 \qquad(x\in\mathbb C[S_5]).
\]
Its inverse is right multiplication by \(\alpha\). Moreover,
\[
 \Phi_\alpha\bigl(\chi_T(\tau)v\bigr)
 =(\tau v)\alpha^{-1}
 =\tau(v\alpha^{-1})
 =\chi_{T'}(\tau)\Phi_\alpha(v)
 \qquad(\tau\in S_5).
\]
This invertible map intertwines the permutation actions, giving
the representation isomorphism
\[
 (V_T,\chi_T)\cong(V_{T'},\chi_{T'}).
\]
Both are concrete models of the type \(V_{(3,2)}\).
The conjugate partition \((2,2,1)\) also has hook-length product \(24\), so
\(\dim V_{(2,2,1)}=5\). The classification of irreducible types nevertheless
gives \(V_{(3,2)}\not\cong V_{(2,2,1)}\): equal dimensions alone do not
give an isomorphism of representations.

\paragraph{Constituents and isotypic subspaces.}
On $W:=V_T\oplus V_{T'}$, consider the direct-sum representation
whose operators $\chi_W(\tau):W\to W$ are
\[
 \chi_W(\tau)(v,w)
 :=\bigl(\chi_T(\tau)v,\chi_{T'}(\tau)w\bigr).
\]
The subspaces \(V_T\oplus\{0\}\) and \(\{0\}\oplus V_{T'}\) are
invariant and carry irreducible subrepresentations, each isomorphic to
\(V_{(3,2)}\). They are the two constituents of the displayed decomposition.
Thus \(W\) is reducible and has dimension \(10\).
The multiplicity of \(V_{(3,2)}\) in \(W\) is \(2\).
Since both constituents have this type, the \((3,2)\)-isotypic
subspace is all of \(W\).

For the regular representation on \(\mathbb C[S_5]\),
Lemma~\ref{lem:pre-regular} gives
\[
 \mathcal P_{(3,2)}\cong V_{(3,2)}\otimes V_{(3,2)},
 \qquad \dim\mathcal P_{(3,2)}=5^2=25.
\]
Under either the range or the domain action alone, this isotypic space
decomposes into five constituents isomorphic to \(V_{(3,2)}\), each of dimension \(5\).
Under the joint \(S_5\times S_5\) action, it~is~irreducible.

\section{Classical Bounds for Collision Problems}
\label{app:classical-bounds}

{
For completeness, we give self-contained proofs of the classical bounds for
\kFind and \kColl. In both cases, non-adaptive sampling is optimal even among
adaptive randomized algorithms.

\begin{proposition}[Classical finding complexity]
\label{prop:classical-finding}
Let $k\ge2$ be fixed, $n=kN$, and $q\ge N$.
The classical randomized query complexity of \kFind with constant
success probability is $\Theta_k(n^{1-1/k})$, with or without adaptivity.
\end{proposition}
\begin{proof}
For the upper bound, fix a promised input and let $B_1,\ldots,B_N$
be its collision blocks. Query a uniformly random set $S\subseteq[n]$
of $t$ positions and output any block contained in $S$.
Let $Z$ count such blocks and write $\mu:=\mathbb EZ=N(t)_k/(n)_k$.
Here $(a)_r=0$ for integers $0\le a<r$.
The indicators of distinct blocks are negatively correlated:
\[
 \Pr[B_y\cup B_z\subseteq S]
 =\frac{(t)_{2k}}{(n)_{2k}}
 \le\left(\frac{(t)_k}{(n)_k}\right)^2
 \qquad(y\ne z).
\]
Consequently, $\mathbb EZ^2\le\mu+\mu^2$.
For $\mu>0$, the second-moment bound gives
\[
 \Pr[Z>0]\ge\frac{(\mathbb EZ)^2}{\mathbb EZ^2}
 \ge\frac{\mu}{1+\mu}.
\]
Choosing $t=\min\{n,\lceil C_k n^{1-1/k}\rceil\}$ for a sufficiently
large constant $C_k$ gives constant success probability.
All queries can be made in one round.

For the lower bound, draw the input uniformly from strings containing
exactly $k$ copies of each of $N$ fixed distinct symbols.
Consider an algorithm making at most $t$ queries, possibly adaptively,
and first suppose $s:=t+k\le n$. Fix its random coins and answer
repeated queries from the transcript. Conditional on any transcript,
every placement of the remaining symbol copies among the unqueried
positions is equally likely. Thus each fresh answer has the
distribution of a uniformly sampled remaining copy, regardless of
how the algorithm chooses the next unqueried position. The fresh
answers are therefore distributed as draws without replacement
from the multiset of symbol copies.

After the algorithm stops, query any valid output positions that have
not yet been queried, then query arbitrary unused positions until
exactly $s$ distinct positions have been queried in total.
Invalid outputs count as failures. Every successful output implies
that these $s$ answers contain all $k$ copies of some symbol.
A union bound over the $N$ symbols bounds the average success
probability $p$ by
\[
 p\le N\frac{(s)_k}{(n)_k}
 \le N\left(\frac{s}{n}\right)^k
 =\frac{(t+k)^k}{k n^{k-1}}.
\]
This bound holds for every fixing of the random coins, so it also
holds for the randomized algorithm. A constant worst-case success
probability therefore requires $t=\Omega_k(n^{1-1/k})$.
For fixed $k$, the remaining regime $t+k>n$ already implies this
lower bound.
\end{proof}

\begin{proposition}[Classical testing complexity]
\label{prop:classical-testing}
Let $k\ge2$ be fixed, $n=kN$, and $q\ge n$.
The classical randomized query complexity of \kColl with constant
success probability more than $1/2$ is $\Theta(\sqrt{n/k})$, with or without adaptivity.
\end{proposition}
\begin{proof}
For the upper bound, query $t$ uniformly chosen distinct positions.
Output ``$k$-to-one'' if two answers agree and ``injective'' otherwise.
This is always correct on an injective input. On an exactly $k$-to-one
input, after $j$ distinct values have been observed, their blocks
contain $j(k-1)$ of the $n-j$ unqueried positions. Hence, for $t\le N$,
\[
 \Pr[\text{no repeated value}]
 =\prod_{j=0}^{t-1}\left(1-\frac{j(k-1)}{n-j}\right)
 \le\exp\!\left(-\frac{(k-1)t(t-1)}{2n}\right).
\]
For $t>N$, a repeated value is certain. Thus
$t=\min\{N+1,\lceil C\sqrt{n/k}\rceil\}$ for a sufficiently large
constant $C$ gives constant bias using one query round.

For the lower bound, choose the two promise cases with equal
probability and draw the input uniformly within the chosen case.
It suffices to consider $t\le N$, since $t>N$ already gives the
claimed lower bound. Fix the algorithm's random coins and answer
repeated queries from its transcript.
In the $k$-to-one case, conditional on observing $j$ distinct values,
the next fresh answer repeats one of them with probability
$j(k-1)/(n-j)$. Conditional on no repetition, the new value is
uniform among the $q-j$ unused alphabet symbols, just as in the
injective case. We can therefore couple the two query transcripts
to agree until the first repeated value, even when query positions
are chosen adaptively. A union bound on the probability $p_{\rm rep}$
of a repetition gives
\[
 p_{\rm rep}
 \le\sum_{j=0}^{t-1}\frac{j(k-1)}{n-j}
 \le\frac{(k-1)t(t-1)}{2(n-t+1)}
 \le\frac{kt^2}{n},
\]
where the last inequality uses $t\le N\le n/2$.
When the coupled transcripts agree, the algorithm makes the same
decision in both cases. Its average success probability is therefore
\[
 p_{\rm succ}\le\frac12+\frac{p_{\rm rep}}2
 \le\frac12+\frac{kt^2}{2n}.
\]
Averaging over the random coins preserves this bound. A constant
worst-case bias consequently requires
$t=\Omega(\sqrt{n/k})$, proving the adaptive lower bound.
\end{proof}
}

\section*{AI methodology}
\label{sec:aimethodology}
\addcontentsline{toc}{section}{AI methodology}

\rev{ChatGPT 6 Astra assisted in developing the harmonic framework for the collision problems and in constructing proofs of the adaptive and one-round quantum lower bounds. Subsequent AI-assisted interactions helped refine the arguments and improve the exposition. The authors independently verified, developed, and simplified all proposed arguments and take full responsibility for the correctness, presentation, and attribution of the work.}

\rev{We further note that an earlier version of this work titled ``\emph{Adaptivity is Necessary for Quantum Speedups in Collision Problems}'', containing only the non-adaptive lower bounds, was submitted to a conference in July 2026. Those proofs used a separate framework based on \cite{belovs2026tight}. In September 2026, we independently established the adaptive lower bound using the harmonic framework through extensive interactions with ChatGPT. The present paper adopts this framework to unify the adaptive and non-adaptive analyses. During a later AI-assisted review of the manuscript, ChatGPT pointed out potential technical overlap with \cite{carolan2026revisited} and \cite{kim2026soundness}; we therefore added a dedicated section comparing our results and techniques with these two works.}

\bibliographystyle{alpha}
\bibliography{ref}
\addcontentsline{toc}{section}{References}

\end{document}